\documentstyle[12pt,epsf,epsfig]{article}
\newfont{\thiplo}{msbm10 scaled\magstep 2}
\newfont{\gothic}{eufb10 scaled\magstep 2}
\newfont{\unc}{eurb10} 
\newskip\humongous \humongous=0pt plus 1000pt minus 1000pt
\def\caja{\mathsurround=0pt}
\def\eqalign#1{\,\vcenter{\openup1\jot \caja
        \ialign{\strut \hfil$\displaystyle{##}$&$
        \displaystyle{{}##}$\hfil\crcr#1\crcr}}\,}
\newif\ifdtup

\def\eqright #1\cr{\noalign{\hfill$\displaystyle{{}#1}$}}
\def\eqleft #1\cr{\noalign{\noindent$\displaystyle{{}#1}$\hfill}}

\def\oldreffmt#1{\rlap{[#1]} \hbox to 2\parindent{}}

\def\figfmt#1{\rlap{Figure {#1}} \hbox to 1in{}}

\def\sectioneq{\def\theequation{\thesection.\arabic{equation}}{\let
\holdsection=\section\def\section{\setcounter{equation}{0}\holdsection}}}%

\newcounter{holdequation}

\def\auto{\eqno(\refstepcounter{equation}\theequation)}
\def\begineq #1\endeq{$$ \refstepcounter{equation}\eqalign{#1}\eqno
	(\theequation) $$}
\def\contlimit{\,{\hbox{$\longrightarrow$}\kern-1.8em\lower1ex
\hbox{${\scriptstyle (a\rightarrow0)}$}}\,}
\def\centeron#1#2{{\setbox0=\hbox{#1}\setbox1=\hbox{#2}\ifdim
\wd1>\wd0\kern.5\wd1\kern-.5\wd0\fi
\copy0\kern-.5\wd0\kern-.5\wd1\copy1\ifdim\wd0>\wd1
\kern.5\wd0\kern-.5\wd1\fi}}
\def\centerover#1#2{\centeron{#1}{\setbox0=\hbox{#1}\setbox
1=\hbox{#2}\raise\ht0\hbox{\raise\dp1\hbox{\copy1}}}}
\def\centerunder#1#2{\centeron{#1}{\setbox0=\hbox{#1}\setbox
1=\hbox{#2}\lower\dp0\hbox{\lower\ht1\hbox{\copy1}}}}
\def\lsim{\;\centeron{\raise.35ex\hbox{$<$}}{\lower.65ex\hbox
{$\sim$}}\;}
\def\gsim{\;\centeron{\raise.35ex\hbox{$>$}}{\lower.65ex\hbox
{$\sim$}}\;}
\def\st#1{\centeron{$#1$}{$/$}}

\def\super#1{\ifmmode \hbox{\textsuper{#1}}\else\textsuper{#1}\fi}
\def\textsuper#1{\newcount\holdspacefactor\holdspacefactor=\spacefactor
$^{#1}$\spacefactor=\holdspacefactor}

\def\getcite#1,{\advance\citenumber by1
\def\getcitearg{#1}\def\lastarg{@}
\ifnum\citenumber=1
\ref{#1}\let\next=\getcite\else\ifx\getcitearg\lastarg\let\next=\relax
\else ,\ref{#1}\let\next=\getcite\fi\fi\next}

\def\pom{{\rm P\kern -0.53em\llap I\,}}
\def\spom{{\rm P\kern -0.36em\llap \small I\,}}
\def\sspom{{\rm P\kern -0.33em\llap \footnotesize I\,}}

\relax
\def\contlimit{\,{\hbox{$\longrightarrow$}\kern-1.8em\lower1ex
\hbox{${\scriptstyle (a\rightarrow0)}$}}\,}
\def\upon #1/#2 {{\textstyle{#1\over #2}}}
\relax
\renewcommand{\thefootnote}{\fnsymbol{footnote}} 

\def\mainhead#1{\setcounter{equation}{0}\addtocounter{section}{1}
  \vbox{\begin{center}\large\bf #1\end{center}}\nobreak\par}
\sectioneq
\def\subhead#1{\bigskip\vbox{\noindent\bf #1}\nobreak\par}

\def\autolabel#1{\auto\label{#1}}
\def\til#1{\centeron{\hbox{$#1$}}{\lower 2ex\hbox{$\char'176$}}}
\def\tild#1{\centeron{\hbox{$\,#1$}}{\lower 2.5ex\hbox{$\char'176$}}}
\def\sumtil{\centeron{\hbox{$\displaystyle\sum$}}{\lower
-1.5ex\hbox{$\widetilde{\phantom{xx}}$}}}

\def\q{\unc q}
\def\p{\unc p}

\newcommand{\bit}{\begin{itemize}}
\newcommand{\eit}{\end{itemize}}

\newcommand{\beq}{\begin{equation}}
\newcommand{\eeq}{\end{equation}}
\newcommand{\beqa}{\begin{eqnarray}}
\newcommand{\eeqa}{\end{eqnarray}}

\begin{document} 
\begin{titlepage} 

\rightline{\vbox{\halign{&#\hfil\cr
&ANL-HEP-PR-99-102\cr
&\today\cr}}} 
\vspace{0.25in} 

\begin{center} 
 
{\large\bf 
THE TRIANGLE ANOMALY IN TRIPLE-REGGE LIMITS }\footnote{Work 
supported by the U.S.
Department of Energy, Division of High Energy Physics, \newline Contracts
W-31-109-ENG-38 and DEFG05-86-ER-40272} 
\medskip

Alan. R. White\footnote{arw@hep.anl.gov }

\vskip 0.6cm

\centerline{High Energy Physics Division}
\centerline{Argonne National Laboratory}
\centerline{9700 South Cass, Il 60439, USA.}
\vspace{0.5cm}

\end{center}

\begin{abstract} 

Reggeized gluon interactions due to a single quark loop are studied in the
full triple-regge limit and in closely related helicity-flip helicity-pole
limits. Triangle diagram reggeon interactions are generated that include local
axial-vector effective vertices. It is shown that the massless quark triangle
anomaly is present as a chirality-violating infra-red divergence
in the interactions generated by maximally non-planar Feynman diagrams. 
An asymptotic dispersion relation formalism is developed which 
provides a systematic counting of anomaly contributions. The asymptotic
amplitude is written as a sum over dispersion integrals of triple
discontinuities, one set of which is unphysical and can produce chirality
transitions. The physical-region anomaly appears in the generalized real parts,
determined by multi-regge theory, of the unphysical discontinuities. The
amplitudes satisfy a signature conservation rule that
implies color parity is not conserved by vertices containing the
anomaly. In the scattering of elementary quarks or gluons
the signature and color parity of the exchanged reggeon states are such that
the anomaly cancels. At lowest-order, it cancels in individual diagrams after
the transverse momentum integrations are performed.

\end{abstract}

\renewcommand{\thefootnote}{\arabic{footnote}} \end{titlepage}

\mainhead{1. INTRODUCTION}

Multi-regge limits within QCD have the virtue that they
are, a-priori, close to perturbation theory at large transverse momentum
while in the infra-red transverse momentum region very strong constraints 
imposed by analyticity and $t$-channel unitarity must also be
satisfied\cite{gpt,arw1}. Many calculations\cite{fkl}-\cite{arw2} have shown
that if gluons and quarks are given a mass via spontaneous symmetry breaking
the unitarity constraints are satisfied perturbatively, in an
elegant and minimal manner, by reggeon diagrams containing only reggeized 
gluons and quarks. If there are circumstances in which the symmetry breaking 
can be removed smoothly we may hope to see an accompanying transition
to reggeon diagrams containing hadrons and the pomeron with, ideally, a
connection to perturbation theory maintained at large transverse momentum.

The purpose of this paper is to demonstrate that when quarks are massless 
many 
high-order reggeized gluon interactions contain an infra-red divergence
that can be understood as the infra-red appearance\cite{cg} of the U(1)
quark anomaly. Although, of course, QCD
contains only vector interactions, in multi-regge limits 
effective vertices are generated by quark loops which involve products of 
$\gamma$-matrices. The full triple-regge limit\footnote{This is a limit of 
three-to-three scattering amplitudes\cite{gw}, not to be confused with the
incorrectly named ``triple-regge'' limit of the one-particle inclusive 
cross-section that is actually a ``non-flip helicity-pole'' limit.} is
sufficiently intricate (as are the helicity-flip helicity-pole limits that 
we also study) that both the axial-vector couplings and the orthogonal momenta
needed to generate the triangle anomaly are present. Since triple-regge
vertices appear as components in a wide array of multi-regge reggeon
diagrams\cite{arw98} this is a new manifestation of the U(1) anomaly which we
believe plays a crucial dynamical role in producing a transition to 
hadron and pomeron reggeon diagrams.

We present direct calculations\footnote{In a companion paper\cite{arw99} we
present an abbreviated version of the central calculation together with a very
brief overview of other arguments in this paper.}
showing that the anomaly is present in 
the triple-regge six-reggeon interaction vertex obtained from the 
``maximally non-planar'' Feynman diagrams that appear in three-to-three
quark scattering. These diagrams contain a single quark loop and the anomaly 
appears because an unphysical singularity combination 
in which every quark propagator in the loop is on-shell approaches the 
asymptotic region. The configuration in which one quark in the loop
carries zero momentum and undergoes a chirality transition produces the 
(infra-red) anomaly. There are, however, many obvious possibilities for a
cancelation. We have to sum over the different choices for 
the quark that carries zero momentum, 
over all diagrams of this kind, and
finally, over all other kinds of diagrams as well.
While non-planar quark loop diagrams provide the essential
analytic structure of regge cut couplings, other diagrams are needed 
for the reggeon Ward identity cancelations\cite{arw98} 
that (indirectly) reflect the underlying gauge
invariance. A-priori such cancelations might be expected to include
cancelation of the anomaly. However, the reggeon Ward identities 
include gluon self-interaction contributions that 
can not produce the chirality transition involved in the anomaly divergence.
As a result, reggeon
Ward identities are violated by the anomaly and do not prevent it's 
occurrence. (In an 
abelian theory the corresponding Ward identities do produce a cancelation.)

Chirality transitions are well-known to be produced by non-perturbative
interactions, such as those due to instantons. From the point of view of the
dispersion theory on which we ultimately base our analysis, the anomaly
appears in the generalized real parts that multi-regge theory provides for
perturbatively calculated unphysical asymptotic discontinuities. It is just
because the multiple discontinuities are unphysical that they can contain
chirality transitions.
Our hope is to eventually show that, in appropriate circumstances,
the chirality violating processes dominate the 
soft background to a hard scattering process and, in doing so, provide 
a fundamental origin for the parton model outside of 
leading-twist perturbation theory. However, a more immediate property that 
must first be established is that the chirality violation produced by a single
reggeon interaction cancels in elementary scattering processes where it 
clearly should not appear. This certainly includes 
helicity conserving processes that have
only perturbative QCD ingredients for accompanying interactions and may well
extend to any process where the chirality violation can not be linked to
(reggeized) gluon configurations with the quantum numbers of the winding-number
current. 

Since multi-reggeon ``states'' are virtual, exchanged, configurations that do
not directly produce particle states, the chirality and reggeon Ward identity
violation associated with
the anomaly does not produce any fundamental conflict that requires a
cancelation within a reggeon vertex. 
Rather such cancelations are secondary
effects within the full scattering process that have to be traced. The number
of Feynman diagrams contributing to even the lowest-order three-to-three quark
scattering processes of the kind we study is very large 
($O(100)$) and some diagrams, the maximally non-planar diagrams in particular,
produce several anomaly contributions.
 Therefore, even though we make no attempt to calculate 
the full reggeon interaction vertex, understanding
diagrammatically when the anomaly occurs and how and when the 
necessary cancelations take place would be very difficult. 

Fortunately, we are able to systematically count all anomaly contributions
by using the asymptotic dispersion relation formalism developed
in \cite{arw1} and \cite{sw}. In this formalism the
full asymptotic amplitude is constructed as a relatively simple sum over
dispersion integrals of multiple discontinuities.
Multi-regge theory then allows the multiple discontinuities to be converted to
amplitudes containing generalized real parts by introducing appropriate
signature factors and, for the lowest-order 
amplitudes we consider, the signature factors have a particularly trivial form.
A very important feature of the asymptotic dispersion relation we use, which 
is not present in the simpler case of multi-regge production processes, is 
that there is a set of unphysical triple discontinuities that contribute.
Indeed we find that the anomaly appears only in the amplitudes given by 
multiple discontinuities of this kind 
obtained from the maximally non-planar diagrams and, also, the closely
related diagrams required by reggeon Ward identities. Once the 
discontinuities involved have been isolated, the study of
cancelations reduces to a discussion of signature and the symmetry
properties of color factors.

The amplitudes that contain the anomaly have special analytic properties.
In particular, they satisfy a very important
signature conservation rule. (Although we do not discuss it in this paper,
we expect this rule to lead to the even signature of the pomeron in hadronic
reggeon diagrams.) The signature rule implies that the anomaly
chirality transition be accompanied by a color parity violation and, most
likely, requires that all reggeon states coupling to the anomaly carry
anomalous color parity (not equal to the signature). 
For color zero reggeon states, anomalous color parity implies the quantum 
numbers of the winding-number current for either the complete state or
a sub-component. In addition, color parity violation by the anomaly vertex 
requires a symmetric $d$-tensor and so requires at 
least SU(3) for the gauge group.  
When the 
external scattering states are elementary quarks (or gluons) anomalous 
color parity reggeon states can not appear and the anomaly cancels.
At lowest-order, it cancels in individual diagrams after
the transverse momentum integrations are performed.

Far more important, of course, is determining when the anomaly does not cancel. 
For high-order multi-regge amplitudes that have 
clusters of particles in initial and final states there is, as we briefly
elaborate in Section 7, no reason for anomalous color parity reggeon states
and the anomalous reggeon interactions to
cancel and they are likely to be a pervasive phenomenon. However,
the infra-red divergences are then suppressed by Ward identity zeroes of
the accompanying interactions. Nevertheless, the associated 
ultra-violet effects of the anomaly should not be
suppressed and we expect the consequence to
be a power (rather than a logarithmic) violation of unitarity bounds.

Avoiding the violation of unitarity by the anomaly is, we believe, the 
core problem in finding the full multi-regge S-Matrix of QCD.
Our proposal, outlined in \cite{arw1}, is that this can be achieved by
enhancing the anomaly in the infra-red region 
so that the ultra-violet effects are dominated by infra-red divergences that 
can be absorbed into the definition of reggeon
states. To achieve this is very subtle.
The anomaly does produce infra-red divergences if an 
anomalous color parity ``reggeon condensate'' (with the quantum numbers of the 
winding-number current\footnote{This ``condensate''
is actually a ``wee-parton'' contribution in a physical reggeon state, 
rather than a vacuum condensate and so need not be parity
violating, as a true vacuum winding-number condensate surely would be.})
is introduced. In the program outline we gave previously\cite{arw98} 
we demonstrated that in 
a color superconducting phase of QCD (with
the gauge symmetry broken from SU(3) to SU(2)) such a condensate can be
consistently reproduced in all reggeon states by anomaly infra-red
divergences. We also showed how the perturbative reggeon diagrams are replaced
by diagrams containing hadrons and a Reggeon Field Theory (RFT) supercritical
pomeron\cite{arw1}, with restoration of the full SU(3) symmetry 
producing, in principle, the RFT critical pomeron\cite{cri}.

In \cite{arw98} we assumed the existence  of the anomaly.
While the properties we assumed 
were essentially correct there are significant differences. 
Having understood the full structure of the anomaly we hope to implement our
previously outlined program in detail in future papers.
If we successfully obtain a unitary (reggeon) S-Matrix as
we hope, it will be very close to perturbation theory, and the connection with
the parton model should be clear. In effect, the non-perturbative properties
of confinement and chiral symmetry breaking will be obtained as a consequence of
regulating the anomaly so that unitarity is satisfied in the regge region.

In this paper, apart from brief discussions in Sections 2 and 7, 
we will not enlarge on what we believe to be the
dynamical role of the anomaly divergences. Instead we will focus
entirely on the technical problem of studying the asymptotic behavior of
Feynman diagrams, setting up the necessary multi-regge formalism, and
isolating the occurrence of the anomaly. 
We have organized the paper in a manner that we hope will allow a reader to
extract some general understanding of our results without necessarily
absorbing all of the underlying multi-regge theory. Section 2 is a general 
outline of the purpose of the paper and a summary of it's contents that, as 
far as possible, avoids technical language. Section 3 describes the
triple-regge and related helicity-pole limits in terms of light-cone
variables. Section 4 is devoted to the calculation, using light-cone
co-ordinates, of triple-regge contributions from three specific diagrams.
This allows us to illustrate how the anomaly occurs as an infra-red
divergence of reggeon vertices. We concentrate on the kinematic structure of
diagrams and ignore color factors until we have set up the necessary
machinery to discuss cancelations. We study one diagram that obviously does
not contain the anomaly, one that might have anomaly contributions 
but actually does not and one, a maximally non-planar diagram, that does. At
the end of the Section we discuss how the anomaly contributions from
maximally non-planar diagrams cancel. In Section 5 we develop the asymptotic
dispersion relation and multi-regge formalism that ultimately allows us to
systematically discuss all anomaly contributions. In Section 6 we study the
complete set of double discontinuities and conclude that only those
originating from maximally non-planar diagrams, and diagrams closely related 
by reggeon Ward identities, 
give amplitudes that contain the anomaly. We finally discuss the role of color
factors in cancelations in Section 7. We then briefly discuss diagrams which
give anomaly contributions that we do not expect to be canceled.

\newpage

\mainhead{2. OUTLINE AND SUMMARY} 

The triple-regge limit (and closely related helicity-flip
helicity-pole limits\footnote{A helicity-pole limit isolates the leading 
helicity amplitude that ultimately gives a physical particle amplitude.}) 
can be formulated as the high-energy, near-forward, scattering of 
three particles carrying light-like 
momenta $P_1,P_2$ and $P_3$ whose spacelike components are orthogonal to
each other. 
This limit (defined precisely in the next Section)  
is discussed in \cite{arw98} for some simple diagrams
but otherwise has not been discussed in QCD. 
In this paper we will study Feynman diagrams of the kind illustrated in 
Fig.~2.1 in which the three particles scatter via gluon 
interactions involving a single quark loop - the solid circle. 
\begin{center}
\leavevmode
\epsfxsize=2.7in
\epsffile{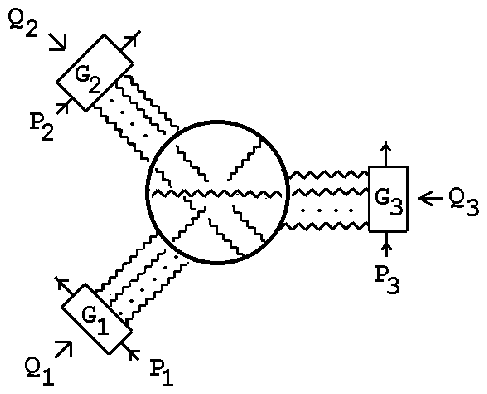}

Fig.~2.1 The Class of Feynman Diagrams Studied
\end{center}
In most of our discussion the scattering particles will be 
single quarks and the 
couplings $G_1$, $G_2$ and $G_3$ will be the lowest-order elementary 
couplings. However, in discussing anomaly cancelations we will also allow 
these couplings to have more general properties, including the production
and/or absorption of additional particles.

The quark loop initially contains a sufficiently large
number of quark propagators that there are no ultra-violet divergences.
At finite momentum, this loop also has no infra-red divergences, even 
when the quark mass is zero. If the gluons are massive, the gluon loops
also have no divergence problems. For most of our analysis we will, for
simplicity, set the gluon mass to zero. This means that the diagrams we study
will formally have infra-red divergences at zero gluon transverse momentum, 
just where the anomaly divergence occurs. Ultimately the interplay
between these divergences is crucial and has to be discussed in detail. 
(It is well-known that the gluon infra-red divergences cancel for 
reggeon states carrying zero $t$-channel color but do not produce confinement.)
In Section 4 we will briefly mention using gluon mass(es) to avoid 
anomaly cancelations. In the main body of the paper we 
simply ignore the divergences due to the zero mass of the gluon.

In the limits we consider the most important contributions come from 
regions of the gluon loop integrations where a number of the 
propagators in the quark loop and the scattering quark systems
are either on-shell or close to on-shell. 
We will be particularly interested 
in diagrams for which, with 
appropriate quantum numbers in the $t_i~(= Q_i^2)$ channels, 
all the relevant quark lines are precisely on-shell in the leading 
contribution. (We discuss below which propagators are involved.) 
If the propagator poles are used to carry out
light-cone longitudinal momentum integrations the integrals over gluon loop
momenta reduce to two-dimensional ``transverse momentum'' integrals. 
The leading contribution then has the form
$$
\eqalign{ ~~~~~P_{1^+}~ P_{2^+}~ P_{3^+}~
\prod_{i=1}^3 \int & { d^2 k_{i1}d^2 k_{i2}  
~ \cdots \delta^2 (Q_i -  k_{i1} -  k_{i2} - \cdots)~G_i(k_{i1},k_{i2},\cdots)
\over k_{ i1}^2  k_{i2}^2 ~\cdots }\cr &~~~~~~~~\times ~ R(Q_1,Q_2,Q_3,
k_{11}, k_{12}, \cdots )} \auto \label{211}
$$
(Note that, in
contrast to simpler multi-regge limits, the transverse momenta in
each integral can not be taken to be in a common plane.)

Provided that $\alpha_i = 1 +~O(g^2), ~i=1,2,3$ we can write 
$$
\eqalign{~~~~~~~P_{1^+}~& P_{2^+}~ P_{3^+}~\sim ~S_{12}^{1/2}~S_{23}^{1/2}~
S_{31}^{1/2}~\cr
&= ~(s_{13})^{(\alpha_1+\alpha_3-\alpha_2)/2}
(s_{23})^{(\alpha_2+\alpha_3-\alpha_1)/2}
(s_{12})^{(\alpha_1+\alpha_2 -\alpha_3)/2} ~~+~ O(g^2) }
\auto\label{212}
$$
where $S_{ij}=(P_i + P_j)^2$. This is the 
lowest-order triple-regge behavior 
for the amplitudes that interest us (and, in particular, potentially contain
the anomaly). Consequently, the transverse momentum 
integrations, together with the gluon propagators and the external couplings 
$G_i$, are straightforwardly interpreted as the leading-order contribution of 
multi-reggeon states in which each gluon is regarded as a lowest-order 
reggeon. As illustrated in Fig.~2.2, 
\begin{center}
\leavevmode
\epsfxsize=4.5in
\epsffile{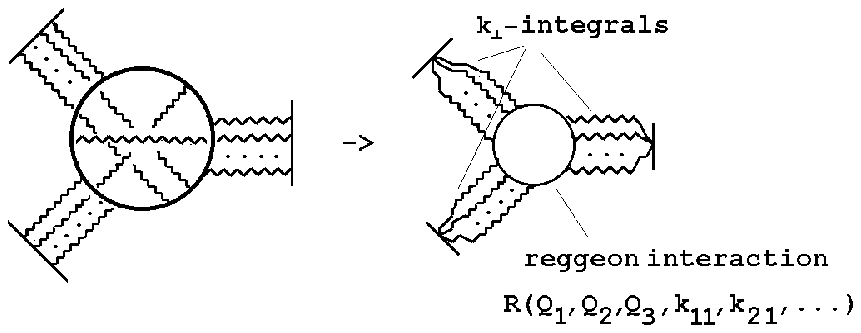}

Fig.~2.2 Generation of a Reggeon Vertex 
\end{center}
$R(Q_1,Q_2,Q_3, k_{11},k_{21},\cdots)$ can then 
be extracted as a ``reggeon interaction vertex''. In general, the lowest-order
contribution to this vertex will survive as higher-order corrections add
reggeization effects to the exchanged gluons and modify the $G_i$ couplings.
In particular if there is an infra-red divergence in the lowest-order vertex 
this would also be expected to survive as higher-order effects are added.
Reggeon vertices appear in 
the reggeon diagrams describing a 
wide range of high-energy multi-regge processes\cite{arw98}. Therefore the
general structure of such vertices is very important. In particular any
divergences that are present may have a dynamical significance going far
beyond the low-order circumstances in which we initially discover them.
 
Since one propagator in the quark
loop is placed on-shell for each gluon loop integration, only three of the
original loop propagators are off-shell. The effective vertices produced by
the longitudinal integrations contain, in general, both local and non-local
components. By (our) definition, the local components are products of $\gamma$
matrices that in some cases reduce to $\gamma_5\gamma$ couplings. Clearly,
if there is an odd number of $\gamma_5 $'s then, a-priori, the U(1) triangle
anomaly could be present in the reduced loop. Intrinsically,
reggeon diagrams are most unambiguously defined at low transverse momentum.
Therefore, we look for the infra-red manifestation of the anomaly as a
divergence that is present when the quark mass vanishes\cite{cg}. This 
divergence requires that the remaining three off-shell quarks go on-shell
(producing a complete loop of on-shell quarks).
We will not be able to identify the full Lorentz structure but we will find
the characteristic chirality violation. (At the end of Section 5 we identify the
origin of the unphysical singularity configuration of on-shell quarks that is
able to produce the chirality violation.) Note that analyticity properties of
reggeon vertices imply that if the anomaly is present in the infra-red then it
should also be present as an ultra-violet effect. We briefly discuss the nature
and potential significance of such ultra-violet effects in Section 7.

Gauge-invariance relates diagrams of the form of Fig.~2.1 
to other diagrams involving the triple-gluon coupling. 
We will make brief references to such diagrams in the context of reggeon Ward
identity cancelations. However, we do not 
attempt to calculate reggeon 
vertices corresponding to all diagrams of a fixed-order. 
Rather we concentrate on demonstrating the 
presence of the anomaly in contributions from particular 
diagrams and on determining when and how such contributions 
cancel. The infra-red divergence we are looking for 
requires\cite{cg} a quark triangle Landau singularity and so
diagrams of the kind we have isolated are the important ones.
Most of our discussion will be concerned with the lowest-order 
diagrams, illustrated in Fig.~2.3, in which the scattering states are quarks
and there are just two-gluons exchanged in each $t$-channel. 
This simplest set already contains $O(100)$ diagrams and so 
counting all possibilities will be 
a very difficult thing to do unless we have a very systematic procedure.
\begin{center}
\leavevmode
\epsfxsize=4.5in
\epsffile{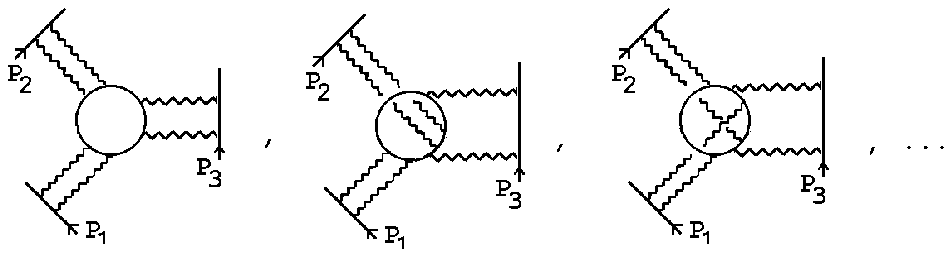}

Fig.~2.3 Quark Scattering Diagrams with Two Gluons in each $t_i$-channel. 
\end{center}

Two conceptually distinct calculational methods can be used to arrive at 
(\ref{211}). The arguments for placing propagators on-shell are related
but differ in important ways that we want to emphasize. The most popular
calculational method is applied directly to Feynman diagrams and utilises
light-cone co-ordinates (or Sudakov parameters). The large light-cone momenta
are routed through a diagram and if a large momentum is carried by a propagator 
it must be on-shell, or close to on-shell, if it is not to (power) suppress the
asymptotic behavior. If there is no corresponding intermediate state 
in which the propagator is on-shell then, in a leading-log
calculation, only the close to on-shell configuration contributes and a real
logarithm is generated. In high-orders a careful discussion of the closing of
longitudinal integration contours in the complex plane is required, to make
sure that the propagator pole involved can not be avoided by the distortion of
an integration contour. In general, the contribution of a real, close to
on-shell, configuration reflects the presence of a cross-channel branch-cut.

The second, much less intuitive, calculation method employs a dispersion
relation\cite{fkl} which contains discontinuities in which the relevant lines
are specifically on-shell. Real amplitudes involving logarithms corresponding
to close to on-shell configurations in a particular channel are reproduced 
by dispersion integrals over the intermediate states in the cross-channel
that they are related to. The
dispersion relation formalism generally has the advantage (particularly in a
gauge theory) that fewer diagrams need to be calculated. In higher-orders, in
principle, dispersion relations for production amplitudes must also 
be introduced. However, the simple relationship between the signature factors
and discontinuities for regge amplitudes can be exploited\cite{bar} to
short-cut, at least part of, such calculations.
The dispersion relation approach sometimes has the disadvantage that a
cancelation which manifests itself via the closing of a contour in the direct
diagrammatic approach can appear as a more elaborate cancellation between
discontinuities, dependent on signature and quantum number properties.
For our purposes, however, the crucial feature of the dispersion relation 
method is that the ambiguity of which 
on-shell configurations contribute is resolved by 
the unambiguous process of taking the necessary
discontinuities. 

As we already implied, because we are interested in the low-order behavior of
a large number of relatively complicated diagrams we will not attempt
a complete diagrammatic analysis. In fact,
although there are three distinct large
momenta to be routed through diagrams, in the configuration that interests us
the crucial quark loop carries finite momentum. This makes the ambiguity
as to which quark propagators should be placed on-shell particularly
serious. 
Fortunately, the asymptotic dispersion relation formalism developed 
in \cite{arw1} and \cite{sw} provides a fundamental basis for 
calculating triple discontinuities and assembling them 
to form the complete asymptotic amplitude. We will see that the structure of
multiple discontinuities, although involving subtleties crucial for the
emergence of the anomaly, is relatively simple and that the problem of
counting contributions from all diagrams becomes straightforward. Indeed,
when the amplitude is regge-behaved,  the relationship between
discontinuities and the full amplitude is such that
reggeon interaction vertices can be extracted from multiple discontinuities
directly. The most important subtlety, for our purposes, is that the 
dispersion relation includes unphysical discontinuities that can contain the
chirality transitions necessary for the anomaly to appear. In fact this feature
can be regarded as the main consequence of the increased complexity
of the triple-regge limit, compared to the multi-regge limits previously
studied.

To illustrate the general idea behind using multi-regge theory to obtain
amplitudes from multiple discontinuities we note that 
when the leading-order amplitude has the form of (\ref{211}) 
discontinuities can be taken trivially using (\ref{212})
$$
[Disc]_{s_{12}} ~~\sim ~ (s_{12})^{1/2}~-~(e^{-2\pi i}s_{12})^{1/2}~
=~2(s_{12})^{1/2}
\auto\label{211dc}
$$
Indeed if (\ref{211}) were derived\footnote{This particular multiple
discontinuity is forbidden by the Steinmann relations, but for pedagogical
reasons we ignore this for the moment.} as an asymptotic  multiple 
discontinuity in $s_{12}, s_{23}$ and $s_{31}$ 
and the momentum behavior interpreted using
(\ref{212}), the asymptotic result has a trivial
extension away from the discontinuity by including the phases due to the
square-root branch-cuts in each of $s_{12}, s_{23}$ and $s_{31}$. In particular
the amplitude can be extended 
to negative values of the invariants where the amplitude is real and there
are no discontinuities. 
Of course, (\ref{212}) is only one possible way to 
write the large momentum factor ($P_{1^+}P_{2^+}P_{3^+}$) in terms of large 
invariants. To justify this particular choice 
it is necessary to calculate higher-order corrections and see
the appropriate reggeization effects appear. 
The asymptotic dispersion relation provides a
sum over all allowed possibilities and multi-regge
theory incorporates the higher-order corrections and 
generalizes the extension of the amplitude away from the discontinuities
via the introduction of phases and signature factors.

As part of our effort to organize the paper to provide some benefit for a 
general reader we begin, in Section 3, by formulating the 
triple-regge limits we discuss in terms of light-cone kinematics. As a 
result, in Section 4 we are able to initially discuss some diagrams 
directly in terms of light-cone co-ordinate calculations without developing 
the multiple discontinuity formalism. This allow us to 
illustrate how the anomaly occurs. We study all three diagrams shown 
explicitly in Fig.~2.3. 
As described in Appendix A, 
the anomaly appears as an infra-red divergence of the triangle
diagram when an odd number of axial-vector couplings
is present and a lightlike momentum and orthogonal
spacelike momenta flow through
the diagram - with the spacelike momentum scaled to zero. 
The first diagram of Fig.~2.3, fairly obviously, does not 
contain the anomaly since it generates only vector effective vertices. 
The 
second diagram contains a $\gamma_5$ effective vertex, but the necessary 
light-like momentum can not flow through the diagram. This illustrates the
general point that, while several diagrams generate quark triangle reggeon
vertices with the necessary effective axial-vector couplings, in most cases
the longitudinal integrations produce additional effects that either prevent
the occurence of the divergence, or lead to a cancelation. 

The third diagram of Fig.~2.3 actually gives more than one 
reggeon interaction contribution containing the anomaly. 
When this diagram is 
redrawn as in Fig.~2.4, 
\begin{center}
\leavevmode
\epsfxsize=3.3in
\epsffile{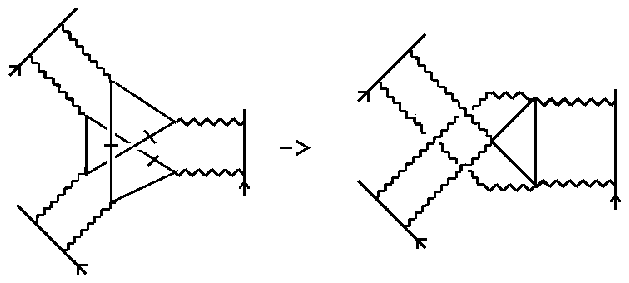}

Fig.~2.4 A Diagram with an Anomaly Contribution. 
\end{center}
the ``maximally non-planar'' property is apparent.
(The couplings to the quark loop by the two gluons in the same $t$-channel 
are separated, in both directions around the loop, by couplings to gluons in 
the other two $t$-channels.)
As we already alluded to above, this non-planarity property 
ensures that such diagrams unambiguously contribute to regge cut vertices. 
When the
hatched lines are placed on-shell by the gluon loop longitudinal
integrations a triangle diagram reggeon interaction is generated as shown. 
The local coupling component is shown in Fig.~2.5. 
\begin{center}
\leavevmode
\epsfxsize=4.5in
\epsffile{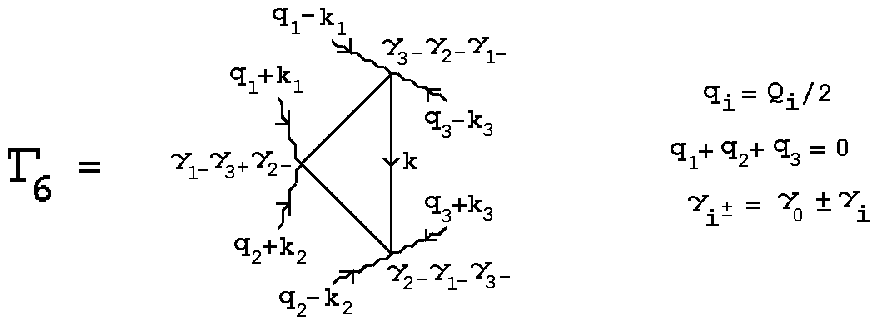}

Fig.~2.5 A Triangle Diagram Reggeon Interaction. 
\end{center}
In obtaining these local couplings we have used the 
special light-cone co-ordinates discussed in Appendix B. 
It is straightforward to show that the necessary 
$\gamma_5$ couplings are present within the products of $\gamma$-matrices
shown.

To illustrate how the appropriate momentum configuration for the anomaly 
appears we first define the light-like vector
$$
\underline{n}_{lc}~=~ (1, cos {\theta}_{lc}~, sin {\theta}_{lc}~, 0)
\auto\label{lcm1}
$$ 
and the orthogonal space-like vector
$$
\underline{n}_{lc\perp}~=~ (0, - sin {\theta}_{lc}~,cos {\theta}_{lc}~,  0)
\auto\label{olcm1}
$$ 
We then take 
$$
\eqalign {q_1 + k_1 + q_2 + k_2~&=~O(\hbox{\q})~\underline{n}_{lc\perp} \cr
q_2 -k_2+q_3 +k_3 ~&= ~l~\underline{n}_{lc} ~+~
O(\hbox{\q})~\underline{n}_{lc\perp} \cr
q_1 -k_1+q_3 -k_3~&= ~- l~\underline{n}_{lc} ~+~
O(\hbox{\q})~\underline{n}_{lc\perp} }
\auto\label{250}
$$
We also take the loop momentum $k \sim O(\hbox{\q})$ 
and let $\hbox{\q} \to 0 $ with 
$$
(q_1 - k_1) ~\to~-2 l~(1,1,0,0)~, ~~~~~(q_2- k_2) ~\to~2 l~(1,0,1,0)~,
\auto\label{2510}
$$
and 
$$
q_3 ~\to ~l~(0,1-1,0)~, ~~~~~k_3~\to~l~(0,1-2cos {\theta}_{lc}~, 
1-2 sin {\theta}_{lc}~, 0)
\auto\label{2511}
$$ 
In the limiting configuration 
the momenta corresponding to the hatched lines 
of Fig.~2.4 are on-shell. Also  
$$
q_1^2~=~q_2^2~= ~k_1^2 ~=~k_2^2
\auto\label{lcon}
$$
and only 
the lightlike momentum $k_{lc}=l~ \underline{n}_{lc}$
flows through the triangle graph of Fig.~2.5. As a result, the anomaly 
divergence appears and gives
$$
\Gamma_6 ~~\sim  ~~ {(1 - cos {\theta}_{lc} - sin {\theta}_{lc})^2
~l^2 \over \hbox{\q} } 
\auto\label{5847}
$$

The momentum configuration (\ref{2510}) and (\ref{2511}) describes the 
physical scattering process illustrated in Fig.~2.6(a), where the time axis is 
vertical on the page.
Fig.~2.6(a) is the basic process associated with 
the anomaly in the reggeon vertices obtained from the 
lowest-order graphs. 
The dashed lines indicate light-like (``wee
parton'') gluons, one incoming produced by an incoming quark and one outgoing
that is absorbed by
an outgoing quark. A zero-momentum quark (indicated by the open line) is emitted
by the incoming wee-parton gluon, undergoes a chirality transition, and then 
is absorbed by the outgoing wee-parton gluon. The accompanying antiquark has
it's incoming light-like momentum pointed along $\underline{n}_{lc}$ by
scattering off a spacelike gluon. It then forward scatters off two more
spacelike gluons before another
scattering points it's lightlike momentum in the outgoing direction.
\begin{center}
\leavevmode
\epsfxsize=4in
\epsffile{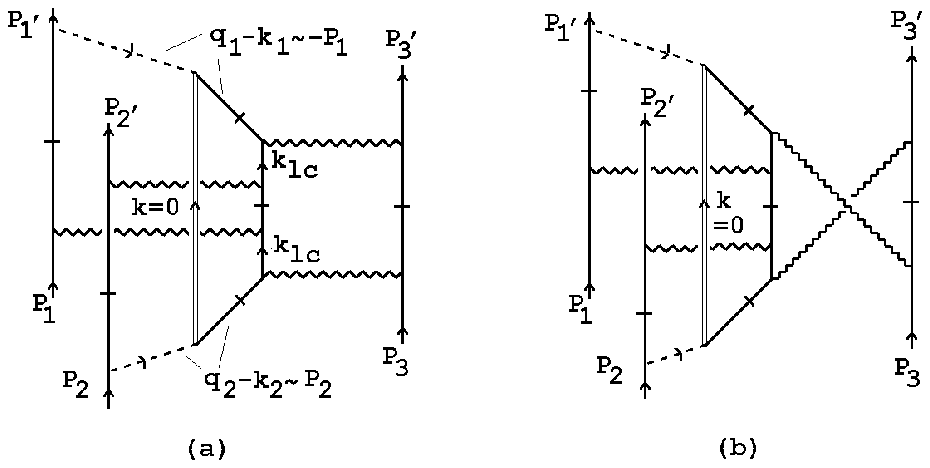}

Fig.~2.6 Physical Scattering Processes Involving the Anomaly
\end{center}c

The hatched lines are put on-shell
in extracting the reggeon interaction. In the momentum configuration we have 
given, the hatched light-like antiquark states necessarily give the
$\gamma$-matrix couplings shown in Fig.~2.5. 
However, it is clear from
Fig.~2.6(a) that there is no physical intermediate state in the scattering
process to which the corresponding 
on-shell propagators can contribute. It is essential that the 
amplitude with these lines on-shell is obtained by using multi-regge theory 
to extrapolate an amplitude away from an (unphysical) 
multiple discontinuity as we discussed 
above. That the initial and 
final quark/antiquark pair can have the net chirality due to
effective axial couplings can be regarded as a consequence of the hatched 
lines having been placed on-shell in an unphysical region.

That the divergence occurs at $ (q_1 - k_1)^2~ =~(q_2- k_2)^2 ~=~0~$
implies that it mixes with several other effects. Firstly, 
it mixes with the infra-red divergences due to the 
zero mass of the gluon that, in this paper, we are ignoring. More interesting
are reggeon Ward identity zeroes
that should occur in the reggeon vertex.
The derivation\cite{arw98} of the
identities giving these zeroes combines regge pole factorization with 
the underlying gauge invariance Ward identities. Provided there are no
transverse momentum singularities of the reggeon vertex involved, the zeroes
should occur at $ (q_1 - k_1)^2 =0$ or $(q_2- k_2)^2 =0$, independent of 
any light-like momenta that are present. The zeroes result
from cancelations that might be expected to also cancel the anomaly. 
As we have emphasized, the
anomaly divergence can only be canceled by interactions that also contain the
same quark loop divergence. 

As we discuss in more detail in Section 6, the diagram giving 
the space-time scattering of Fig.~2.6(b) also  
contains the anomaly,
and in an abelian theory would produce the Ward identity
zeroes. Without color factors these two diagrams cancel, as required, 
at $ (q_1 - k_1)^2~ =~0$ and $(q_2- k_2)^2 ~=~0~$. Since the two diagrams have
identical kinematic structure, this cancelation will include
the anomaly. With color factors, however, 
the two wee-parton gluons can form a color octet state and in this case the
two diagrams of Fig.~2.6 do not cancel. As illustrated in Fig.~C6, the desired
reggeon Ward identity is produced by the addition of a diagram
in which an interaction involving the triple gluon coupling replaces the zero
momentum quark exchange. The triple gluon diagram can not, however, produce the
chirality transition of the anomaly and so the reggeon Ward identity
necessarily fails when the light-like momentum configuration giving the anomaly
dominates. When the external
couplings $G_1$,$G_2$ and $G_3$ involve multiparticle states they also contain
Ward identity zeroes that (unless they also contain the anomaly) will not be
sensitive to the light-like momenta. We briefly discuss in Section 7 how, in
this case, it can be that the anomaly does not cancel but the 
zeroes in the $G_i$ suppress the infra-red divergence effects. The effects
of the anomaly are then, primarily, in the ultra-violet transverse momentum
region.

In lowest-order, although the reggeon Ward-identity fails so that
the anomaly is present in the reggeon vertex, there is actually a 
relatively simple cancelation
within each maximally non-planar Feynman diagram. In the diagram of Fig.~2.4,
we could equally well place on-shell the lines that are hatched in the diagram
of Fig.~2.7. 
\begin{center}
\leavevmode
\epsfxsize=3in
\epsffile{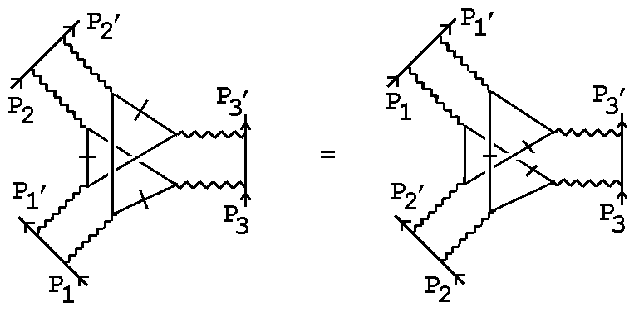}

Fig.~2.7 An Alternative Set of On-shell Lines. 
\end{center}
As we have shown in the figure, this new set of hatched lines can
be related back to the original set simply by interchanging the role of $P_1$
and $P_2$, together with $k_1 \to -k_1$ and $k_2 \to -k_2$. 
In the process of Fig.~2.6 the incoming and outgoing wee parton
gluons are correspondingly interchanged.
As we discuss in Section 4, the interchange
of the roles of $1$ and $2$ (together with $k_1 \to -k_1$ and $k_2 \to -k_2$) 
can be viewed as a parity transformation
that produces a change of sign of the anomaly. However, the rest of the 
reggeon diagram is kinematically insensitive to this
transformation. The large momentum factors also do not change sign and so
if the color factors are appropriately symmetric,
after the transverse momentum integrations are performed, there will be a
cancelation between the two anomaly contributions. 
More generally, a cancelation is obtained after all
maximally non-planar diagrams are added together. If the anomaly is canceled
in the maximally non-planar diagrams then, as we discuss further in Section 6,
the reggeon Ward identities ensure it's complete cancelation.

We can give an illustration of the physical amplitudes that 
we anticipate the anomaly divergence 
ultimately produces, and the relation to the parton model, 
as follows. The divergence (\ref{5847}) appears in the full triple-regge
vertex. If amplitudes involving triple-regge vertices of this
kind are first selected as reggeon amplitudes, with the anomaly divergence
factored off, particle amplitudes will
be obtained (essentially) by taking an additional (helicity-flip helicity-pole)
limit in which $l \to 0$. In this case all the quarks and gluons in Fig.~2.6,
apart from those exchanged in the central scattering, carry zero momentum.
In higher-orders, and in appropriate circumstances, combinations of 
scatterings of the kind illustrated in Fig.~2.8 are produced.
A finite momentum ``parton scattering'' takes place in the background of a zero
momentum scattering via the anomaly. The parton scattering will be momentum
conserving but will be (a component of) a Lorentz vector. 
\begin{center}
\leavevmode
\epsfxsize=3.5in
\epsffile{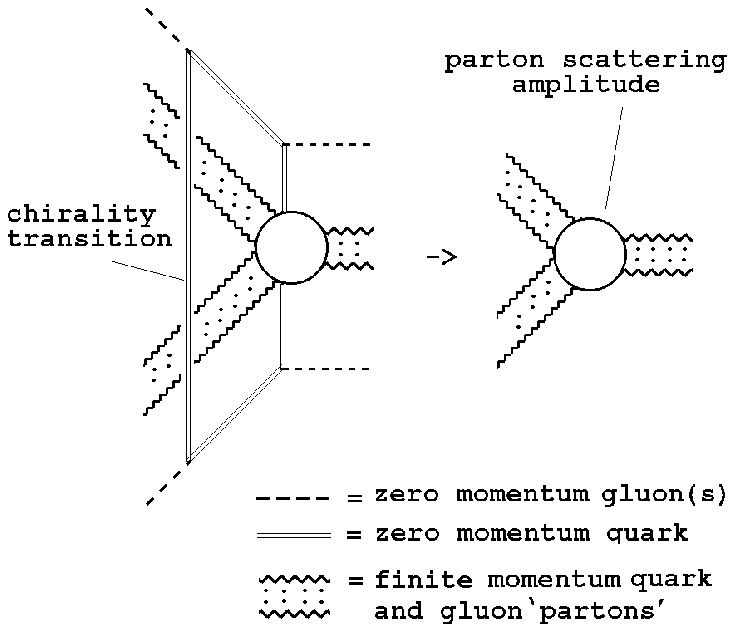}

Fig.~2.8 Parton Scattering Within the Anomaly
\end{center}
To truly isolate such scatterings it is necessary to break the SU(3) gauge
symmetry of QCD to SU(2). The zero momentum scattering should lie in the
unbroken part of the gauge group while the parton scattering lies in the
broken part. In addition the zero momentum gluons should carry the quantum
numbers of the winding number current in each $t$-channel. That the reggeon
diagrams of unbroken QCD can be constructed from such a starting point is the
potential origin of the parton model referred to in the Introduction. In the
unbroken theory the direction, within the gauge group, of the zero momentum
scattering will be averaged over.

In Section 5 we describe the asymptotic dispersion relation that holds in 
the triple-regge limit.
The physical-region triple discontinuities that appear are
relatively simple. Using tree diagrams in which an internal line
represents a channel discontinuity, the triple discontinuities are of three
kinds. 
\begin{center}
\leavevmode
\epsfxsize=2.2in
\epsffile{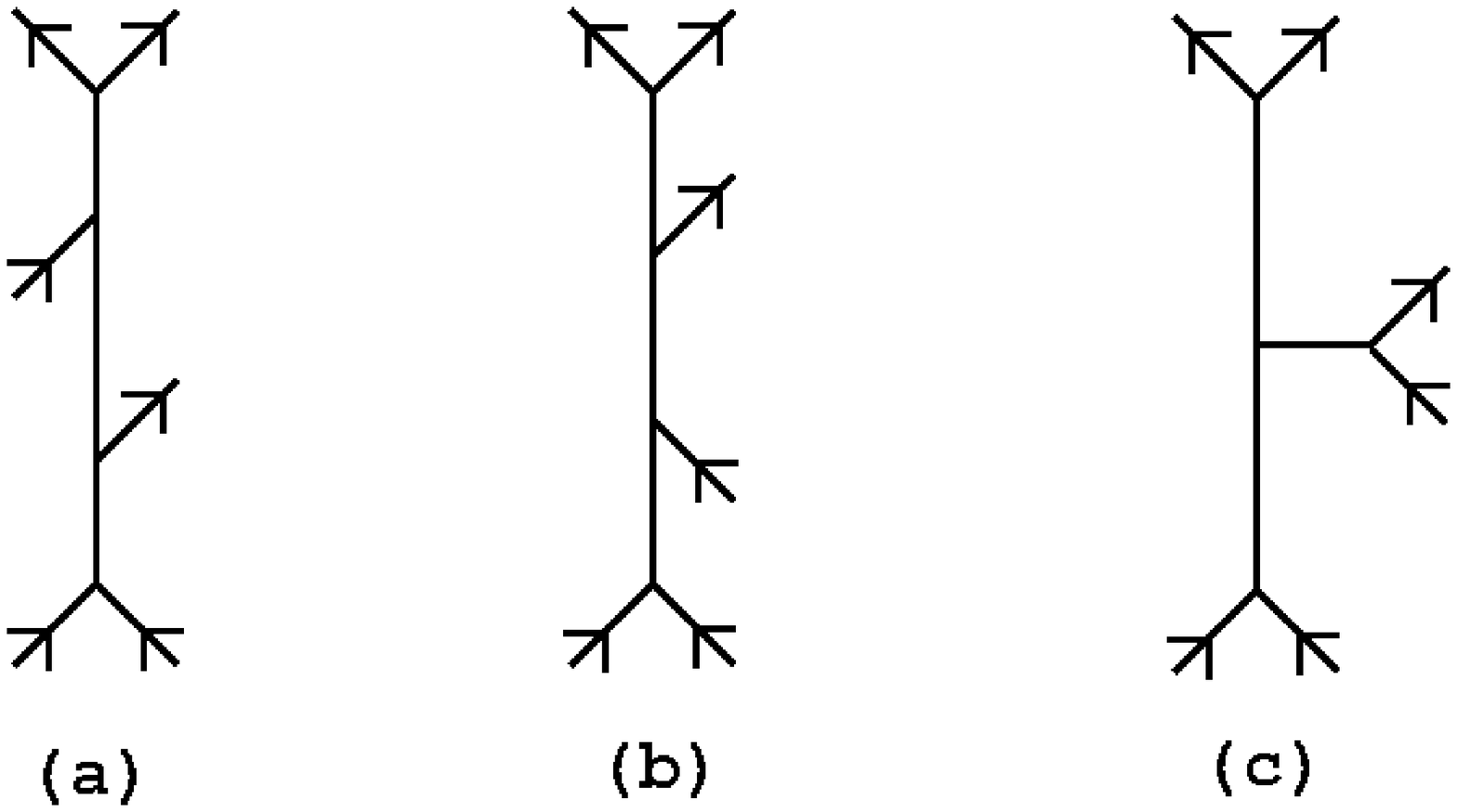}

Fig.~2.9 Tree Diagrams for Triple Discontinuities
\end{center}
There are 24 of the first kind, illustrated in Fig.~2.9(a), that are
related to one-particle inclusive cross-sections (via optical theorems).
There are 12 contributions of the form of Fig.~2.9(b). 
The asymptotic dispersion relation also contains 12 triple
discontinuities of the
form of Fig.~2.9(c) which, unlike those of Fig.~2.9(a) and (b),
do not occur in any of the physical
regions. However, they contribute in a crucial way to the asymptotic behavior
and are essential for the appearance of the anomaly.

The multi-regge Sommerfeld-Watson 
representations are quite different for the triple 
discontinuities of Fig.~2.9(a) and those of Fig.~2.9(b) and (c). 
The 24 combinations of the first kind break up
into three sets in which the eight making up a set combine to form signatured
amplitudes with three possible signatures. Each of the 
sets of Fig.~2.9(b) and (c) provide 
only four distinct signatured amplitudes leading to a 
``signature conservation'' rule. This 
rule is superficially the same as the usual Gribov signature rule for the
triple-regge vertices appearing in elastic scattering 
reggeon diagrams. However, it's origin is
quite different. We anticipate that this signature rule will ultimately lead to
the even signature property of 
the pomeron when we finally extract the physical S-Matrix from reggeon 
diagrams. 

A major part of Section 5 is devoted to the hexagraph notation for counting 
the contribution of multiple discontinuities to the asymptotic dispersion 
relation. The hatched amplitudes of 
Fig.~2.4 and Fig.~2.7 correspond to distinct double
discontinuities and hence to distinct hexagraphs. 
Section 6 has a very simple purpose and result. We study all contributions of 
diagrams of the form of Fig.~2.3 to a particular hexagraph (i.e. to a 
particular physical region double 
discontinuity) and look for the anomaly in the contributions to six-reggeon 
interactions. We show that only the double discontinuities originating from 
a maximally non-planar diagram give
an amplitude that contains the anomaly, apart from the diagram 
that is closely related to the non-planar diagram by reggeon Ward identities.
This implies that to
fully discuss the cancelation of the anomaly we only have 
to add a relatively simple discussion of color factors to our discussion at 
the end of Section 4. This we do in Section 7. 
We also discuss processes in which the anomaly does not cancel but rather
gives predominant ultra-violet effects.

\newpage

\mainhead{3. KINEMATICS - TRIPLE-REGGE LIMITS }

In order to extract the asymptotic behavior of 
Feynman diagrams using familiar light-cone techniques, 
we begin by formulating the triple-regge limits we study using 
light-cone momenta. In Section 5 we will relate this formulation of limits to 
the usual description of multi-regge limits in terms of angular variables. 

\subhead{3.1 Light-Cone Description of the Triple-Regge Limit}

We consider the three-to-three scattering process illustrated in Fig.~3.1(a)
and define momentum transfers $Q_1, Q_2$ and $Q_3$ as in Fig.~3.1(b).

\begin{center}

\leavevmode 
\epsfxsize=4in
\epsffile{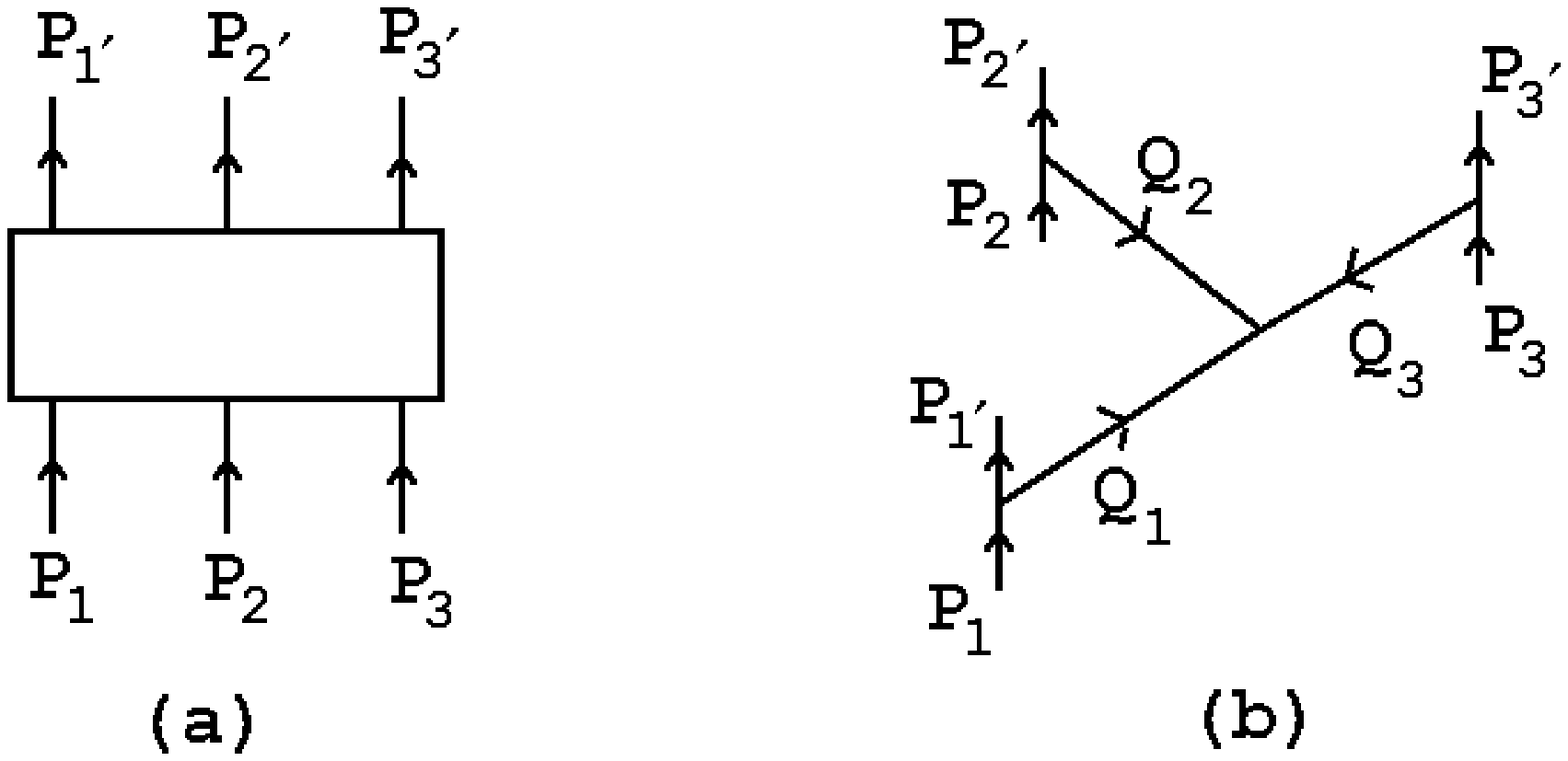}

Fig.~3.1 Three-to-Three Scattering.

\end{center}
Consider first the ``full triple-regge limit'' in which
each of $P_1,~P_2$ and $P_3$ are taken large 
along distinct light-cones, with $Q_1, Q_2$ and $Q_3$ fixed, i.e.
\newline \parbox{3in}{ 
$$
\eqalign{ P_1~\to&~ P_1^+~= ~(p_1,p_1,0,0)~,~~p_1 \to \infty \cr
P_2~\to&~ P_2^+~= ~(p_2,0,p_2,0)~,~~p_2 \to \infty \cr
P_3~\to&~ P_3^+~= ~(p_3,0,0,p_3)~,~~p_3 \to \infty  }
$$}
\parbox{3in}{
$$ \eqalign{
~~~Q_1~\to&~~ (\hat{q}_1,\hat{q}_1,q_{12},q_{13})\cr
~~~Q_2~\to&~ ~(\hat{q}_2,q_{21},\hat{q}_2,q_{23})\cr
~~~Q_3~\to&~~(\hat{q}_3,q_{31},q_{32},\hat{q}_3)}
\auto\label{np3}
$$}
Momentum conservation requires that 
$$
\hat{q}_1 + \hat{q}_2 + \hat{q}_3 = 0,~ \hat{q}_1 + q_{21} 
+q_{31}=0,~ \hat{q}_2 + q_{12} 
+q_{32}=0,~ \hat{q}_3 + q_{13} 
+q_{23}=0
\auto
$$
and so there are a total of five independent $q$ variables which, along  
with $P_1, P_2$ and $P_3$, give the necessary eight variables. 
(Obviously $P_{i'} = P_i - Q_i, ~ i=1,2,3$. Also we omit light-cone components
of both the $P_i$ and the $Q_i$ that go to zero asymptotically, but are
necessary to put the initial and final particles on mass-shell.) 

In terms of invariants, writing $s_{ij} = (P_i + P_j)^2$, $s_{ij'} = 
(P_i-P_{j'})^2$ and 
\newline $s_{i'j'} = (P_{i'} + P_{j'})^2$, the limit (\ref{np3}) gives 
$$
\eqalign{& s_{12}~\sim~s_{1'2'}~ \sim~ - s_{12'} ~\sim ~- s_{1'2} ~
\to ~2p_1p_2~, \cr
& s_{23}~ \sim ~- s_{2'3} ~\to ~2p_2p_3~,~~~
s_{31}~\sim ~s_{3'1} ~\to ~2p_3p_1~, }
\auto\label{npl01}
$$
while for invariants of the form $s_{122'} = (P_1 +Q_2)^2 = (P_1 +P_{2} - 
P_{2'})^2 $
$$
\eqalign{& s_{122'}~ \sim ~2P_1 . Q_2 ~\to ~2 p_1(\hat{q}_2 - q_{21})~, \cr
& s_{133'}~ \sim ~2 P_1 . Q_3 ~\to ~2p_1(\hat{q}_3 - q_{31})~, \cr
& s_{233'}~\to ~2 p_2(\hat{q}_3 - q_{32})~,
~~~s_{311'}~\to ~2 p_3(\hat{q}_1 - q_{13})~, ~ \cdots }
\auto\label{npl1}
$$
Note that there is no constraint on 
the relative magnitudes of the $Q_i$. They can lie in either a spacelike 
plane ($s-s$ in the notation of \cite{gw} and of Appendix D) 
or in a plane with a timelike component ($s-t$ in the same notation).

\subhead{3.2 Light-Cone Description of Helicity-Flip Helicity-Pole 
Limits}

We can take a ``helicity-flip helicity-pole limit'', 
in addition to the triple-regge limit, by also taking
\beqa
{s_{31} \over s_{133'}~s_{311'}} ~~ ~ &\sim& ~~~ { 1 \over 
(\hat{q}_3 - q_{31})(\hat{q}_1 - q_{13})} ~~\to~~\infty \label{npl2} \\
{s_{32} \over s_{233'}~s_{322'}} ~~ ~ &\sim& ~~~ { 1 \over 
(\hat{q}_3 - q_{32})(\hat{q}_2 - q_{23})} ~~~ \to ~ \infty \label{npl3}
\eeqa
Introducing the notation of Appendix B, this limit is therefore equivalent to 
taking 
\beqa
q_{21^-} ~&=&~\hat{q}_2  - q_{21} ~= ~q_{31} - \hat{q}_3~ \to~0 \label{npl4} 
\\
q_{12^-} ~&=&~\hat{q}_1  - q_{12} ~= ~q_{32} - \hat{q}_3~ \to~0 \label{npl33}
\eeqa
With this additional limit taken
\newline \parbox{3in}{ 
$$
\eqalign{ P_1~\to&~ P_1^+~= ~(p_1,p_1,0,0)~,~~p_1 \to \infty \cr
P_2~\to&~ P_2^+~= ~(p_2,0,p_2,0)~,~~p_2 \to \infty \cr
P_3~\to&~ P_3^+~= ~(p_3,0,0,p_3)~,~~p_3 \to \infty  }
$$}
\parbox{3in}{
$$ \eqalign{
~~~Q_1~\to&~~ (q_{112-},q_{112-},q_{112-},q_{13})\cr
~~~Q_2~\to&~ ~(q_{212-},q_{212-},q_{212-},q_{23})\cr
~~~Q_3~\to&~~(q_{33},q_{33},q_{33},q_{33})}
\auto\label{np31}
$$}
where now the constraints of momentum conservation are 
$$
q_{112-} + q_{212-}+ q_{33}= 0~, ~~q_{13} + q_{23} + q_{33}=0
\auto
$$
giving three independent $q$ variables. As our notation indicates, the
helicity-flip  
limit is naturally expressed in terms of the light-cone variables introduced 
in Appendix B.

We can obviously also define additional helicity-pole limits by taking 
\beqa
q_{32^-} ~&=&~\hat{q}_3  - q_{32} ~= ~q_{23} - \hat{q}_1~ \to~0 
~~~~~~~~~ \label{npl41} \\
q_{23^-} ~&=&~\hat{q}_2  - q_{23} ~= ~q_{13} - \hat{q}_1~ \to~0 
~~~~~~~~~ \label{npl331}
\eeqa
or
\beqa
q_{13^-} ~&=&~\hat{q}_1  - q_{13} ~= ~q_{23} - \hat{q}_2~ \to~0 
~~~~~~~~~ \label{npl42} \\
q_{31^-} ~&=&~\hat{q}_3  - q_{31} ~= ~q_{21} - \hat{q}_2~ \to~0 
~~~~~~~~~ \label{npl332}
\eeqa
corresponding to further sets of light-cone co-ordinates defined as in 
Appendix B. Note that for all three helicity-pole limits,
the $Q_i$ must lie in the $s-s$ region, i.e. a 
spacelike plane in which the euclidean constraint
$$
|Q_i|~+~|Q_j|~\geq~|Q_k|~~~~~~~~~ \forall ~i,j,k
\auto\label{euc}
$$
is satisfied. This is necessary for the helicity-flip limit to be a physical 
region limit. In fact the $Q_i$ lie in the $s-s$ region provided only that
$q_{ij^-} $ are sufficiently small.
  
\newpage \mainhead{4. Calculation of Feynman Diagrams}

In this Section we calculate directly contributions to the triple-regge limit
from selected diagrams. We will not attempt to be complete in our discussion
and will not include color factors. The counting of all contributions from all
diagrams has to be done via the multiple discontinuity asymptotic dispersion
relation formalism that we develop in Section 5. In anticipation of this
formalism we first look for on-shell configurations that form
intermediate states in the scattering and extract the corresponding reggeon 
interaction. We will show that the necessary axial triangle diagram 
appears in the contributions of a maximally non-planar diagram. However, when we
subsequently locate the momentum configuration in which the anomaly 
divergence actually occurs we find that it is in a region where the 
discontinuities we have potentially evaluated are no longer present. 

\subhead{4.1 The Simple Planar Diagram}

We begin with the first diagram of Fig.~2.3, which is also discussed in
\cite{arw98}. This planar diagram (almost obviously) contains no anomaly 
and, as we 
discuss shortly, will not contribute at all to the six-reggeon interaction
if all-orders reggeization of quarks and gluons is exploited. Nevertheless,
we begin with it
since it is the simplest to evaluate and to use to illustrate our general
methods. The notation we use is illustrated in Fig.~4.1. 
\begin{center}
\leavevmode
\epsfxsize=2.5in
\epsffile{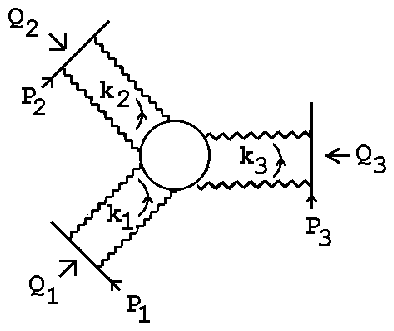}

Fig.~4.1 The Simple Planar Diagram.
\end{center}
We will not specify the direction of the quark line but rather sum over both 
possibilities in the diagrams we discuss. 
Because we are interested in infra-red
contributions from the central quark loop we can suppose that large 
momenta do not flow through this loop. We can then use directly, for each 
scattering quark, arguments (reviewed in Appendix C) that apply when a fast 
quark scatters off a slow system. 

We begin by reducing each loop integral 
involving gluon propagators to a ``transverse momentum'' integral by 
carrying out longitudinal integrations. For a general diagram there will be
an ambiguity as to which light-cone co-ordinates to use and also which quark 
propagators to use to perform longitudinal integrations. For Fig.~4.1,
however, each gluon loop momentum naturally passes through only one line of
the quark loop and there is no ambiguity as to how to proceed.

If we draw Fig.~4.1 as describing a physical scattering process with time in
the upward vertical direction, as in Fig.~4.2, 
\begin{center}
\leavevmode
\epsfxsize=2in
\epsffile{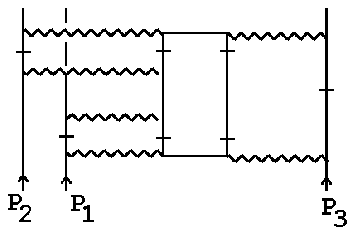}

Fig.~4.2 A Physical Scattering Process. 
\end{center}
it is clear that each of the
quark propagators marked with a hatch can be naturally close to mass-shell 
and contribute to intermediate states as
part of the scattering process. Since we are only looking for interesting 
contributions in this Section, we will not give a complete contour-closing 
argument as to whether a particular on-shell configuration is definitively
present asymptotically. Rather if propagators
are close to mass-shell during a scattering process we will take this as 
an indication that a
leading asymptotic contribution may be obtained if these propagators are put
on-shell by performing corresponding longitudinal momentum integrations.

Each loop integral has the form $I_i$ illustrated in Fig.~4.3.
\begin{center}
\leavevmode
\epsfxsize=2.5in
\epsffile{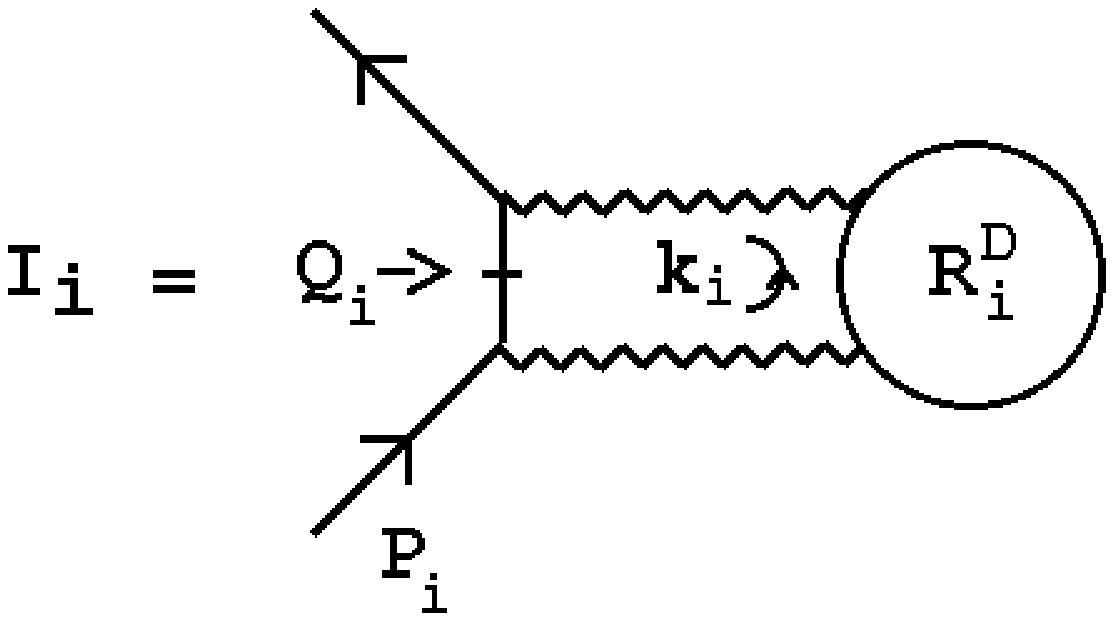}

Fig.~4.3 The One Loop Integral $I_i$.

\end{center}
$R^D_i $ denotes the remainder of the diagram besides the two gluon 
propagators shown.
We choose a combination of conventional light-cone 
co-ordinates for each loop, i.e. (in the notation of Appendix B)
$k_{ii^+},k_{ii^-}$ and $k_{i\perp}$, 
$i=1,2,3$. 
In the limit $P_{i^+} \to \infty$,  we can use (\ref{rlc0}) - (\ref{rlc21}) 
to approximate the
initial and final state spinors by $\st{p}_{ii^+} /m$ and so write 
each of the three $I_i$ (in Feynman gauge) in the form 
$$
I_i ~=~ g^2~ \int d^4k_i ~
\biggl[{\st{p}_{ii^+} \over m} \biggr]~ \gamma_{\mu}~ [ \st{p}_i - \st{k_i} + m 
]^{-1} ~\gamma_{\nu}~ \biggl[{\st{p}_{ii^+} \over m} \biggr]
~\biggl[ {g_{\mu \alpha} \over
k_i^2 } \biggr] \bigg[ { g_{\nu \beta} \over (Q_i - k_i)^2 }\biggr]  
~R^D_{i\alpha \beta}
$$
$$
\to ~~~ ~g^2~{p_{i^+} \over m} ~ \int d^4k_i ~
\delta\biggl( k_{ii^-} - (k_{ii\perp}^2 - m^2)/ 2 p_{ii^+} \biggr)~ 
{ 1 \over
k_i^2 ~  (Q_i - k_i)^2 }~~
R^D_{i~ i^-,i^-}
\auto\label{rd1}
$$
As discussed above, we have replaced the hatched quark propagator of
Fig.~4.3 by a $\delta$-function. 
Using this $\delta$-function to perform the $k_{ii^-}$ integration we obtain
$$
I_i ~=~~g^2 ~{p_{ii^+} \over m} 
~\int d^2k_{ii\perp}  dk_{ii^+} ~ ~
{ 1 \over
k_{ii\perp}^2 ~ (Q_{ii\perp} - k_{ii\perp})^2 }~ ~
R^D_{i~i^-,i^-}
\auto\label{rd2}
$$
where we have used $k_{ii^-} \sim 1 /p_{ii^+}~\to 0 $, together with 
$Q_{ii^-}=0$,
to eliminate the longitudinal momentum components in the gluon propagators.

For Fig.~4.1 the remaining $k_{ii^+}$ integrations can be performed very
simply. As illustrated in Fig.~4.4, 
\begin{center}
\leavevmode
\epsfxsize=4in
\epsffile{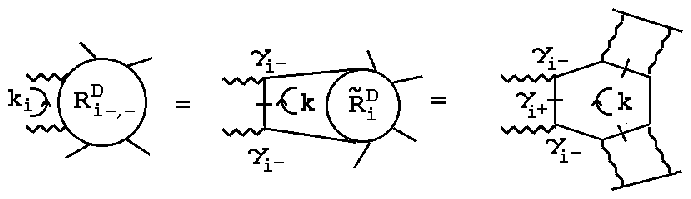}

Fig.~4.4 $~~R^D_{i~-,-}$ for Fig.~4.1.
\end{center}
the two gluons in $I_i$ are separated by the single quark
line within $R^D_i$ that carries the only dependence on $k_{ii^+}$.
We again replace the hatched propagator for this line by the corresponding
$\delta$-function and use it to carry out the $k_{ii+}$ integration, i.e. we
write 
$$
\eqalign{\int dk_{ii^+}~ R^D_{i~-,-}~ &=~ g^2~\int dk_{ii^+} ~
\delta\biggl( (k_{i} + k)^2 - m^2\biggr)
~\gamma_{i^-}~[~\gamma 
\cdot (k_i + k) ~ ]~\gamma_{i^-}~\tilde{R}^D_{i} \cr
&=~ g^2 \int dk_{ii^+}~ 
\delta\biggl(k_{ii^+}k_{i^-} + k_{i^+}k_{i^-} - (k_{i\perp} + k_{ii\perp})^2
-m^2\biggr) \cr
&~~~~~~~~ \times ~\gamma_{i^-}~[\gamma_{i^+} 
\cdot k_{i^-}  +  \cdots ]~\gamma_{i^-}~\tilde{R}^D_{i} \cr
&=~ g^2~ \gamma_{i^-} ~ \gamma_{i^+}~  \gamma_{i^-}  
~\tilde{R}^D_{i~} ~=~g^2~\gamma_{i^-}  ~\tilde{R}^D_{i~}
}
\auto\label{rd3}
$$

Using (\ref{rd1}) - (\ref{rd3}) for each of the $k_i$ integrations, 
Fig.~4.1 gives the asymptotic amplitude
$$
g^{12}~{p_{11^+}p_{22^+}p_{33^+} \over m^3}~ J_1(Q_1^2)
J_1(Q_2^2)~J_1(Q_3^2)~ \Gamma^v_{1^-2^-3^-}(Q_1,Q_2,Q_3)
\auto\label{rd4}
$$
where $J_1(Q^2)$ is the familiar two-dimensional integral (\ref{J1}) and
$\Gamma^v_{1^-2^-3^-}(Q_1,Q_2,Q_3)$ can 
be identified with a particular component of the tensor that the
triangle diagram
contributes to the three-point function of three vector currents, i.e. 
$$
\Gamma^v_{1^-2^-3^-}(Q_1,Q_2,Q_3,m) = i\int {  d^4 k~ Tr \{ 
\gamma_{1^-} ~(\st{k} +m) \gamma_{2^-} ( \st{k} + \st{Q}_2 +m) 
\gamma_{3^-} (\st{k} + \st{Q}_1 +m  ) \} 
\over  (k^2 - m^2) ([k + Q_2 ]^2 - m^2) 
 ([k + Q_1]^2 - m^2) }
\auto\label{rd5}
$$
Note that (\ref{rd4}) has been derived in the full triple Regge limit
(\ref{np3}) in which the $Q_i$ do not lie entirely in the $k_{i\perp}$ 
plane. While the $J_1$ factors depend only on 
the corresponding $Q^2_i$, the triangle diagram factor
$\hat{\Gamma}^v_{1^-2^-3^-}(Q_1,Q_2,Q_3,m)$ will have a 
dependence on the light-like momenta $\hat{q}_1, \hat{q}_2$ and
$ \hat{q}_3$ of (\ref{np3}). In the helicity-pole limit (\ref{np31}) 
the magnitudes of the light-like 
momenta are identified with one of the spacelike components of the $Q_i$. 
This limit can clearly be taken smoothly within $\hat{\Gamma}^v$. 

A-priori, it is straightforward to choose the quantum numbers of the 
$Q_i$-channels so that the lowest-order contribution is associated with 
two-reggeon exchange in each channel (color zero would be the simplest).
In this case, the $J_1(Q_i^2)$ factors
would be associated with the two-reggeon
state. However, as we remarked at the beginning of this sub-section, since 
Fig.~4.1 is planar, we expect that if reggeization effects are added it
ultimately does not provide a coupling for two-reggeon states. 
We can briefly describe how this happens for Fig.~4.1 as follows.

We performed the $k_{11^+}$-integration by using the hatched propagator 
contained in $R^D_{1-,-}$ as illustrated in Fig.~4.4. This integration can 
instead be written as an integral over the ``missing mass'' cross-energy
$$
M^2~=~(k_1 + Q_2)^2
\auto\label{mm}
$$ 
The singularities of $R^D_{1-,-}$ are all on the positive axis in the 
$M^2$-plane and so the contour integration over $M^2$ could be closed to zero
if the large $M^2$ behavior were appropriate. In the lowest-order diagram we 
are discussing this is provided by quark-antiquark exchange which is just 
divergent enough to prevent the contour closing. Therefore, 
the reggeon vertex can be written as an integral at infinity and 
it's main role will be to provide contributions that cancel
less planar diagrams and produce reggeon Ward identity zeroes at zero 
transverse momentum points. It is well-known from studies\cite{fgl} of QED 
that planar diagrams give contributions with very little analytic structure
that, typically, can be written as contour integrals at infinity or
equivalently as a subtraction in a dispersion relation. The presence of the
planar diagram contributions can, in fact, be deduced by studying the
non-planar contributions and demanding that the Ward identity constraints of
gauge invariance be satisfied.

In higher-orders the 
quark-antiquark reggeization illustrated in Fig.~4.5 will appear.
\begin{center}
\leavevmode
\epsfxsize=3in
\epsffile{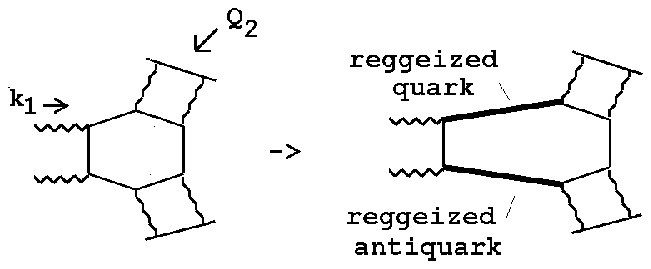}

Fig.~4.5 Quark-Antiquark Reggeization
\end{center}
If the reggeization effects are summed to all orders (which does not
destroy the validity of the low-order approximation) the reduced power behavior
will allow the closing of the $M^2$ contour to give zero. Consequently the
analytic structure of the triangle reggeon interaction we have extracted will
disappear as higher-order contributions are included (in parallel with the
well-known AFS cancelation\cite{hs}). We anticipate that only diagrams with
sufficient non-planarity to prevent any possible contour closing will survive
as reggeization effects are included. The reggeon Ward identities will 
appear as properties of these diagrams that follow from the shifting of 
integration variables that becomes allowed once reggeization effects are 
included. 

The presence of the anomaly would, of course, be expected to interfere with
integration shifts and could be found in the ultra-violet region this way. 
However, it is just because we expect this to be a very subtle issue 
that we have focussed on finding the anomaly in the infra-red region.
While (\ref{rd4}) and (\ref{rd5}) demonstrate how, as a lowest-order
approximation, the full four-dimensional triangle diagram can appear as an
effective interaction in the triple-Regge limit, since only
vector couplings, i.e. the $\gamma_{i^-}$,
appear there is no possibility for the anomaly infra-red
divergence. 

We must proceed further to find diagrams that generate a reggeon interaction
containing the effective $\gamma_5$ coupling necessary to produce the
anomaly. For the next diagram we study a $\gamma_5$ coupling does appear. 
However, we then find that the correct tensor and momentum structure for the
full anomaly divergence is still absent. 

\subhead{4.2 A Diagram With Some Non-planarity.}

In all other diagrams besides that of Fig.~4.1 (apart from those that 
are simply twisted versions of this diagram) one or more of the gluon 
loop momenta flows through more than one line of the quark loop. This 
introduces an extra complexity in carrying out the integrations over
the longitudinal gluon momenta. 
The next diagram we consider, the second shown in Fig.~2.3, introduces the 
minimal complexity of this kind. This diagram
can be redrawn as in Fig.~4.6(a), or as in Fig.~4.6(b).
There is
just one gluon loop momentum, i.e. $k_3$, that 
flows through more than one line of the internal quark loop. 
The $k_1$ and $k_2$ longitudinal integrations are straightforward and 
can be performed in the same way as we did for the longitudinal integrations 
of Fig.~4.1. For $k_3$ there are two possible routes. The first is shown in
Fig.~4.6(a). The second would be that shown in 
Fig.~4.6(b) if the external momentum flow was kept as in Fig.~4.1(a).
\begin{center}
\leavevmode
\epsfxsize=4.2in
\epsffile{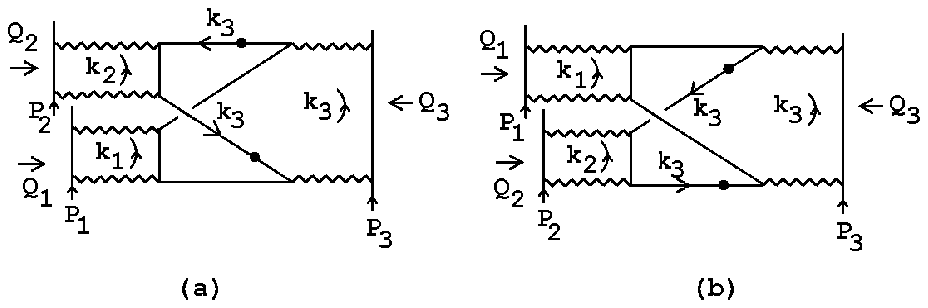}

Fig.~4.6 (a) A Diagram With Some Non-planarity (b) The Same Diagram Redrawn.
\end{center}

If we route $k_3$ as in Fig.~4.6(a), then there are two possible quark 
propagators, each marked with a dot, that could be used to perform the 
$k_{3^+}$-integration. If the time-ordering of the scattering process is
essentially represented by Fig.~4.6(a), then it would appear that only the
lower dotted propagator gives a quark state that can be part of an on
mass-shell intermediate state. The upper dotted propagator appears to describe
a virtual exchange that will be a long way from mass-shell. However, when the
diagram is redrawn as in Fig.~4.6(b), in the scattering process now described,
the role of the two propagators is interchanged. It is now the lower 
propagator that is virtual and far from mass-shell. Clearly 
the two contributions obtained by using the two possible propagators to 
perform the $k_{3^+}$-integration must be added.

Consider first the contribution of the upper dotted propagator in 
Fig.~4.6(a). After the longitudinal $k_1$ and $k_2$ integrations have been 
performed we will be left with the box-diagram integral illustrated in 
Fig.~4.7. 
\begin{center}
\leavevmode
\epsfxsize=5in
\epsffile{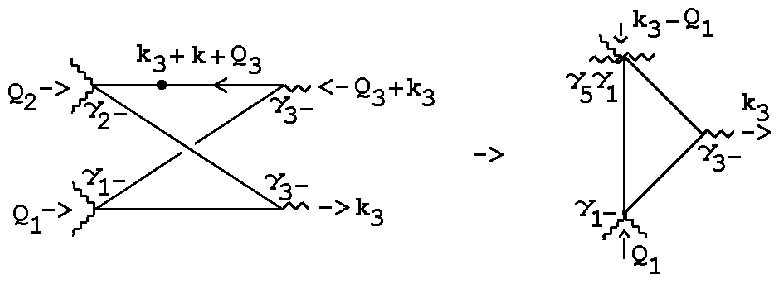}

Fig.~4.7 The Box and Triangle Diagrams Generated by Fig.~4.6.
\end{center}
The $k_{33^-}$ integration can be done by again utilising the evaluation of
Fig.~4.3. With the dotted 
propagator replaced by a $\delta$-function, the relevant factors in the
$k_{33^+}$ integration are
$$
\eqalign{ &\int d k_{33^+}~ \delta\biggl( (k_3 + k +Q_3)^2 - m^2 \biggr)
~\gamma_{2^-} ~\biggl( (k_3 + k  + Q_3 ) \cdot 
\gamma + m \biggr) ~ \gamma_{3^-} \cr
&~~~~=~\int d k_{33^+} ~\delta\biggl( ( k_{3^-} + Q_{33^-} )k_{33^+}
 + \cdots \biggr) \cr
& ~~~~~~~~~~~~~ \times \gamma_{2^-} ~
\biggl( ( k_{3^-} + Q_{33^-} ) \cdot 
\gamma_{3^+} + \cdots \biggr) ~
\gamma_{3^-}~ \cr
&~~~~= ~~ \gamma_{2^-} \gamma_{3^+} \gamma_{3^-} ~~ + ~~\cdots \cr
&~~~~= ~~ 2~(\gamma_0 - \gamma_2 - \gamma_3)
~+~2~i\gamma_5 \gamma_1  ~~ + ~~\cdots 
}
\auto\label{46}   
$$
where, in the last line, we have used (\ref{3ga}) to show that an axial 
$\gamma_5$-coupling is produced by the product of three $\gamma$-matrices.
The omitted terms 
generate only what we call ``non-local
couplings''. As elaborated in Appendix C, a non-local coupling is 
generated whenever the momentum dependence of the integrated propagator does 
not simply scale out of the integral, as it does for the part of 
(\ref{46}) that we have written explicitly.   

Focussing on the $\gamma_5$-interaction produced by (\ref{46}), 
the asymptotic amplitude, with all longitudinal integrations 
performed, can be written as 
$$
\eqalign{& g^{12} ~{p_{11^+}p_{22^+}p_{33^+} \over m^3} ~ 
 \int { d^2 \underline{k}_{11\perp} \over
\underline{k}_{11\perp}^2 (Q_1 - \underline{k}_{11\perp})^2}
~\int {d^2  \underline{k}_{22\perp} \over
\underline{k}_{22\perp}^2 (Q_2 - \underline{k}_{22\perp})^2}
~\int  {d^2  \underline{k}_{33\perp} \over
\underline{k}_{33\perp}^2 (Q_3 - \underline{k}_{33\perp})^2}
\cr  
& ~~ \int d^4 k ~~{ Tr \{ \gamma_{1^-} (\st{k}+ \st{Q}_1 +m) 
\gamma_5\gamma_{1} ~(\st{k}+ \st{k}_3 +m) 
\gamma_{3^-} ( \st{k} +m) \} \over 
(k - m^2) 
([k + Q_1 ]^2 - m^2) 
 ([k + k_3]^2 - m^2)} ~~~~+ ~~ \cdots }
\auto\label{461}
$$
where $k_{33^-}=0$ and $k_{33^+}$  is 
determined from the $\delta$-function used in (\ref{46}), and that part of the
amplitude not written explicitly now 
contains  either a vector coupling or a non-local coupling in place of the 
$\gamma_5\gamma_1$ coupling. If we again remove 
the $k_{ii\perp}$ integrations, the gluon propagators, and the $p_{ii^+}$
dependence that are all associated with the 
three two-reggeon states, (\ref{461}) 
gives a six-reggeon interaction containing the triangle diagram of Fig.~4.7,
i.e.
$$
\Gamma^a(Q_1,Q_3, k_{33\perp}) 
~=~\int d^4 k { Tr \{ \gamma_{1^-} (\st{k}+ \st{Q}_1 +m) 
\gamma_5\gamma_{1} ~(\st{k}+ \st{k}_3 +m) 
\gamma_{3^-} ( \st{k} +m) \} \over 
(k - m^2) 
([k + Q_1 ]^2 - m^2) 
 ([k + k_3]^2 - m^2)} ~~~~+ ~~ \cdots 
\auto\label{462}
$$ 
where we still have $k_{33^-}=0$ and $k_{33^+}$ is to be determined by 
the mass-shell constraint of (\ref{46}), thus giving the 
$Q_3$-dependence of (\ref{462}).

Since $\Gamma^a(Q_1,Q_3, k_{33\perp})$ contains a $\gamma_5$ coupling, 
it is
straightforward to identify it with a component of the triangle diagram 
tensor for an axial current and two vector currents. However, the 
maximal singularity
associated with the anomaly requires specific momenta and tensor components 
to be present. We must have two tensor components that can project on to 
the same light-cone component - this would have to be $\gamma_{1^-}$ and 
$\gamma_5 \gamma_1$. The third vertex must then carry spacelike momentum of
$O(\hbox{\q}) \to 0$, implying that we should take $k_3 \sim \hbox{\q} \to 0$. 
Finally a finite light-like momentum 
parallel to $\underline{n}_{1^-}$ must enter at the $\gamma_5$ vertex. But 
the light-cone component of $Q_1$ is orthogonal to $\underline{n}_{1^-}$. 
This conflict implies that a kinematical configuration producing 
the maximal anomaly divergence can not occur.

\subhead{4.3 Alternative Light-Cone 
Co-ordinates and Absence of the Anomaly}

We can give a direct argument that there is no anomaly
in the full reggeon interaction produced by Fig.~4.6(a). This 
argument will be important for the general analysis of discontinuities in
Section 6.

In Appendix B we have shown that light-cone co-ordinates
and associated $\gamma$-matrices can be introduced using any two light-like
momenta whose space components are orthogonal. The regge limit calculations
of Appendix C demonstrate that equivalent results are obtained using such 
co-ordinates and we use various co-ordinates of this form elsewhere in the
paper (including the discussion of helicity-pole
limits in the previous Section.) In particular let us 
repeat our evaluation of Fig.~4.6(a), with $k_3$ routed as shown, 
but for the $k_3$ longitudinal integration use 
co-ordinates in which $\underline{n}_{2^+}$ and $\underline{n}_{3^+}$ 
are the basic light-like momenta. Our $k_3$ co-ordinates are now
$$
k_{33^-}= k_{30}-k_{33}~, ~~k_{32^-}= k_{30}-k_{32}~, ~~
k_{31}~, ~~ \tilde{k}_{23}= k_{32}+k_{33} - k_{30}
\auto\label{467}
$$
The $k_{33^-}$ integration can again be performed using the evaluation of 
Fig.~4.3. However, instead of (\ref{46}), the $k_{32^-}$ integration of the
upper dotted propagator of Fig.~4.6(a) gives
$$
\eqalign{ &\int d k_{32^-}~ \delta\biggl( (k_3 + k +Q_3)^2 - m^2 \biggr)
~\gamma_{2^-} ~\biggl( (k_3 + k  + Q_3 ) \cdot 
\gamma + m \biggr) ~ \gamma_{3^-} \cr
&~~~~=~\int d k_{32^-} ~\delta\biggl( ( k_{3^-} + Q_{33^-} )k_{32^-}
 + \cdots \biggr) \cr
& ~~~~~~~~~~~~~ \times \gamma_{2^-} ~
\biggl( ( k_{3^-} + Q_{33^-} ) \cdot 
\gamma_{2^-} + \cdots \biggr) ~
\gamma_{3^-}~ \cr
&~~~~= ~~ \gamma_{2^-} \gamma_{2^-} \gamma_{3^-} ~~ + ~~\cdots \cr
&~~~~= ~~ 0~~ + ~~\cdots 
}
\auto\label{468}   
$$
Now no local terms appear. Only omitted ``non-local'' terms are generated.
Evaluation of the contribution from the lower 
dotted propagator in Fig.~4.6(a) will similarly give no local terms. While 
a distinction between the contributions of local and non-local couplings 
may be difficult to maintain in general (after integration), if we assume 
that the anomaly can appear only when an appropriate local $\gamma_5$ coupling
is present. then we have demonstrated it's absence in the diagram of
Fig.~4.6.

\subhead{4.4 A Maximally Non-Planar Diagram}

Finally we study the third diagram of Fig.~2.3. In this case there will be 
clear anomaly contributions which, when general external couplings are present,
cancel only after all diagrams of this kind are
summed. As illustrated in Fig.~2.4, this diagram has a ``maximally
non-planar'' property - which produce a ``maximal complexity''
in terms of evaluating the longitudinal gluon momentum integrations. 

The first point we note is that there is no natural choice for routing the
gluon loop momenta through the internal quark loop. Each momentum flows
through three quark propagators no matter in which direction we send it. 
As to which 
combinations of propagators can be simultaneously close to mass-shell,
we note that if we draw the scattering as in Fig.~4.8(a) then 
\begin{center}
\leavevmode
\epsfxsize=4in
\epsffile{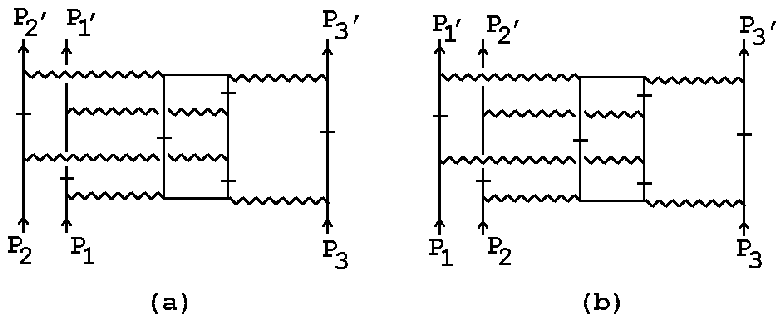}

Fig.~4.8 Two Scattering Processes Described by the Diagram of Fig.~2.4
\end{center}
the hatched propagators can obviously be simultaneously close to mass-shell
and produce intermediate states. However, if we redraw the diagram as in
Fig.~4.8(b), an alternative set of hatched lines is naturally chosen. It is
easy to check that the second set corresponds to the three loop propagators
not hatched in Fig.~4.8(a). (These two contributions were already recognized
in Section 2.) Note that the hatched lines in Fig.~4.8(a) correspond to taking
a double discontinuity in $s_{13}$ and $s_{2'3'}$ while the hatched lines in
Fig.~4.8(a) correspond to taking a double discontinuity in $s_{23}$ and
$s_{1'3'}$. This will be an important distinction in the following.

There are further scattering processes 
described by the diagram we are discussing that involve 
interchanging ingoing and outgoing particles. For example, the 
processes of Fig.~4.9.
Such contributions will be included separately in the multiple discontinuity 
formalism of the next two Sections. (In the language of the next 
Section, one maximally non-planar Feynman diagram 
contains discontinuities associated with several different hexagraphs.) 
In this sub-section we will discuss only the contribution of Fig.~4.8(a) in 
detail. After we have discussed Fig.~4.8(a), it will be obvious that the 
discussion immediately extends to Fig.~4.8(b) and that it also generalises 
to the corresponding contributions from the diagrams of Fig.~4.9.
\begin{center}
\leavevmode
\epsfxsize=4in
\epsffile{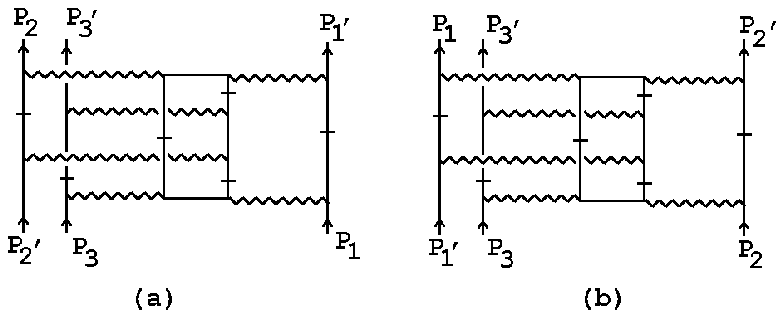}

Fig.~4.9 Further Scattering Processes Described by the Diagram of Fig.~2.4
\end{center}

For reasons that will become apparent, it will be desirable to keep as much 
symmetry as possible in our kinematic analysis, even at the cost of using a 
more complicated labeling for momenta flowing along the quark loop lines.
Therefore, we label the momentum flow into the internal
quark loop of Fig.~4.8(a) as in Fig.~4.10, where 
the $\gamma$ matrices that contribute in the triple-regge limit are also 
shown. 
\begin{center}
\leavevmode
\epsfxsize=3.6in
\epsffile{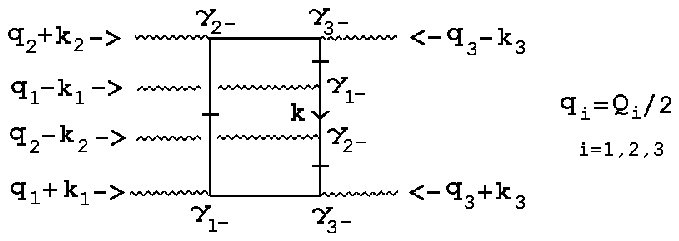}

Fig.~4.10 The Quark Loop in Fig.~4.8(a).
\end{center}
This time we use the light-cone co-ordinates ($k_{i1^-},k_{i2^-},
\underline{\tilde{k}}_{i\perp}$) to perform
the $k_1$ and $k_2$ integrations and to evaluate the $\gamma$-matrix trace
associated with the quark loop. For the $k_3$ integration we use 
conventional light-cone co-ordinates. The evaluation of the integral 
$I_i$ of Fig.~4.3 can be used to 
perform the $k_{11^-},k_{22^-}$ and $k_{3^-}$ integrations.
The remaining longitudinal integrations have to be carried out using 
$\delta$-functions for propagators belonging to 
the internal quark loop, A-priori there are six 
different options for choosing the longitudinal $k_i$ to be used to
put the hatched lines in Fig.~4.10 on-shell. These possibilities 
are indicated schematically in Fig.~4.11 
\begin{center}
\leavevmode
\epsfxsize=5in
\epsffile{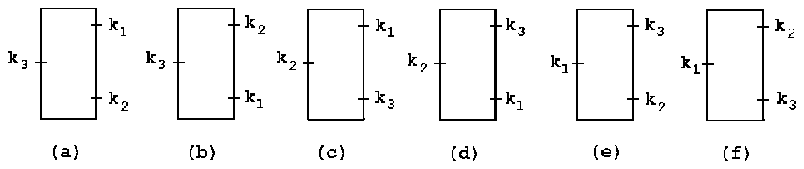}

Fig.~4.11 Possible Choices of $\delta$-function Integrations for Fig.~4.10.
\end{center}

We consider first which of the possibilities in Fig.~4.11 can generate the
necessary local couplings.
For the $\delta$-function assignment of Fig.~4.11(b), we note that both the
$k_{12^-}$ and the $k_{21^-}$ integrations will be analagous to (\ref{468})
in that the potential point-coupling,
involving the $\gamma$-matrix (within the numerator of the 
on-shell quark) that is multiplied by the momentum 
scaling the $\delta$-function momentum,
is eliminated by one of the adjacent $\gamma$-matrices.
Hence no local coupling is produced.
For Fig.~4.11(c) the $k_{21^-}$ integration similarly produces no local 
coupling. For Fig.~4.11(d) it is the
$k_{12^-}$ and the $k_{21^-}$ integrations, for Fig.~4.11(e) the $k_{12^-}$ 
integration, and for Fig.~4.11(f) both the 
$k_{12^-}$ and $k_{21^-}$ integrations, that produce no local coupling.
Consequently, with the light-cone 
co-ordinates we have chosen, only the $\delta$-function assignment of 
Fig.~4.11(a) gives a contribution with local couplings from all three
integrations.

For the $\delta$-function assignment of Fig.~4.11(a) we route momenta through
the quark loop of Fig.~4.10 as illustrated in Fig.~4.12
\begin{center}
\leavevmode
\epsfxsize=2.5in
\epsffile{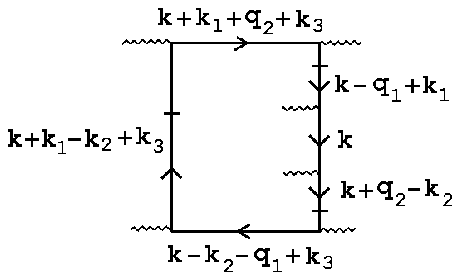}

Fig.~4.12 Momentum Flow for the $\delta$-function Assignment of
Fig.~4.11(a).
	\end{center}
We calculate the local couplings generated as follows. 
$$
\eqalign{ &\int d k_{12^-} ~\delta\biggl( (k_1 + k - q_1)^2 - m^2 \biggr)
~\gamma_{3^-} ~\biggl( (k_1 +k - q_1) \cdot 
\gamma + m \biggr) ~ \gamma_{1^-} \cr
&~~~~= ~\int d k_{12^-} ~\delta\biggl( k_{1^-}~ k_{12^-}
 + \cdots \biggr) \cr
& ~~~~~~~~~~~~~ \times \gamma_{3^-} ~
\biggl( k_{1^-}  \cdot 
\gamma_{2^-} + \cdots \biggr) ~
\gamma_{1^-}\cr
&~~~~ = ~~ \gamma_{3^-} \gamma_{2^-} \gamma_{1^-} ~~ + ~~\cdots \cr
}
\auto\label{571}   
$$
$$
\eqalign{&\int d k_{21^-} \delta\biggl( (k_2 - k -q_2)^2 - m^2 \biggr)
~\gamma_{2^-} ~\biggl( (k_2 -k - q_2 ) \cdot 
\gamma + m \biggr) ~ \gamma_{3^-} \cr
&~~~~= ~\int d k_{21^-} ~\delta\biggl( k_{2^-} k_{21^-}
 + \cdots \biggr) \cr
& ~~~~~~~~~~~~~ \times \gamma_{2^-} ~
\biggl( k_{2^-}  \cdot 
\gamma_{1^-} + \cdots \biggr) ~
\gamma_{3^-} ~~~~~~~~~~~~~~~~~~~~~~~~~~~~~~~~ \cr
&~~~~= ~~ \gamma_{2^-} \gamma_{1^-} \gamma_{3^-} ~~ + ~~\cdots 
}
\auto\label{572}   
$$
$$
\eqalign{ &\int d k_{33^+}~ \delta\biggl( (k_3 + k +k_1 -k_2)^2 - m^2 
\biggr)
~\gamma_{1^-} ~\biggl( (k_3 + k  + k_1 -k_2) \cdot 
\gamma + m \biggr) ~ \gamma_{2^-} \cr
&~~~~=~\int d k_{33^+} ~\delta\biggl( (k_{3^-} + k_{13^-} 
- k_{23^-} )k_{33^+}
 + \cdots \biggr) \cr
& ~~~~~~~~~~~~~ \times \gamma_{1^-} ~
\biggl( ( k_{3^-} + k_{13^-} 
- k_{23^-} ) \cdot 
\gamma_{3^+} + \cdots \biggr) ~
\gamma_{2^-}\cr
&~~~~= ~~ \gamma_{1^-} \gamma_{3^+} \gamma_{2^-} ~~ + ~~\cdots \cr
}
\auto\label{573}   
$$
In each case the dots indicate the contribution of additional non-local 
couplings. We defer the 
evaluation of the $\delta$-functions for the moment. 
The triangle diagram structure of the local couplings is illustrated in 
Fig.~4.13,
\begin{center}
\leavevmode
\epsfxsize=4in
\epsffile{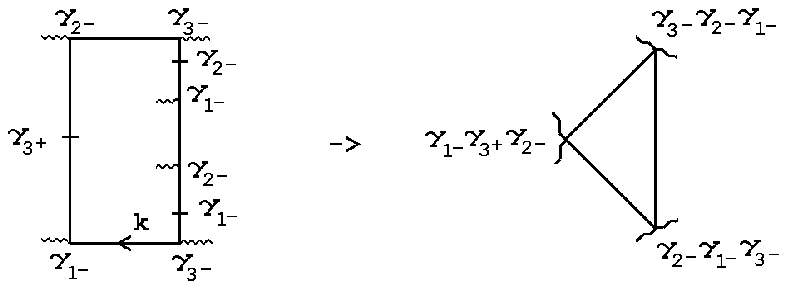}

Fig.~4.13 Local couplings generated by the $\delta$-function Assignment of
Fig.~4.11(a).
\end{center}                             

With all longitudinal integrations performed, the asymptotic amplitude 
obtained from Fig.~4.8(a) can be written as 
$$
\eqalign{& ~~~~~~~~~~~~~ g^{12} ~~{p_{11^+}~p_{22^+}~p_{3^+} \over m^3} 
~~~ \times \cr 
&\int { d^2 \underline{k}_{112+} \over
(q_1 + \underline{k}_{112+})^2(q_1 - \underline{k}_{112+})^2}
~\int {d^2  \underline{k}_{212} \over
(q_2 + \underline{k}_{212+})^2(q_2 - \underline{k}_{212+})^2}
~\int  {d^2  \underline{k}_{33\perp} \over
(q_3 + \underline{k}_{33\perp})^2(q_3 - \underline{k}_{33\perp})^2} 
\cr  
&\int d^4 k~{ Tr \{ \hat{\gamma}_{12} (\st{k}+ \st{k}_1 
+ \st{q}_2 + \st{k}_3 +m) 
\hat{\gamma}_{31}( \st{k} +m) 
\hat{\gamma}_{23} (\st{k}- \st{k}_2 + \st{q}_1 
+ \st{k}_3 +m) \} \over 
([k + k_1 + q_2 + k_3]^2 - m^2) 
(k^2 - m^2) 
 ([k - k_2 + q_1 +k_3]^2 - m^2)} ~+ ~ \cdots }
\auto\label{578}
$$
where 
$$
\hat{\gamma}_{31}= \gamma_{3^-} \gamma_{2^-} \gamma_{1^-} ~,~~~
\hat{\gamma}_{23} = \gamma_{2^-} \gamma_{1^-} \gamma_{3^-} ~,~~~
\hat{\gamma}_{12} = \gamma_{1^-} \gamma_{3^+} \gamma_{2^-} 
\auto\label{5781}
$$
and $k_{11^-}= k_{22^-} = k_{33^-}=0$, with $k_{12^-}, k_{21^-}$ and $k_{33^+}$ 
still to be determined by $\delta$-function constraints. That part of the
amplitude not shown explicitly in (\ref{578})
contains  non-local couplings at
one, or more, vertices of the triangle diagram. 

Before extracting a reggeon interaction from (\ref{578})
we first separate out a potential anomaly generating part.
Using, again, the identity (\ref{3ga}), we can write
$$
\hat{\gamma}_{31} ~=~\gamma^{-,+,-}~+~ i~
\gamma^{-,-,-} ~\gamma_5
\auto\label{574}
$$
$$
\hat{\gamma}_{23} ~=~\gamma^{+,-,-}~+~ i~
\gamma^{-,-,-}  ~\gamma_5
\auto\label{575}
$$
$$
\hat{\gamma}_{12} ~=~\gamma^{-,-,-}~-~ i~
\gamma^{-,-,+}  ~\gamma_5
\auto\label{576}
$$
where 
$$
\gamma^{\pm,\pm,\pm} ~=~ \gamma^{\mu}\cdot n^{\pm,\pm,\pm}_{ \mu} ~,~~~~
n^{\pm,\pm,\pm \mu} ~= ~ (1,\pm1,\pm1,\pm1)
\auto\label{5760}
$$
To obtain the divergence (\ref{a3}) when $m =0$, 
we must have a component of the axial-vector triangle diagram
tensor  $\Gamma^{\mu\nu\lambda}$  with
$\mu= \nu $ having a lightlike projection and $\lambda $ 
having an orthogonal spacelike projection.
Since $\hat{\gamma}_{31}$
and $\hat{\gamma}_{23}$ have the same $\gamma_5$ component, 
this requirement is met if 
we choose the 
$\gamma_5$ component from all three of the $\hat{\gamma}_{ij}$. 
The finite light-like momentum involved must simply have a projection on 
$n^{-,-,- \mu}$. (We will discuss how this occurs shortly.)
$n^{-,-,+ \mu}$ provides a distinct spacelike component in the 
$\underline{n}_3$ direction. 
The anomaly infra-red divergence (like the 
ultra-violet anomaly) is also present in the corresponding tensor 
component of the triangle diagram for one axial and two vector currents.
Since the vector part of $\hat{\gamma}_{12}$ is identical 
to the $\gamma_5$ 
component of $\hat{\gamma}_{31}$ and $\hat{\gamma}_{23}$, 
the requirements for this case are partially 
met when we take the $\gamma_5$ part 
of either $\hat{\gamma}_{31}$ or $\hat{\gamma}_{23}$ together with the 
vector parts of the remaining two $\hat{\gamma}_{ij}$. However, 
the necessary distinct spacelike component in the 
$\underline{n}_3$ direction is not present. 

The three $\gamma_5$ couplings give the ($m=0$) reggeon interaction 
$$
\eqalign{ &\Gamma_6(q_1,q_2,q_3,
\tilde{\underline{k}}_1,\tilde{\underline{k}}_2, 
\underline{k}_{3\perp},0) ~=\cr
& \int d^4 k  {  Tr \{ 
\gamma_5 \gamma^{-,-,+} (\st{k}+ \st{k}_1 + \st{q}_2 +\st{k}_3) 
\gamma_5 \gamma^{-,-,-} ~\st{k}~ 
\gamma_5 \gamma^{-,-,-}(\st{k}- \st{k}_2 + \st{q}_1 + \st{k}_3 ) 
\over  (k + k_1 + q_2 + k_3 )^2  
~k^2 ~
 (k - k_2 + q_1 + k_3)^2 }  ~+ ~ \cdots }
\auto\label{580}
$$
where, again, we note that $k_{11^-}=
k_{22^-} = k_{33^-}=0$ and that $k_{12^-}, k_{21^-}$ and $k_{33^+}$ remain
to be determined by the $\delta$-functions of (\ref{571}) - (\ref{573}). 
(\ref{580}) corresponds to the triangle diagram 
illustrated  in Fig.~4.14. 
\begin{center}
\leavevmode
\epsfxsize=1.8in
\epsffile{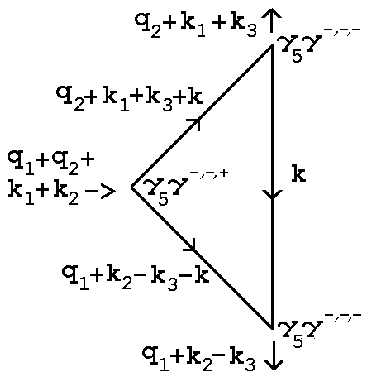}

Fig.~4.14 The Triangle Diagram Corresponding to (\ref{580})
\end{center}

To see the maximal anomaly divergence we must be able to take the limit
$$
(k_1 + q_2 +k_3)^2 \sim (q_1 + q_2 + 
k_1 +k_2)^2 \sim (k_2 + q_1 - k_3)^2 \sim 
\hbox{\q}^2 \to 0
\auto\label{5801}
$$
of (\ref{580}) with a finite light-like momentum flowing through the diagram
that has a projection on $n^{-,-,- \mu}$. This momentum flow 
must also be consistent with the three mass-shell constraints 
determining $k_{12^-}, k_{21^-}$ and $k_{33^+}$ respectively, i.e. 
\beqa  (k - q_1 + k_1)^2~&=&~0 \label{5851} \\
(k + q_2 - k_2)^2 ~&=&~0 \label{5852} \\
(k+k_1 -k_2 + k_3)^2~&=&~0 \label{5853}
\eeqa
To find momenta satisfying all of the required constraints, we first consider
the limiting configuration in which $\hbox{\q}=0$ and ask whether this can
be realized with the loop momentum 
$k \sim \hbox{\q} = 0 $ (as discussed in Appendix A). 
It will be straightforward to subsequently add momenta that are $O(\hbox{\q})$.

We identify Fig.~4.14  with Fig.~A2 by identifying $q_1 + k_2 - k_3$ with 
$\hbox{\q}_1$ and $q_1 + q_2 + k_1 + k_2$ with $\hbox{\q}_2$. This requires 
that (in the limit $\hbox{\q} \to 0$)
$$
q_1 + q_2 + k_1 + k_2 ~= ~0 
\auto\label{5841}
$$
To satisfy (\ref{5851}) and (\ref{5852}) we take
$q_1-k_1$ and $q_2-k_2$ lightlike, i.e. 
$$
\eqalign{ &q_1-k_1~=~(2l_{2^-},2l_{2^-},0,0)~, ~~~
q_2-k_2~=~(2l_{1^-},0,2l_{1^-},0) \cr
&~~~=> ~~~~ \underline{q}_{112+} ~=~ \underline{k}_{112+} ~, ~~~
\underline{q}_{212+} ~=~ \underline{k}_{212+} }
\auto\label{5843}
$$ 
(using again the co-ordinates of Appendix B). Since the light-cone components
of $q_1 + k_1$ and $ q_2 +k_2 $ can not cancel, satisfying (\ref{5841}) 
requires that
$$
q_{12^-} ~=~-~k_{12^-} ~= ~l_{2^-}~, ~~~ q_{21^-} ~=~-~k_{21^-} 
~=~ l_{1^-}~~~
\underline{q}_{112+} ~=~- \underline{q}_{212+}
\auto\label{584300}
$$
We then have
$$
q_3~=~- (q_1+q_2)~= ~-(l_{2^-} + l_{1^-},l_{2^-},l_{1^-},0)
\auto\label{58430}
$$
and so for $q_3$ to have the form (\ref{np3}), we must have 
$$
l_{2^-}~ =~ - ~
l_{1^-}~ = ~l ~ ~~=>~~~Q_3^2~ =~ 4~ q_3^2~ = ~ -  8 ~l^2
\auto\label{58434}
$$
We choose $l$ to be positive, we will discuss the implications of
this shortly.

The most general light-cone momentum form for $q_1 + k_2 -k_3$
that has a projection on
$n^{-,-,- \mu}$ and is orthogonal to $\underline{n}_3$ is
$$
q_1 + k_2 -k_3 ~\sim ~\underline{n}_{lc}~=~ 
(1, cos {\theta}_{lc}~, sin {\theta}_{lc}~, 0)
\auto\label{glcm}
$$ 
where $\theta_{lc}$ is arbitrary. Since $\underline{q}_{112+} 
+ ~\underline{k}_{212+} = 0$, this requires that 
$$
\eqalign{ &(l,l,0,0) ~+ ~ (l,0,l,0) 
~-~ (k_{33^+}, k_{31},k_{32},k_{33^+}) ~\sim ~ 
\underline{n}_{lc} \cr
& ~~~ => ~~~~~ k_{3} = ~l ~
(0, 1 -2~ cos {\theta}_{lc}~, 1 - 2~sin {\theta}_{lc}~, 0) }
\auto\label{584301}
$$
so that
$$
(q_3 - k_3)^2 ~=~4 q_3^2 ~(1 - cos {\theta}_{lc}) ~, 
~~~~~ (q_3 + k_3)^2 ~= ~ 4 q_3^2 ~(1 -  sin {\theta}_{lc})
\auto\label{584302}
$$

Finally, we must satisfy the last mass-shell condition (\ref{5853}). Writing
$$
k_1 - k_2 + k_3 ~=~ - q_1 - k_2 + k_3 ~+ ~
2 ~\underline{q}_{112+}
\auto\label{584303}
$$
and using (\ref{glcm}) together with (\ref{584301}) we obtain
$$
k_1 - k_2 + k_3 ~=~- 2l(1,cos {\theta}_{lc}~, sin {\theta}_{lc}~,0) + 
2q_{112-}(1,1,1,0) +2q_{13}(0,0,0,1)
\auto\label{584304}
$$
and so (\ref{5853}) becomes
$$
2~l~q_{112-}~(1 - cos {\theta}_{lc} - sin {\theta}_{lc})  
=~- q_1^2~=~\underline{q}_{112+}^2~= ~q_{112-}^2 + q_{13}^2 
\auto\label{584305}
$$
or, equivalently, 
$$
\eqalign{ (q_1+k_1)^2~&=~4 ~q_1^2~=~Q_1^2~=~(q_2+k_2)^2~=~4q_2^2~=~Q_2^2 \cr
&=~ - 8 ~l~q_{112-}~(1 - cos {\theta}_{lc} - sin {\theta}_{lc})  }
\auto\label{5843050}
$$ 
Note that as $Q_3^2~\sim~l^2 ~\to~0$ then also $(q_1+k_1)^2 
~\sim~(q_2+k_2)^2~\to~0$

(\ref{584305}) can apparently be satisfied for arbitrary $q_1^2$ by choosing
$q_{112-}$ and $q_{13}$ appropriately. However, there is a subtlety. 
To give light-like intermediate states in which the scattering takes place
as in Fig.~4.8(a) we must take $l >0$. The time component of $k_1 - k_2 + k_3$,
i.e.
$$
-2l~+~2q_{112-}
\auto\label{tco}
$$
should also be positive. This requires $q_{112-} > l >0$, but then
(\ref{584305}) can only be satisfied if 
$$
q_{112-} ~< ~2 l (cos {\theta}_{lc} + sin {\theta}_{lc} -1)~<~2l(\sqrt{2} 
-1)~<~ l
\auto\label{stly}
$$
so that (\ref{tco}) can not be positive. 
We conclude that if the mass-shell constraints that we have imposed are to 
be associated with taking discontinuities, then
the anomaly can only occur simultaneously if the scattering process takes
place in an unphysical region. As 
we will discuss at length in the next Section, there is indeed an unphysical
region where the mass-shell constraints can give discontinuities and the anomaly
can occur. That unphysical scattering processes must play a fundamental role
in the occurrence of the anomaly should not be a surprise because of the
chirality transition that has to be involved. 

In fact, for $l<0$ the momentum configuration we have arrived at does 
describe the scattering 
illustrated in Fig.~4.15(a), if this is interpreted as a space-time
diagram with time directed up the page. This 
is the process already illustrated in Fig.~2.6(a) and it clearly is 
physical.
The dashed lines carry light-like momentum while the open line is the
quark carrying zero momentum ($k=0$) in the anomaly configuration - this is
the quark that, as explained in Appendix B, undergoes the chirality
transformation. $l > 0$ gives the process illustrated in Fig.~4.15(b) which,
however, is unphysical in that a spacelike gluon appears for longer than
on-shell particles.
\begin{center}
\leavevmode
\epsfxsize=4in
\epsffile{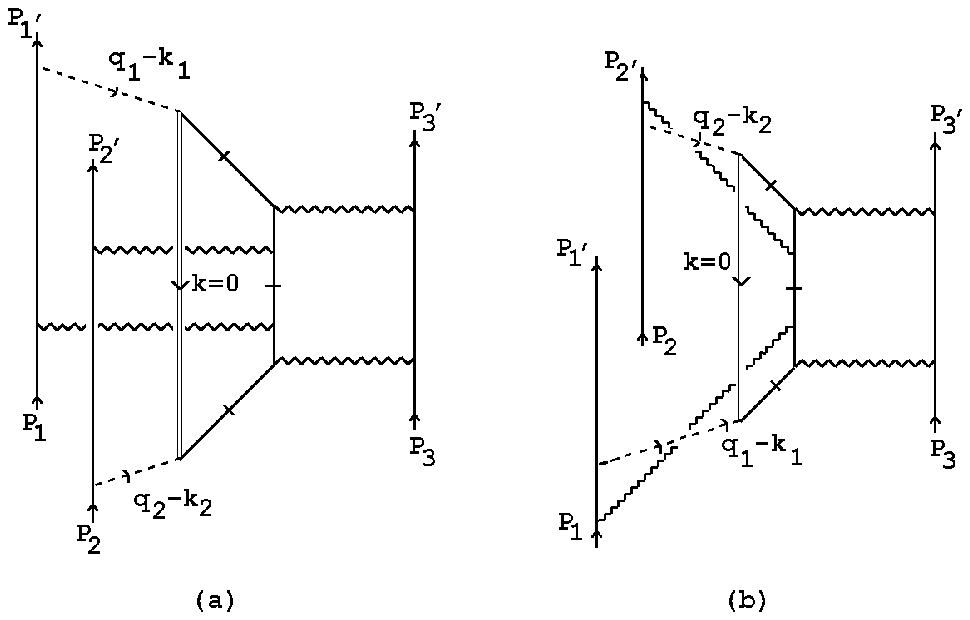}

Fig.~4.15 Physical Scattering Processes Involving the Anomaly
\end{center}

The hatched lines in Fig.~4.15(a) are again on-shell and are the same lines
that are hatched in Fig.~4.8(a). However, the on-shell lines can not give
discontinuities in this configuration. Rather the asymptotic amplitude 
has to be interpreted as extrapolated away from wherever the 
discontinuities were taken, as discussed briefly in Section 2. As we will
discuss further in the next Section, the
dispersion and multi-regge theory that we develop will ultimately
justify the initial calculation of a multiple discontinuity
corresponding to Fig.~4.8(a) in one part of the physical region
and then searching for the anomaly configuration
in a different physical region 
(or different part of the same physical region) with the propagators giving
the discontinuity kept on-shell.
The important point, at this stage, is that the appropriate momentum
configuration for the anomaly divergence does occur and the amplitude involved
has reality properties.

To explicitly see the anomaly divergence we add a spacelike momentum of 
$O(\hbox{\q})$ orthogonal to both $\underline{n}_{lc}$ and $\underline{n}_3$ 
and let $\hbox{\q}\to 0$ with $l$ fixed. For fixed $\underline{n}_{lc}$ we 
can choose 
$$
\underline{\hbox{\q}} ~ \sim ~ \pm ~\underline{n}_{lc\perp} 
~=~ \pm~ (0, - sin {\theta}_{lc}~, cos {\theta}_{lc}~, 0)
\auto\label{olcm}
$$
It is the component 
of $q_1 + q_2 + k_1 + k_2$ in this direction that contributes ($k_3$ can also
have a component, but it does not contribute).
To evaluate the form of the reggeon interaction (\ref{580}) with $\hbox{\q}\to
0$ as in (\ref{olcm}), we work in the frame in which the light cone momentum
(\ref{glcm}) can be identified with ${\hbox{\q}}_1^+$ in (\ref{200}).
Projecting each of the ${\gamma}^{-,-,- \mu}$ on $\underline{n}_{lc}$ 
gives a factor
$$
(1 - cos {\theta}_{lc} - sin {\theta}_{lc})^2
\auto\label{pfac}
$$
and so using (\ref{a3}) we obtain
$$
\Gamma_6 ~~\sim  ~~ {(1 - cos {\theta}_{lc} - sin {\theta}_{lc})^2~
l^2~ (q_1 + q_2 + k_1 + k_2)\cdot \underline{n}_{lc\perp}    
\over \hbox{\q}^2 } 
~\sim ~{Q_3^2 \over \hbox{\q} } 
\auto\label{05847}
$$
which manifestly changes sign when $\underline{n}_{lc\perp} \to -
\underline{n}_{lc\perp}$ with $\underline{n}_{lc}$ fixed.

Since contributions with both signs for $\underline{n}_{lc\perp}$
are present in the integral (\ref{578}) of the reggeon interaction, the
crucial question is whether there is a cancelation.
It is fundamental that, since we integrate over $\theta_{lc}$, the two 
possibilities 
are related by a parity transformation interchanging the $1$ and
$2$ axes. The two possibilities are also related by 
reversing the space component in
$\underline{n}_{lc}$ and keeping (\ref{olcm}) as the orthogonal spacelike
momentum. As we discuss in Appendix A, from either perspective the result is
the same, the sign of the anomaly contribution is reversed. 

In the lowest-order diagrams we have discussed the transverse momentum 
integrations are sufficient to produce a cancelation. We first integrate over
$k_3$ so that there is a symmetry under $k_3 \leftrightarrow - k_3$. For 
$l \neq 0$ this is not sufficient to produce a symmetry under  
$\underline{n}_{lc\perp} \to - \underline{n}_{lc\perp}$. However, 
if we also integrate
over $k_1$ and $k_2$ then since the external couplings are simple constants
there will also be symmetry under $q_i + k_i \leftrightarrow q_i - k_i, ~
i=1,2$.
If we then add the two contributions we have discussed from Fig.~4.8(a) and
(b) then all contributions to the amplitude, apart from the anomaly,
will be completely
symmetric under $1 \leftrightarrow 2$.  The
antisymmetry of the anomaly then requires that it cancel.

In higher-orders the external reggeon couplings $G_{h}(q_i,k_i)$ 
aquire non-trivial momentum dependence.
For fixed helicity $h$ these couplings need not be symmetric under 
$q_i + k_i \leftrightarrow q_i - k_i$. To discuss cancelations in this case
it is necessary to add the contributions from the twisted diagrams of Fig.~4.16
\begin{center}
\leavevmode
\epsfxsize=5in
\epsffile{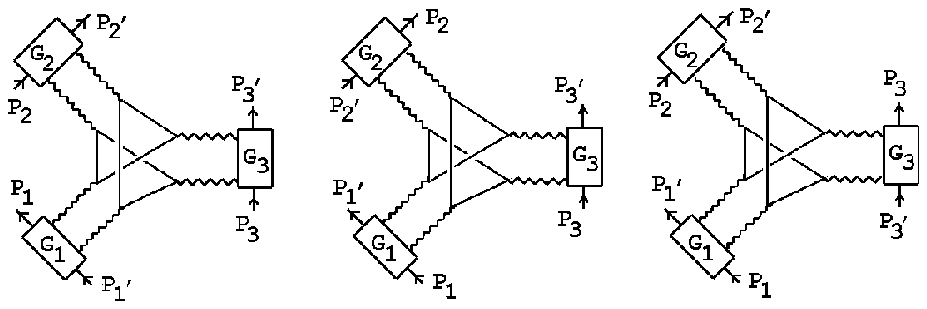}

Fig.~4.16 Twisted Diagrams
\end{center}
and also to discuss the signature properties of the
reggeon states. If the external couplings have sufficiently asymmetric 
transverse momentum dependence as might be expected, 
for example, if they contain the chirality violation produced by an
instanton interaction, then the anomaly need not cancel. 
We will reserve a more extensive 
discussion of this for Section 7. Here we simply remark that,
for elementary quark scattering, 
both the color factors and all three of the $k_i$-integrations are symmetric
also in higher-orders when we sum over all diagrams of this form 
and so the anomaly cancelation continues to hold. 

Finally, we note that a relatively trivial way to break the transverse
momentum symmetries that cancel the anomaly in an individual diagram 
would be to introduce masses for
some, or all, of the gluons. In particular, it would be sufficient 
if one of each of the pairs of
gluons were massive and the other massless. The cancelation would be restored
after the summation over 
all the diagrams of Fig.~4.16 unless the two gluons were distinguished by 
additional quantum numbers besides their mass. Introducing a gluon mass,
of course, requires spontaneous symmetry breaking which would extend our
analysis considerably. In fact, the symmetry breaking of the QCD gauge
symmetry from SU(3) to SU(2), discussed at several points in this paper, does
introduce a massive SU(2) singlet gluon and the required transverse momentum
asymmetry is produced when this combines, not with a single massless gluon,
but with three massless gluons carrying the quantum numbers of the
winding-number current.

\newpage 

\mainhead{5. Multi-Regge Theory}

In this Section we describe the asymptotic dispersion relation formalism,
together with the multi-regge theory based on it, that is needed to 
systematically study the contribution of quark loops to triple-regge 
vertices in QCD.  There will be some overlap with Section 4 
of \cite{arw98}. However, the treatment we gave in \cite{arw98}
is missing several crucial elements that we discuss here. As a result,
we have made the following essentially self-contained.

\subhead{5.1 Angular Variables: $s$- and $t$-channel Physical Regions}

We begin with the introduction of the angular variables that provide the link
between asymptotic limits taken in a direct, or ``$s$-channel'', 
and partial-wave analysis in various ``cross-channels''
or ``$t$-channels''.   We use the variables 
$z_1,z_2,z_3, u_{12}, u_{23}$ and $u_{31}$ 
corresponding to the ``Toller Diagram'' of Fig.~5.1~. 
\begin{center}
\leavevmode
\epsfxsize=3.5in
\epsffile{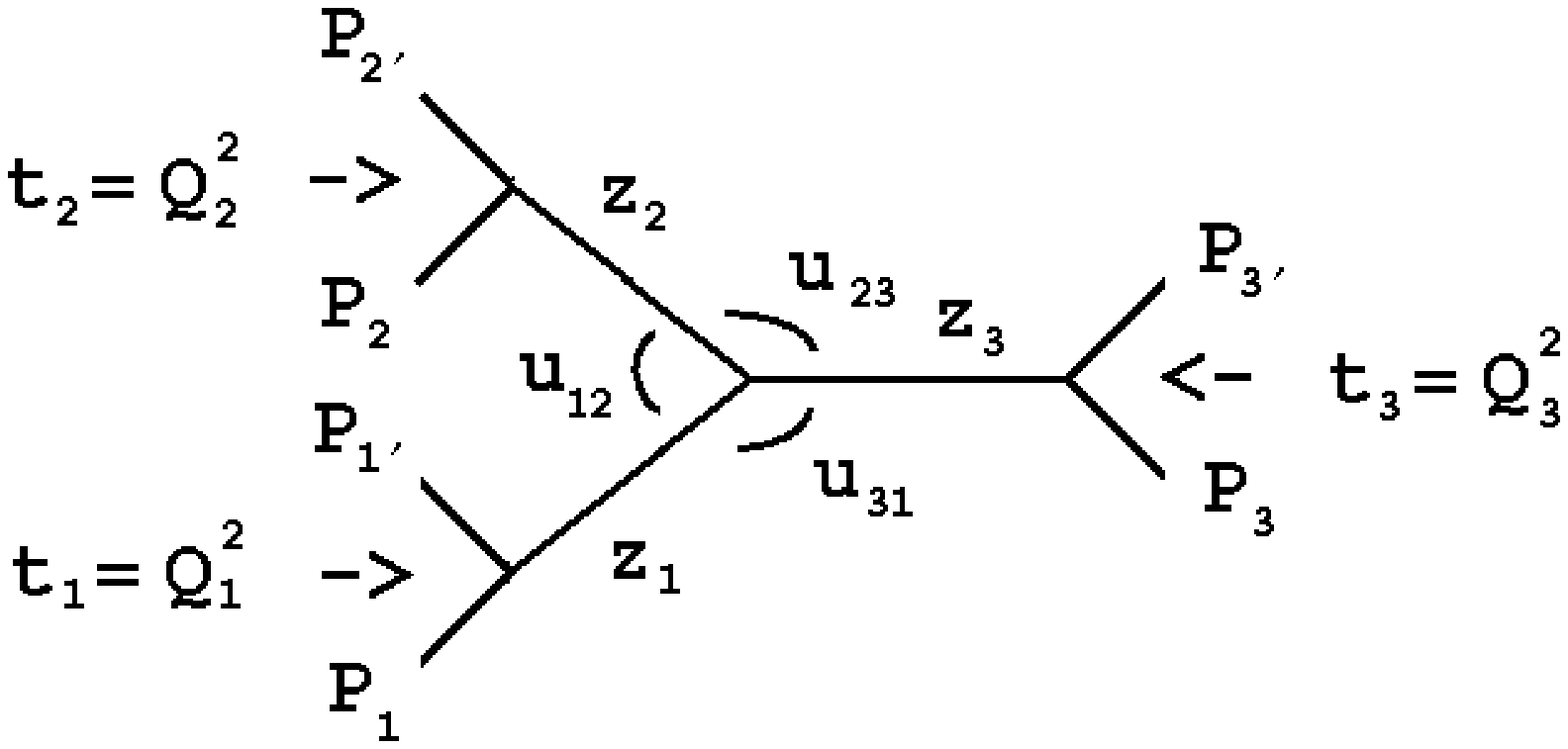}

Fig.~5.1 A Toller Diagram for the Six-Particle Amplitude
\end{center}
The definition of these variables via standard Lorentz frames 
as well as complete expressions for
invariant variables in terms of them, is given in Appendix D. 
We discuss their definition in
three $t$-channel physical regions where the $t_i$ are positive 
and in four $s$-channel physical regions where the $t_i$ are
negative. The variables introduced in the different regions are related by
analytic continuation. 
The $z_i$ (and the $t_i$) are independent variables but since 
$u_{ij}= e^{~i\omega_{ij}} = e^{~i(\nu_i -\nu_j)} $ where each $\nu_i$ is an 
azimuthal angle in the $t_i$-channel, we have 
$$
u_{12}~u_{23}~u_{31}~=~e^{~i(\nu_1 -\nu_2)}~ e^{~i(\nu_2 -\nu_3)} ~
e^{~i(\nu_3 -\nu_1)} ~=~ 1
\auto\label{51}
$$
Choosing any two of the $u_{ij}$, together with the $t_i$ and the $z_i$, gives 
the appropriate eight independent variables. In the following we will take 
$u_{31}$ and $u_{32} = u^{-1}_{23}$ as our independent variables and, for 
simplicity, relabel them as $u_1=u_{31}$ and $u_2=u_{32}$. (\ref{51}) then
gives
$$
u_{12}~ = ~{ u_2 \over u_1 }
\auto\label{52}
$$
This choice of $u_{ij}$ variables is appropriate for discussing the
particular multiple discontinuities or, in the classification we introduce
below, the particular hexagraphs that we focus on in the following. As we will
indicate, the alternative choices are appropriate for other hexagraphs.

In principle, since we will be considering the scattering of particles with 
spin (i.e. quarks) we should 
add additional azimuthal angles to describe the rotation of helicities. 
However, as we already saw in the last Section, for the lowest-order
quark-gluon couplings there is ($s$-channel) helicity-conservation.
It is trivial to carry out helicity projections and show that the
lowest-order couplings are also helicity-independent. In discussing anomaly
cancelations at the end of the last Section we saw that the higher-order
helicity dependence of these couplings is important. However, we will 
discuss the consequences of this only qualitatively in Section 7. To keep 
our discussion in this Section as simple as possible we will treat the
scattering quarks kinematically as if they were scalar particles. Also,
since this Section will be concerned with abstract kinematics and
analyticity properties, we will continue to ignore color (and any other)
quantum numbers. In Section 7, both color quantum numbers and the
existence of two helicities will be important 
when we discuss discrete $C$, $P$ and $T$
transformations on amplitudes. In this Section we will refer only 
to the $CPT$ combination. Nevertheless the presence of both helicity and 
color should be kept in mind. 

As we show in Appendix D, all invariants are polynomial functions of
the $z_i= cos~\theta_i$, the $sin~\theta_i$ and the $cos~\omega_{ij}$ 
($u_{ij}=e^{i~\omega_{ij}}$). The following approximations give 
the leading behavior when all the $z_i$ are large and are easily derived
from (\ref{inv33}) and (\ref{inv330}). These approximations will be sufficient 
for us to describe the behavior of invariants in the limits we discuss. 
\beqa
s_{122'} ~\sim~ s_{1'3'3}~\sim ~- s_{1'22'} ~\sim~ - s_{13'3} 
~~&\to& ~2\biggl({t_1 - 4m^2 \over t_1} \biggr)^{1 \over 2}
\lambda^{ 1 \over 2}(t_1,t_2,t_3) ~z_1 \nonumber\\ 
&\sim& ~~~z_1 \label{ap1} \\
s_{233'} ~\sim~ s_{2'1'1}~\sim ~- s_{2'33'} ~\sim~ - s_{21'1} 
~~&\to& ~2\biggl({t_2 - 4m^2 \over t_2} \biggr)^{1 \over 2}
\lambda^{ 1 \over 2}(t_1,t_2,t_3) ~z_2 \nonumber \\
&\sim& ~~~ z_2\label{ap2} \\
s_{311'} ~\sim~ s_{3'2'2}~\sim ~- s_{3'11'} ~\sim~- s_{32'2} 
~~&\to& ~2\biggl({t_3 - 4m^2 \over t_3} \biggr)^{1 \over 2}
\lambda^{ 1 \over 2}(t_1,t_2,t_3) ~z_3 \nonumber \\
&\sim& ~~~ z_3 \label{ap3} 
\eeqa
\newpage 
\beqa
s_{13} ~&\sim&~ s_{1'3'}~\sim~ -s_{13'}~\sim~ -s_{1'3} ~~~\to \nonumber \\
&-& 4 (t_1 - 4m^2 )^{1 \over 2}(t_3 - 4m^2 )^{1 \over 2}
\biggl[(1-z_1^2)^{1 \over 2}(1-z_3^2)^{1 \over 2} \biggl(u_1 + {1 \over u_1}
\biggr) +  {t_3 + t_1 - t_2 \over \sqrt{t_1} \sqrt{ t_3} } z_1 
z_3 \biggl] \nonumber \\
&\sim& ~~ z_1~ z_3 ~u_{1}~[ 1  + O({1 \over u_1})]   \label{ap4} \\
s_{23} ~&\sim& ~s_{2'3'}~\sim~ -s_{23'}~\sim ~-s_{2'3} ~~~\to \nonumber \\
&-& 4 (t_2 - 4m^2 )^{1 \over 2}(t_3 - 4m^2 )^{1 \over 2}
\biggl[(1-z_2^2)^{1 \over 2}(1-z_3^2)^{1 \over 2} \biggl(u_2 + {1 \over u_2}
\biggr) +  {t_3 + t_2 - t_1 \over \sqrt{t_2} \sqrt{ t_3} } z_2 
z_3 \biggl] \nonumber \\
&\sim& ~~ z_2~ z_3~ u_2^{-1}~[1  + O(u_2)]  \label{ap5} \\
s_{12} ~&\sim& ~s_{1'2'} ~\sim~ -s_{12'}~ \sim~ -s_{1'2} ~~~\to \nonumber \\
&-& 4 (t_1 - 4m^2 )^{1 \over 2}(t_2 - 4m^2 )^{1 \over 2}
\biggl[(1-z_1^2)^{1 \over 2}(1-z_2^2)^{1 \over 2} \biggl({u_1 \over u_2} + 
{u_2 \over u_1} \biggr) +  {t_3 - t_1 - t_2 \over \sqrt{t_1} \sqrt{ t_2} } z_1 
z_2 \biggl] 
\nonumber \\
&\sim& ~~ z_1~ z_2~ (u_1 / u_2 )~[ 1 + O( u_2 / u_1)]
\label{ap6}  
\eeqa
$\lambda(t_1,t_2,t_3)$ is the familiar triangle function defined 
explicitly in (\ref{lam}). The branch-points at $\lambda(t_1,t_2,t_3) = 0 $ 
in (\ref{ap1}) - (\ref{ap3}) play an important role when analytic continuation
is discussed, both between the different physical regions and within a fixed
$s$-channel physical region.
Note that all invariants are unchanged when $u_1 \to 1/u_1, u_2 \to 1/u_2$. 

In each of the three $t$-channels $t_i= Q^2_i > 4m^2$ and 
$\lambda(t_1,t_2,t_3) > 0$. Also, in the particular channel that we refer to as
the $t_i$ channel
$$
|Q_i| > |Q_j| + |Q_k|
\auto\label{tchi}
$$
The choice of signs in (\ref{ap1})-(\ref{ap3}) is a convention which is 
irrelevant in the $t$-channels since 
$$ 
-1 ~\leq ~z_i ~\leq ~1, ~~~|u_i| ~=~1 
\auto\label{tch}
$$
i.e. the $z_i$ take both positive and negative values. Note, however, that 
the sign convention can be reversed by, for example, interchanging the role 
of $Q_1$ and $Q_2$ when introducing the variables via standard frames 
defined in the $t_3$-channel. This is discussed in Appendix D.

In each $s$-channel physical region there are four sub-regions\cite{gw} 
distinguished 
by the relative value of the $t_i$. There are three ``$s-t$'' 
regions in which $\lambda(t_1,t_2,t_3) >0 $ and one of the three $t$-channel
constraints (\ref{tchi}) is satisfied.
The ``$s-s$'' region is the remaining part of the physical region 
in which $\lambda(t_1,t_2,t_3) < 0 $. 
The relationship between the three $t$-channels and the sub-regions of one 
$s$-channel is illustrated topographically in Fig.~D2. For our analysis of
anomaly cancelations it is important 
that a change of sign of all the $z_i$ is equivalent to a change of sign
of $\lambda^{1 \over 2}(t_1,t_2,t_3) $. This is apparent from (\ref{ap1}) -
(\ref{ap6}). 

Each of the three $s-t$ sub-regions of the 
four $s$-channel physical regions also 
has two distinct parts, in one of which 
the $z_i$ each have a certain sign and the other in which they all have the 
opposite sign. This removes the antisymmetry of the signs in
(\ref{ap1})-(\ref{ap3}). Which part of the physical region corresponds to a 
particular set of signs of the $z_i$ is a convention determined by the choice 
of sign for $\lambda^{1 \over 2}(t_1,t_2,t_3) $.
When the $t_i$ satisfy the $s-t$ sub-region constraint the 
four $s$-channel physical regions are
$$
\eqalign{ ~~~~~~~~~~
i)& ~~~ z_1,~z_2,~z_3~ \geq ~1 ~, ~~~~~ z_1,~z_2,~z_3~ \leq ~-1 
~, ~~~|u_i| ~=~1 \cr
&\hbox{the initial particles carry momenta $P_1,P_2$ and $P_3$} \cr 
ii)& ~~~ - z_1,~z_2,~z_3~ \geq ~1 ~, ~~~~~ - z_1,~z_2,~z_3~ \leq ~-1
~, ~~~|u_i| ~=~1 \cr
&\hbox{the initial particles carry momenta $P_{1'},P_2$ and $P_3$} \cr 
iii)& ~~~ z_1,~- z_2,~z_3~ \geq ~1 ~, ~~~~~ z_1,~- z_2,~z_3~ \leq ~-1
~, ~~~|u_i| ~=~1 \cr
&\hbox{the initial particles carry momenta $P_1,P_{2'}$ and $P_3$} \cr 
iv)& ~~~ z_1,~z_2,~- z_3~ \geq ~1 ~, ~~~~~ z_1,~z_2,~- z_3~ \leq ~-1
~, ~~~|u_i| ~=~1  \cr
&\hbox{the initial particles carry momenta $P_1,P_2$ and $P_{3'}$} }
\auto\label{zprs}
$$
We will encounter subtleties associated with the doubling of 
the range of the $z_i$ in individual physical regions at several points in the
following. The physical region $i)$ is that in which the limits of Section 2
are defined and in which the diagrammatic analysis of the last Section is
carried out. That the discontinuities apparently evaluated in Fig.~4.8(a) are
no longer present in the region where the anomaly occurs is a particular 
consequence of the separation of the physical region into two distinct parts.

In the $s-s$ sub-region the physical range of the 
$z_i$ and $u_i$ in the same four physical regions is 
$$
\eqalign{~~~~~~~~~~ i)& ~~~ - \infty ~< ~ i z_i ~ < \infty ~, ~~~~ 0 ~\leq ~  
u_1,~u_2 ~<~ \infty ~~~~~~~~~~~ \cr
ii)& ~~~ - \infty ~< ~ i z_i ~ < \infty ~, ~~~~ 0 ~\leq~  - u_1,~u_2 ~
<~ \infty \cr
iii)& ~~~ - \infty ~< ~ i z_i ~ < \infty ~, ~~~~ 0 ~\leq~  u_1,~- u_2 ~
<~ \infty  \cr
iv)& ~~~ - \infty ~< ~ i z_i ~ < \infty ~, ~~~~ 0 ~\leq ~ - 
u_1,~- u_2 ~ <~ \infty \cr }
\auto\label{uprs}
$$
Clearly the kinematic structure of the $s$-channel physical regions is 
quite complicated. This is reflected in the variety of kinematic limits that
can be taken. Nevertheless, all the limits are described by the same multi-regge
theory.

\subhead{5.2 Definition of Limits via Angular Variables}

The full triple-regge limit is defined to be 
$$
z_1,~z_2,~z_3 ~\rightarrow ~\infty~, ~~~~t_1,t_2,t_3,u_{1},u_{2}~~fixed
\auto\label{trl}
$$
Helicity-pole limits are those in which one or 
two of the $u_{ij}$ are taken either large or small. This can be, but need 
not be, combined with taking one or more of the $z_i$ large.

We will distinguish two distinct helicity-pole limits involving
$u_1$ and $u_2$. The first is 
$$ 
z_3,~u_{1},~ u_{2}~ \rightarrow ~\infty \ ~
({\rm or}\ u_{1}, u_{2} \rightarrow 0) ~~~~t_1,t_2,t_3,z_{1},z_{2}~~fixed
\auto\label{hp1} 
$$
When applied to the relevant discontinuity (with  $P_1$ and $P_2$, and 
$P_1'$ and $P_2'$, respectively 
identified) this limit coincides with the familiar
(incorrectly named) ``triple-regge'' limit of the one-particle inclusive
cross-section. The second helicity-pole limit is 
$$ 
z_3,~u_{1},~ u_{2}^{-1}~ \rightarrow ~\infty \ ~
({\rm or}\ u_{1}, u_{2}^{-1} \rightarrow 0) ~~~~t_1,t_2,t_3,z_{1},z_{2}~~fixed
\auto\label{hp2}
$$
For reasons that will soon become apparent, we refer to the limit 
(\ref{hp1}) as 
the ``non-flip limit'' and the limit (\ref{hp2}) 
as the ``helicity-flip limit''. The helicity-pole limits are formulated in 
terms of the $u_{ij}$ variables we have chosen and, as we discuss further below,
they are controled by singularities in corresponding complex helicity
planes directly related to angular momentum plane Regge singularitites. 
Clearly we can define corresponding limits for any choice of the $u_{ij}$.

In an $s-t$ part of an $s$-channel physical region the triple-regge limit is
a physical limit but the helicity-pole limits are not. 
In the $s-s$ region both the triple-regge limit 
and the helicity-pole limits are physical. 
With the approximations (\ref{ap1}) - (\ref{ap6}) the inter-relation
between the helicity-pole limits (\ref{hp1}) and (\ref{hp2})
and the triple-Regge limit (\ref{trl}) is apparent. It is also
straightforward, using (\ref{npl1}), 
to identify the triple-Regge limit (\ref{trl}) and the light-cone 
formulated limit (\ref{np3}). Clearly 
$$
P_1 ~\sim ~z_1 ~, ~~~~P_2 ~\sim ~z_2 ~, ~~~~P_3 ~\sim ~z_3 ~.
\auto\label{trl1}
$$
For the helicity non-flip limit (\ref{hp1}) we have, from
(\ref{ap1}) - (\ref{ap6}), 
\beqa  s_{12}~&\sim&~s_{1'2'}~ \sim~ - s_{12'} ~\sim ~- s_{1'2} ~
\st{\to} ~\infty \label{hp011} \\
 s_{23}~&\sim& ~- s_{2'3} ~\sim ~z_3~u_2
~, ~~~ s_{31}~\sim ~s_{3'1} ~\sim ~z_3~u_1~ \label{hp012} 
\eeqa
$$
s_{122'}~,~ s_{133'}~,~ s_{233'}~\st{\to} ~\infty ~.
~~~~~s_{311'}~\sim s_{3'22'}~\sim ~z_3 
\auto\label{hp013}
$$
Because $s_{12}~\st{\to}~\infty$ we can not reproduce the non-flip
limit (\ref{hp1}) with
the light-cone variables of (\ref{np3}). However, for the helicity limit 
(\ref{hp2}), the behavior (\ref{hp011}) is replaced by
$$
s_{12}~\sim~s_{1'2'}~ \sim~ - s_{12'} ~\sim ~- s_{1'2} ~
\sim ~u_1~u_2^{-1}
\auto\label{hp02}
$$
Therefore, we can formulate the helicity-flip limit 
in terms of the variables of (\ref{np3})  as 
$$
P_1 ~\sim ~u_1 ~, ~~~~P_2 ~\sim ~u_2^{-1} ~, ~~~~P_3 ~\sim ~z_3 ~.
\auto\label{hp03}
$$
together with
$$
\eqalign{q_{12^-}=~\hat{q}_1 - q_{12}~=q_{32} - \hat{q}_3 
~& \to 0~,~~~~q_{21^-}=~ \hat{q}_2 - q_{21}~= q_{31} - \hat{q}_3 
~ \to \cr  
\hat{q}_1 - q_{13} ~&\st{\to} ~0~, ~~~ \hat{q}_2 - q_{23} ~\st{\to}~0 }
\auto\label{hp04}
$$

We can also take the helicity-flip limit 
(\ref{hp2}) in conjunction with the triple-regge limit so that
$$
z_1,~z_2,~z_3,~u_1~u_2^{-1}~ \rightarrow ~\infty~, ~~~~t_1,t_2,t_3,~~fixed
\auto\label{trh1}
$$
Note, from (\ref{ap4})-(\ref{ap6}), that in this last limit
$$
 s_{23}~\sim~ z_3~u_2^{-1} ~, ~~~ s_{31}~\sim ~z_3~u_1~, ~~~s_{12} ~\sim ~
z_1~u_1~z_2~u_2^{-1} 
\auto\label{trh11}
$$
and so if $z_1\sim z_2 \sim z_3$ then 
$$
s_{23},~s_{13}~ ~<< ~~s_{12}
\auto\label{trh12}
$$

It is important to 
keep account of the $t_1$ and  $t_2$ dependance in (\ref{ap1})-(\ref{ap6}) 
when the limit (\ref{trh1}) is taken with $t_1,t_2 \sim 0$. Therefore, 
using (\ref{npl2}) and (\ref{npl3}) we can write (for small $q_{21^-},q_{12^-}$)
$$ 
u_1~\sqrt{t_1} ~ +~ O(1)  ~\sim ~{s_{13} \over s_{133'}~ s_{311'}} 
~~=>~~ u_1 
~\sim ~{1 \over \sqrt{t_1}~ ~q_{21^-} }
\auto\label{trh2}
$$
$$ 
u_2^{-1} ~\sqrt{t_2} ~ +~ O(1)  ~\sim ~{s_{23} \over s_{233'}~ s_{322'}} 
~~=>~~ u_2^{-1} 
~\sim ~{1 \over \sqrt{t_2} ~~q_{12^-} }
\auto\label{trh3}
$$

As we will see below, the 
helicity-flip limit selects leading (flipped) helicities from the full
triple-Regge vertex. (The non-flip limit similarly selects non-flipped 
leading helicities). (\ref{trh2}) and (\ref{trh3}) imply that while the 
helicity-flip is directly expressed as 
$$
q_{12^-}~,~ q_{21^-}~\to~0~, ~~~~~~ t_1=q_1^2,~t_2=q_2^2,~ fixed
\auto\label{trh4}
$$
this limit is also reached if 
$$
q_1^2~,~ q_2^2~\to~0~, ~~~~~~ q_{12^-},~q_{21^-},~t_3 ~~fixed
\auto\label{trh5}
$$

\subhead{5.3 Dispersion Theory and Asymptotic Cut Structure}

A fundamental ingredient for our multi-regge analysis is the
existence\cite{arw1,sw} of an ``asymptotic dispersion relation'' that breaks
the full triple-regge asymptotic amplitude up into components that each have
a distinct set of asymptotic cuts. The dispersion relation is written in 
$z_1,z_2$ and $z_3$ with $t_1,t_2,t_3,u_1$ and $u_2$ kept fixed. It is 
initially written with all the $t_i < 0$ and with
$\lambda(t_1,t_2,t_3) >0 $ so that physical contributions are obtained from an
$s-t$ region of each of the four $s$-channels. However, we expect the form of
the dispersion relation to remain unchanged as we continue between $s$-channel
sub-regions and also to the $t_i$-channels.

The most important feature\cite{arw1,sw} is
that the asymptotic cut structure of multiple discontinuities can
be treated as if there were only normal threshold cuts satisfying the
Steinmann relations. i.e. no double discontinuities in overlapping channels.
This asymptotic structure, in turn, matches naturally with the asymptotic
formulae obtained from multi-regge theory. At a fundamental level, this match
is presumably a consequence of the close relationship between multi-regge
analyticity and the primitive analyticity domains of field theory, i.e. the
simple off-shell analyticity properties of field theory survive asymptotically
on-shell\cite{arw00}. In physical regions the asymptotic validity of the
Steinmann relations can be derived within S-Matrix Theory by showing that the
``bad boundary-values'' in which a variety of complications due to
higher-order singularities appear, are hidden in multi-regge limits. In the
particular dispersion relation that we use in this paper an additional
fundamental complication arises in that there are essential contributions from
non-physical triple discontinuities.

The discontinuities involved are physical in two-four
scattering processes, but not in the three-three processes we discuss.
As a matter of principle, the presence of discontinuities outside the physical
region, as well as their explicit form, can not be discussed directly from
an S-Matrix starting-point. However, as we alluded to above, 
the dispersion relation can also be based on 
the field theory formalism of Generalised Retarded Functions, by starting 
with spacelike masses and utilising the primitive analyticity 
domains\cite{arw00,cs}. The asymptotic structure of the dispersion relation
should persist straightforwardly on-shell, with the standard discontinuity
formulae holding. The triple discontinuities that are unphysical in our
problem will appear directly in the field theory formulation 
just because they satisfy the Steinmann
relations. As we will see, these discontinuities are manifestly present in
Feynman diagrams and they are esssential to obtain
consistent multi-regge behavior. In fact, as we have already noted several 
times, they are crucial for the appearance of the anomaly.

The Steinmann relations are not satisfied by individual Feynman graphs and 
it is a subtle feature that they emerge
asymptotically. As a consequence the matching of
multi-regge behavior with features of the graphs studied in the previous and 
succeeding chapters will also be subtle. In particular, as we noted above,
in our discussion of the anomaly configuration of Fig.~4.15 we found that 
discontinuities present in one part of the physical region, and represented
by Fig.~4.8(a), are no longer present in another part. A property that is 
only possible because of the violation of the Steinmann relations by the 
graphs involved. (The turning off of the branch-cuts involved 
requires the existence of simultaneous singularities in additional variables 
that are forbidden by the Steinmann relations.) 

The dispersion relation gives the leading triple-regge 
behavior (up to powers) as 
a sum over triple discontinuity contributions 
allowed by the Steinmann relations, i.e. we can write 
$$
M(P_1,P_2,P_3,Q_1,Q_2,Q_3)~ =~ 
\sum_{\cal C} M^{\cal C}(P_1,P_2,P_3,Q_1,Q_2,Q_3)
~+~M^0~,\auto\label{dis}
$$
$M^0$ contains all non-leading 
triple-regge behavior, double-regge behavior, etc. and the sum is
over all triplets ${\cal C}$ of three non-overlapping,
asymptotically distinct,  
cuts. For each triplet ${\cal C}$ of cuts 
in invariants, say ${\cal C}= (s_1,s_2,s_3)$, we write 
$$  
\eqalign{M^{\cal C}(P_1,P_2,P_3,Q_1,Q_2,Q_3)~=~{1\over (2\pi i)^{3}}  &~~\int
ds'_1 ds'_2 ds'_{3} ~~{\Delta^{\cal C}(\til{t},
\til{u},s'_1,s'_2,s'_{3})\over
(s'_1-s_1)(s'_2-s_2)(s'_{3}-s_{3})}\cr
&\{s_i > s_{i0},\forall i \} }
\auto\label{dis2}
$$
where $\Delta^{\cal C}$ is the triple discontinuity
$$                                   
\Delta^{\cal C} (\til{t},
\til{u},s_1,s_2,s_{n-3})~=~
\sum_{\epsilon}(-1)^{\epsilon}M(\til{t},
\til{w},s_1 \pm i0,s_2\pm i0,s_{3}\pm i0),
\auto\label{dis4}
$$
The sum over $\epsilon$ is over all combinations of $+$ and $-$ signs in 
(\ref{dis4}) and $(-1)^{\epsilon}$ is positive when the number of $+$ signs 
is even. The integration region in (\ref{dis2})
is bounded by finite, but arbitrary, values $s_{i0}$ of the $s_i$ and the
asymptotic relation between the $z_i,$ and $s_i$ has to be used to change
variables from the $z_i$ back to the $s_i$. 
Because of the validity of the Steinmann relations, $\Delta^{\cal C}$ 
can be expressed in terms of normal phase-space integrals. 
Therefore, we can take multiple discontinuities simply by
putting appropriate lines on mass-shell. (For the 
low-order Feynman diagrams we discuss the subtlety of the 
boundary values for the amplitudes in the discontinuity formulae\cite{sw}  
will not appear.)

\subhead{5.4 Triple Discontinuities}

If we consider only the $z_i$ dependence, the asymptotic relations 
(\ref{ap1})-(\ref{ap6}) reduce to 
\beqa
s_{122'} ~&\sim&~ s_{1'3'3}~\sim ~- s_{1'22'} ~\sim~ - s_{13'3} 
~\sim ~~z_1 \nonumber \\
s_{233'} ~&\sim&~ s_{2'1'1}~\sim ~- s_{2'33'} ~\sim~ - s_{21'1} 
~\sim ~~ z_2 \nonumber \\
s_{311'} ~&\sim&~ s_{3'2'2}~\sim ~- s_{3'11'} ~\sim~- s_{32'2} 
~\sim ~~ z_3 \nonumber \\
s_{13} ~&\sim&~ s_{1'3'}~\sim~ -s_{13'}~\sim~ -s_{1'3} 
~\sim ~~ z_1~ z_3 \nonumber \\
s_{23} ~&\sim&~ s_{2'3'}~\sim~ -s_{23'}~\sim ~-s_{2'3} 
~\sim ~~ z_2~ z_3~ \nonumber \\
s_{12} ~&\sim&~ s_{1'2'} ~\sim~ -s_{12'}~ \sim~ -s_{1'2} 
~\sim ~~ z_1~ z_2~  \label{zap}  
\eeqa
As we discussed in Section 2, the triple discontinuities 
are of three kinds corresponding to the tree diagrams of Figs.~2.9(a), (b) 
and (c). These are the distinct possibilities consistent with the Steinmann 
relations. An example of the Fig.~2.9(a) kind is
$$
{\cal C}_a ~= ~(s_{13},s_{2'3'}, s_{11'3})
\auto\label{cuta}
$$
From (\ref{zap}), the invariants of ${\cal C}_a $
are large and positive when
$$
z_1z_3,~z_2z_3,~z_3~>> 1~~~\leftrightarrow ~~~z_1,~z_2,~z_3~>> 1
\auto\label{zap1}
$$
giving a unique product of (asymptotic) $z_i$ half-axes lying in the first 
part of the $i)$ physical region in (\ref{zprs}). 
All triplets of the Fig.~2.9(a) kind 
similarly correspond to a unique product of (asymptotic) $z_i$ half-axes. 
There are 6 possible combinations of initial and final subenergies so there 
are 6 triple discontinuities of this kind in each physical region, three in 
each part, making a total of 24.

A triplet having the form of Fig.~2.9(b) is
$$
{\cal C}_b ~= ~(s_{13},s_{2'3'}, s_{123}) 
\auto\label{cutb}
$$
Since 
$$
s_{123}~=~ s_{31} ~+~ s_{12} ~+~s_{23} ~-~ 3 m^2
\auto\label{cutb0}
$$
(where $m$ is the mass of the scattering particles) 
the cut in $s_{123}$ will be distinguished from the asymptotic cuts in $s_{31}$ 
and $s_{23}$ only when $s_{12}$ is large. Consequently, the contribution 
of the $s_{123}$ cut, as a distinct asymptotic cut, is effectively as a cut
in $s_{12}$ (particularly in the helicity-pole limit discussed below in which
$s_{12} >> s_{13},s_{23}$). 
In the following, therefore, we will use (\ref{ap6}) also as an
approximation for $s_{123}$. In this case, the invariants of ${\cal C}_b$
are large and positive when
$$
z_1z_3,~z_2z_3,~z_1z_2~>> 1~~~\leftrightarrow ~~~z_1,~z_2,~z_3~>> 1~, ~~
z_1,~z_2,~z_3~<< -1
\auto\label{zap2}
$$
Now there are two regions for each triplet, one in each part of the relevant 
physical region. A closely related complication is that the invariants in 
the triplet
$$
{\cal C}_{b'} ~= ~(s_{1'3'},s_{23}, s_{1'2'3'})
\auto\label{cutb1}
$$
are also real and positive in the two regions of (\ref{zap1}). The two 
sets of cuts ${\cal C}_b$ and ${\cal C}_{b'}$ both satisfy the Steinmann 
relations. Since they are asymptotically equivalent 
the multi-regge representations we derive will not be able to distinguish 
between them. (Their existence, however, is another feature that is crucial 
for the potential appearance of the anomaly.) 
If we start from Feynman graphs 
each triple discontinuity has to be computed separately and added to the
dispersion relation. All triple discontinuities of this kind similarly occur
in equivalent pairs and also appear in both parts of the
physical region involved. Since each triplet is again characterised by an
initial and final subenergy there are 24 in total, or twelve equivalent pairs
- three in each physical region.

A triplet of the Fig.~2.9(c) kind is
$$
{\cal C}_c ~= ~(s_{13},s_{2'3'}, s_{1'2}) 
\auto\label{cutc}
$$
These three invariants are not all positive in any physical region. However,
they are all positive if 
$$
z_1z_3,~z_2z_3,~- z_1z_2~>> 1~~\leftrightarrow ~-iz_1,~-iz_2,~iz_3~>> 1~, 
~~iz_1,~iz_2,~-iz_3~<< -1
\auto\label{0zap2}
$$
Therefore each triplet of this kind gives two unphysical region contributions 
(on the asymptotic imaginary axes) to the asymptotic dispersion relation. For
the discontinuities to have physical intermediate states,
the invariants should satisfy one of the constraints
$$
s_{13}~\geq(\sqrt{s_{2'3'}}+ \sqrt{s_{1'2}})^2~,~~~
s_{2'3'}~\geq(\sqrt{s_{13}}+ \sqrt{s_{1'2}})^2~,~~~
s_{1'2}~\geq(\sqrt{s_{2'3'}}+ \sqrt{s_{13}})^2
\auto\label{prcs}
$$
This separates the triple discontinuity (\ref{cutc}) into three components 
in each of the physical regions. This separation is clearly well-defined in 
the (helicity-pole) limit in which one of the invariants is much larger than the
other two and this will be sufficient for our purposes. The triplet 
$$
{\cal C}_{c'} ~= ~(s_{1'3'},s_{23}, s_{12'}) 
\auto\label{cutc1}
$$
is asymptotically equivalent to the triplet ${\cal C}_{c}$. Indeed, apart 
from their 
occurrence in unphysical regions, the triplets of the Fig.~2.9(c) kind share
all the doubling and pair-wise equivalence properties of the Fig.~2.9(b) kind. 
We will see that the two kinds have closely related asymptotic representations.

Each of the asymptotic equivalences we have discussed identifies initial and
final state discontinuities. Such discontinuities
typically arise from distinct
contributions in individual Feynman graphs. However, the multi-regge theory 
we will develop requires the multiple discontinuities to be closely related.
Note that when the physical range for the $z_i$ is pure imaginary, in 
an $s-s$ sub-region, each unphysical triple discontinuity region asymptotes 
to the real axes on which a physical region
appears in the $s-t$ subregions. This will mean that the standard definition 
of signatured amplitudes can be applied for the unphysical triplets in the
$s-s$ regions.

\subhead{5.5 Hexagraph Notation for Triple Discontinuities}

To develop our multi-regge analysis we introduce a ``hexagraph'' 
notation\cite{arw1,sw}
for classifying the triple discontinuities. The hexagraphs link each triple 
discontinuity to a particular $t$-channel and determines it's contribution to
asymptotic behavior via a Sommerfeld-Watson representation. Our inclusion 
of the unphysical triple discontinuities will involve conventions that may 
appear somewhat arbitrary but it will be clear that the asymptotic
representations we eventually obtain for both the Fig.~2.9(b) and the
Fig.~2.9(c) triple discontinuities are essentially independent of how they are
mapped onto the hexagraph formalism.

The full sum over triple discontinuities in (\ref{dis}) is broken up into
partial sums forming a hexagraph amplitude. Each hexagraph amplitude $M^H$
contains a sum of triple discontinuity integrals, i.e.
$$
M^H=~\sum_{{\cal C} \epsilon H} M^{\cal C}(P_1,P_2,P_3,Q_1,Q_2,Q_3)
,\auto\label{dis1}
$$
where the sum is  
over all triplets ${\cal C}$ of asymptotic cuts in which 
each cut is an ``allowable discontinuity'' of the
hexagraph H.

The hexagraphs associated with a particular 
Toller diagram are obtained by redrawing
the tree diagram in all possible ways (in a plane) with the internal lines
drawn as horizontal lines and the internal vertices drawn separately, with 
relative angles of $120^o$, and joined to the horizontal lines. 
For the Toller diagram of Fig.~5.1 we draw an initial hexagraph, say the 
first graph of Fig.~5.2,  
and then form the additional graphs of Fig.~5.2 by cyclical
rotation of the $t_i$-channels.
\begin{center}
\leavevmode
\epsfxsize=4.8in
\epsffile{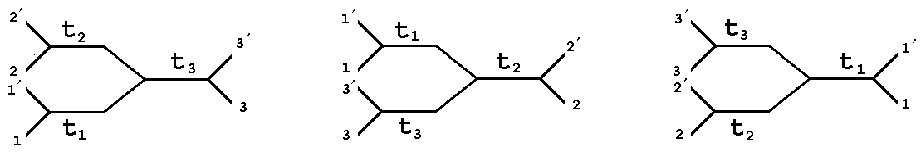}

Fig.~5.2 Hexagraphs Related by Cyclical Rotation of the $t_i$-channels
\end{center}

Hexagraphs have many uses in addition to the classification of discontinuities
that we describe below. One of the simplest is that, as we noted above,
an independent set of angular variables (and their conjugate angular momentum
and helicity variables) can be put in one-to-one correspondence with the lines
of the graph\footnote{Although we will make only minimal reference to it in
this paper, this correspondence plays a vital role in all aspects of the
multi-regge theory for a general multiparticle amplitude.}. The $z_i$ (and
conjugate $J_i$) variables can be associated with the horizontal lines while 
an independent set of the $u_{ij}$ can be associated with the sloping lines.
Helicity-pole limits can then also be associated directly with a hexagraph.
For the first hexagraph of Fig.~5.2 (and all those of Fig.~5.3) the $u_1$ and 
$u_2$ variables used above are naturally associated with the sloping lines, 
while for the other two graphs in Fig.~5.2 one of the alternative choices of 
the $u_{ij}$ is appropriate. 

We form a further set of hexagraphs from each of those shown in 
Fig.~5.2 by making twists (of one half of the graph relative to the other)
about each of the horizontal lines of the graph. 
In Fig.~5.3 we have shown
again the first hexagraph of Fig.~5.2 together with the seven hexagraphs
related to it by twisting.
Twisting also the other two hexagraphs in Fig.~5.2 gives a total of
$(2\times 2\times 2 = 8) \times 3 = 24 $ which is the total number of
hexagraphs associated with the Toller diagram of Fig.~5.1.
\begin{center}
\leavevmode
\epsfxsize=4.5in
\epsffile{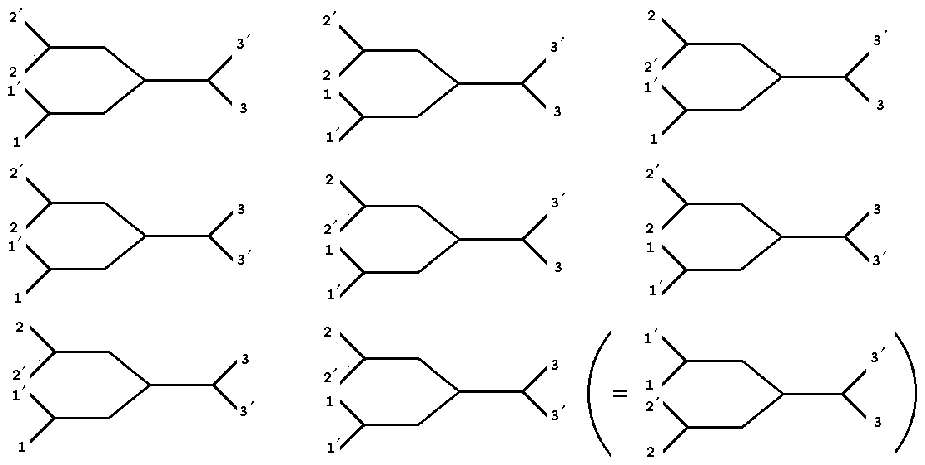}

Fig.~5.3 Eight Hexagraphs Related by Twisting
\end{center}

The multiple discontinuities associated with a particular hexagraph all
appear in (or, in the case of the unphysical discontinuities, 
are associated with) the same part of a particular $s$-channel physical
region, which we therefore associate directly with the graph. The physical
region involved is obtained by regarding the external scattering particles as
entering from the bottom of the hexagraph and exiting at the top. (In a loose
way, the hexagraph represents a time-ordering
in the scattering with the time axis vertical on the page.) Each
hexagraph is also associated with a particular $t$-channel. This channel is 
obtained by interpreting the graph as representing a scattering in which
external particles enter from the left of the hexagraph and exit to the right.
A twist of a hexagraph produces a change of $s$-channel, but leaves the
$t$-channel unchanged. 

Each of the three hexagraphs in Fig.~5.2 is associated with the
same part of the same $s$-channel which, with the conventions we have chosen, 
is $z_1,z_2,z_3 \geq 1$, i.e. the first part of region $i)$ of (\ref{zprs}).
The first is associated with the $t$-channel in which $|Q_3| ~>~ |Q_1|+|Q_2|$,
as are all the graphs of Fig.~5.3. The second and third graphs in Fig.~5.2
are associated, repectively, with the $t$-channels in which 
$|Q_2| ~>~ |Q_1|+|Q_3|$ and $|Q_1| ~>~ |Q_3|+|Q_2|$. Twisting similarly 
generates all the other graphs associated with these $t$-channels giving, 
finally, 3 hexagraphs for each part of each $s$-channel and 8 hexagraphs for
each $t$-channel.

The partial-wave analysis that follows this sub-section, introducing complex
angular momenta and helicities, is carried out in the $t$-channel. 
A twist gives a change of sign of the $z_i$ (and, if it exists, the $u_i$)
associated with the chosen line and so is associated with signature. We do not
distinguish scattering processes related
by a $CPT$ transformation which interchanges all incoming particles with all
outgoing particles. 
We could equally well regard the $s$-channel scattering particles as
entering from the top of the diagram and the $t$-channel scattering particles 
as entering from the right of the diagram. 
(Although, as we noted at the beginning of this Section, 
it is important to note that there are helicities and color quantum numbers 
that distinguish the amplitudes related by a $CPT$ transformation.) 

In Fig.~5.3  we have also shown the $CPT$ reversed version of the last 
hexagraph to emphasize that the associated $s$-channel 
is the same as that of the first hexagraph,
but with the $1$ and $2$ $t$-channels interchanged. 
As a result, 
although our choice of an initial hexagraph appears to treat the 
$1$ and $2$ channels differently, this distinction is removed (up to a sign 
convention) once we have 
formed the complete set of graphs. That the combination of twists in all three
channels does not result in a new $s$-channel is the same phenomenon, already 
noted, that (in an $s-t$ sub-region) there are 
two distinct $z_i$-plane regions in each
$s$-channel in (\ref{zprs}). A change of the sign of all three $z_i$
is equivalent to 
remaining in the same physical region and changing the sign of
$\lambda^{1\over 2}(t_1,t_2,t_3)$ or, equivalently, interchanging the role of
$1$ and $2$ in the $t_3$-channel associated with this set of hexagraphs. (In an
$s-s$ sub-region, the effect is equivalent to the (parity) transformation of
inverting, in space, the triangle formed by $Q_1$, $Q_2$ and $Q_3$.
This is the same parity transformation that we discussed as relevant for anomaly
cancelations in the previous Section.)

The rules for associating cuts with hexagraphs are as follows. A cut 
of a hexagraph is any path drawn through the graph (along internal lines), 
that enters and exits only between non-horizontal lines. This cut defines an 
invariant channel corresponding to all the particles emitted above (or
absorbed below) the cut. An asymptotic cut is an ``allowable discontinuity'' 
of a hexagraph if it is asymptotically equivalent to a cut of the hexagraph.
The cuts of the first hexagraph of Fig.~5.3 are shown in Fig.~5.4, together
with the corresponding allowable discontinuities
To include unphysical discontinuities in the hexagraph formalism we adopt the 
convention that a cut passing through a vertex via only the sloping lines
(the last cut in Fig.~5.4) induces a reversal of incoming and 
outgoing particles on one or other of the two parts of the cut separated by the
vertex. (We believe this convention will generalise appropriately to more 
complicated hexagraps, but we have not studied this in detail.) 

To form a triplet ${\cal C}$ the three cuts must be ``asymptotically
distinct'' from each other and also satisfy the Steinmann relations.
A first triplet formed from the cuts of Fig.~5.4 is
the set ${\cal C}_a$ in (\ref{cuta}) above and a second 
triplet is ${\cal C}_b$ in (\ref{cutb}).
The asymptotically equivalent triplet ${\cal C}_{b'} $
is also present, as are the triplet ${\cal C}_c $ involving 
the unphysical cut $s_{1'2}$ and the asymptotically equivalent triplet 
${\cal C}_{c'}$. Up to asymptotic equivalence 
one of each of the three kinds of triplets identified in Fig.~2.9
is associated with the hexagraph we 
have chosen. For ${\cal C}_{b}$ and ${\cal C}_{b'}$ we associate only the 
triple discontinuity in region $z_1,z_2,z_3 > 1$ with the 
hexagraph. (The triple discontinuity in $z_1,z_2,z_3 < -1$ is associated with 
the hexagraph obtained by three twists.)
For ${\cal C}_{c}$ and ${\cal C}_{c'}$ we associate only the 
triple discontinuity in the region $Im z_1, Im z_2, - Im z_3 > 1$ with the 
hexagraph (with the triple discontinuity in 
$Im z_1, Im z_2, - Im z_3 < - 1$ associated with 
the hexagraph obtained by three twists.)
\begin{center}
\leavevmode
\epsfxsize=5in
\epsffile{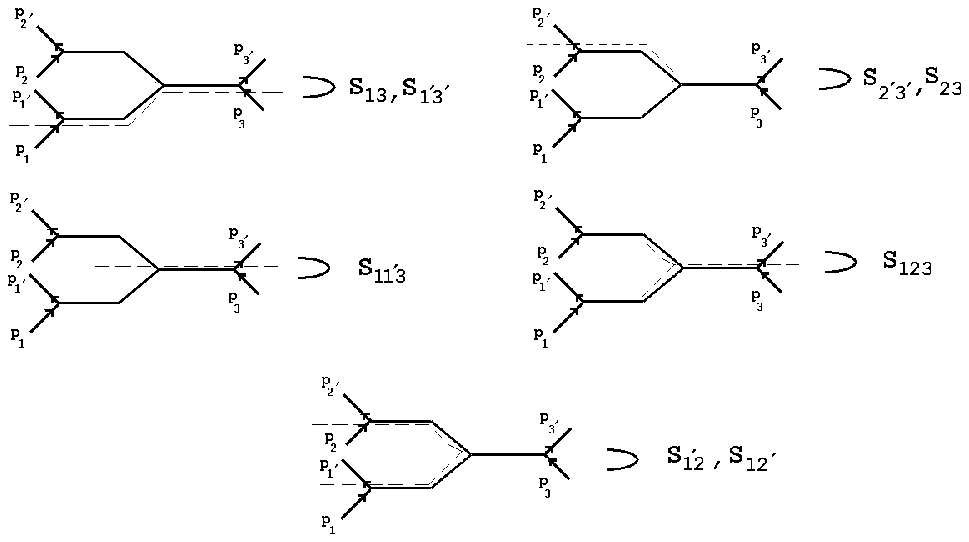}

Fig.~5.4 Cuts and Allowable Discontinuities for the First Hexagraph of Fig.~5.3.
\end{center}

If we now consider the last hexagraph of Fig.~5.3 (associated with the regions
$z_1,z_2,z_3 \leq -1$ and $Im z_1, Im z_2, - Im z_3 < - 1$)
) the triplet ${\cal C}_a$ is replaced by 
$$
{\cal C}_{a'} ~= ~(s_{1'3'},s_{23}, s_{11'3'})
\auto\label{cuta1}
$$
The triplets ${\cal C}_b, {\cal C}_{b'}, {\cal C}_c$ and $ {\cal C}_{c'}$ again
appear. However, as we stated above, we keep only 
the triple discontinuities in $z_1,z_2,z_3 < -1$ and
$Im z_1, Im z_2, - Im z_3 < - 1$. 
With the convention that we have adopted for the unphysical discontinuities,
the initial and final state double discontinuity 
$(s_{13},s_{2'3'})$ 
characterises all three kinds of triplets and 
can therefore be used to identify the hexagraph. We will use this 
identification in the next Section.
Consequently, if the hatched lines of Fig.~4.8(a) are placed on-shell by taking 
discontinuities in $(s_{13}$ and $s_{2'3'})$ then the amplitude obtained
is associated with the hexagraph of Fig.~5.4.

In summary, there are three asymptotically
distinct sets of cuts uniquely associated with each hexagraph.
The triplet having the form of ${\cal C}_a$ determines the product of $z_i$ 
axes to be associated with the hexagraph (the $s_{ijk'}$ invariant involved
must be positive) and also the invariants used to 
describe the remaining cuts. The remaining cuts 
are each pairwise equivalent and appear in two parts of a physical region.
Consequently, from the total 
of 24 hexagraphs there are 48 asymptotically distinct triple discontinuities 
contributing to the dispersion relation.

For $M^{{\cal C}_a}$ we can write straightforwardly 
$$  
\eqalign{M^{{\cal C}_a} ~=~&{1\over (2\pi i)^{3}}
\int {ds'_{13} ds'_{2'3'} ds'_{11'3}\Delta^{{\cal C}_a}(\til{t},
\til{u},s'_{13},s'_{2'3'},s'_{11'3})\over
(s'_{13}-s_{13})(s'_{2'3'}-s_{2'3'})(s'_{11'3}-s_{11'3})} 
}
\auto\label{dis51}
$$
where the integration region can be taken as 
$\{z_1 , z_2 , z_{3} ~ > z_{0}\}$. 
It follows from the representation (\ref{dis51}) 
that for small $s_{13},s_{2',3'}$ and
$s_{11'3}$ , $M^{{\cal C}_a}$ can be expanded as
$$
M^{{\cal C}_a}~=~ \sum_{m,n,r=0}^{\infty}~ c^a_{mnr}~ s_{13}^m~s_{2'3'}^n 
~s_{123}^r
\auto\label{dis6}
$$
where the $c^a_{mnr}$ are functions of the $u_i$ and $t_i$ only. 
The analogues of $M^{{\cal C}_a}$ for each of the hexagraphs have 
analagous representations and expansions.
The expansion (\ref{dis6}) places an important constraint on our 
partial-wave analysis in the next sub-section. 

For $M^{{\cal C}_b}$ we can write 
$$  
~M^{{\cal C}_b} =~{1\over (2\pi i)^{3}}
\int {ds'_{13} ds'_{2'3'} ds'_{123}\Delta^{{\cal C}_b}(\til{t},
\til{u},s'_{13},s'_{2'3'},s'_{123})\over
(s'_{13}-s_{13})(s'_{2'3'}-s_{2'3'})(s'_{123}-s_{123})}
\auto\label{dis22}
$$
but the integration region has two components, 
($\{z_1 , z_2 , z_{3}~ > z_{0}\}$) contributing to one 
hexagraph and ($\{z_1 , z_2 , z_{3}~ < - z_{0}\}$) contributing to the other.
For small $s_{13},s_{2'3'},s_{123}$, (\ref{dis22}) gives the expansion 
$$
M^{{\cal C}_b}~=~ \sum_{m,n,r=0}^{\infty}~ c^b_{mnr}~ s_{13}^m~s_{2'3'}^n 
~s_{11'3}^r
\auto\label{dis7}
$$
This expansion has strong implications for the partial-wave expansions that 
we discuss next. $M^{{\cal C}_{b'}}$ has an identical representation to
$M^{{\cal C}_b}$. $M^{{\cal C}_c}$ and
$M^{{\cal C}_{c'}}$ have analagous representations to 
(\ref{dis22}) but with the integration regions
$\{Im ~z_1 , Im~z_2, - Im~z_{3}~
> Im ~z_{0}\}$ and $\{Im ~z_1 , Im~z_2,  -Im~z_{3}~ <
- Im ~z_{0}\}$. The expansions corresponding to (\ref{dis7}) will have the same
strong implications. 

\subhead{5.6 Partial-Wave Expansions}

To develop the multi-regge theory associated with the Toller diagram of
Fig.~5.1 we begin by writing a partial-wave expansion for each set of hexagraph
amplitudes that are related by twisting and, therefore, have the same
$t$-channel. For the set of amplitudes corresponding to the hexagraphs of
Fig.~5.3 we write
$$ 
\sum_H M^H (z_1,z_2,z_3,u_1,u_2) = 
\sum^{\infty}_{\centerunder{$\scriptstyle J_1,
J_2,J_3=0$}{\raisebox{-5mm}{\centerunder{$\scriptstyle |n_1| 
\leq J_1,|n_2|\leq J_2$}{\raisebox{-5mm}{$\scriptstyle |n_1+n_2|\leq 
J_3$}}}}}
d^{J_1}_{0,n_1} (z_1)\, 
d^{J_2}_{0,n_2} (z_2)\, 
d^{J_3}_{-n_1-n_2,0} (z_3)\,
u^{n_1}_1 u^{n_2}_2 a_{\til{\scriptstyle J}, \til{n}}
\auto\label{pw1}
$$
Our analysis will be focussed on the sub-series in (\ref{pw1}) with $n_1,-n_2 >
0$ and $n_2,-n_1 > 0$. 
In our notation $n_1$ and $-n_2$ are center-of-mass helicities in the
$t_3$-channel. Therefore,  
if $n_1$ and $n_2$ have opposite signs, this corresponds to same sign 
``$t$-channel'' helicities and therefore (at ``zero mass'' $\equiv t_1$ or 
$t_2~=0$)) to
opposite sign ``$s$-channel'' helicities. For this reason, we will refer to 
amplitudes with opposite signs for $n_1$ and $n_2$ as ``helicity-flip''
amplitudes.

Writing (\ref{pw1}) for $M^{{\cal C}_a}$ we can compare the expansion
obtained with the expansion (\ref{dis6}). In the
same leading-power approximation that gives (\ref{ap1}) - (\ref{ap6}), we
can write 
$$
\eqalign{~~~~ d^{J_1}_{0,n_1} (z_1)\, d^{J_2}_{0,n_2} (z_2)\, &
d^{J_3}_{-n_1-n_2,0} (z_3)\,
u^{n_1}_1 u^{n_2}_2 ~~ \sim ~~ z_1^{J_1} z_2^{J_2} z_3^{J_3}  
u^{n_1}_1 u^{n_2}_2 \cr
& = ~(z_1z_3 u_1)^{J_1} (z_2 z_3 u_2^{-1})^{J_2}
z_3^{J_3 - J_1 - J_2}~ u_1^{n_1-J_1} ~u_2^{J_2 - n_2} }
\auto\label{pw2}
$$
Using (\ref{ap1}) - (\ref{ap5}) we can therefore write 
$$
\eqalign{ ~~~~ d^{J_1}_{0,n_1} (z_1)\, d^{J_2}_{0,n_2} (z_2)\, 
d^{J_3}_{-n_1-n_2,0} (z_3)\,
u^{n_1}_1 u^{n_2}_2 ~~ \sim &~~(s_{13})^{J_1} (s_{2'3'})^{J_2}
(s_{11'3})^{J_3 - J_1 - J_2}\cr
&~~~~ \times  c_a(u_1,u_2) }
\auto\label{pw3}
$$
Since we must have a non-negative power of $s_{11'3}$ to obtain a term in 
the expansion (\ref{dis6}), we see that we must have 
$$
J_3 ~\geq~~   J_1 + J_2 
\auto\label{pw4}
$$

We can repeat this last discussion for $M^{{\cal C}_b}$ 
by rewriting (\ref{pw2}) as
$$
\eqalign{ &d^{J_1}_{0,n_1} (z_1)\, d^{J_2}_{0,n_2} (z_2)\, 
d^{J_3}_{-n_1-n_2,0} (z_3)\,
u^{n_1}_1 u^{n_2}_2 \cr
&~~ \sim ~~
(z_1z_3 u_1)^{(J_3 + J_1 - J_2)/2} 
(z_2 z_3 u_2^{-1})^{(J_3 +J_2 - J_1)/2}
(z_1z_2u_1u_2^{-1})^{(J_1 +J_2 - J_3 )/2} u_1^{n_1-J_1} u_2^{J_2 - n_2} \cr
&~~ \sim ~~(s_{13})^{(J_3 + J_1 -J_2)/2} 
(s_{2'3'})^{(J_3 +J_2 - J_1)/2}  (s_{123})^{(J_1 +J_2 - J_3)/2}~~
\times~c_b(u_1,u_2)
}
\auto\label{pw5}
$$
where we have used (\ref{ap6}) for $s_{123}$ ($ \sim s_{12}$)
instead of (\ref{ap3}). Now the requirement 
of a non-negative power for $s_{123}$ implies that 
terms in (\ref{dis7}) can contribute only to those terms in
(\ref{pw1}) with 
$$
J_3 ~\leq~~ J_1 ~ +~J_2
\auto\label{pw6}
$$
Sinc the invariants in $M^{{\cal C}_c}$ have the same asymptotic form (apart 
from a sign) as those in $M^{{\cal C}_b}$ this last argument applies directly
to $M^{{\cal C}_c}$. We see, therefore, that the asymptotic contributions of 
the triple discontinuities $M^{{\cal C}_b}$ and $M^{{\cal C}_c}$ 
($M^{{\cal C}_{b'}}$ and $M^{{\cal C}_{c'}}$) appear in 
a distinct part of the partial-wave expansion (\ref{pw1}) to that of
$M^{{\cal C}_a}$. 
As a result, the Sommerfeld-Watson representation discussed in the next
sub-section is very different in the two cases. 

An additional requirement for (\ref{pw5}) to correspond to a term in 
(\ref{dis6}) is that the powers of the invariants in (\ref{pw5})
must be integer. This places a further restriction on the partial-waves
that $M^{{\cal C}_b}$ and $M^{{\cal C}_c}$ can contribute to.
In fact, if we constrain
$ J_1 +  J_2  - J_3$ to be an even integer, then no further constraint on 
the $J_i$ is required (other than that they be positive integers). This 
is equivalent to the signature constraint
$$
\tau_1\tau_2\tau_3~=~1
\auto\label{pw60}
$$
where at this stage $\tau_i = \pm 1$ when $J_i$ is $even/odd$. 
For this signature constraint on partial-wave amplitudes 
to be matched by the definition of signature via triple discontinuities that
we give below, it must be that sum of the triple discontinuities 
$M^{{\cal C}_b}$ and $M^{{\cal C}_{b'}}$ is symmetric with respect to 
the two parts of the 
physical region where they appear. Similarly the sum of
$M^{{\cal C}_c}$ and $M^{{\cal C}_{c'}}$ must be symmetric with respect to 
the two regions in which they appear.

Note that if we consider leading helicity physical amplitudes (i.e. with
$|n_i|= J_i,~i=1,2$) then if $n_1, n_2
>0$, necessarily 
$$
J_3 ~ \geq ~n_1~+~n_2 ~= ~J_1~+~J_2 
\auto\label{pw7}
$$
Consequently $J_3 < J_1 + J_2 $ 
is only possible for (leading) helicity-flip amplitudes.
It was observed by Detar and Weis\cite{bdw}, in their study many years ago 
of the dual-model triple-Regge vertex, that the terms in the vertex with 
(essentially) the
sets of cuts $M^{{\cal C}_b}$ and $M^{{\cal C}_c}$ 
contribute to partial-wave amplitudes satisfying
inequalities of the form (\ref{pw6}). However, the S-W formalism that we use
was not developed at the time of the Detar and Weis paper. 

\subhead{5.7 Signature and the Sommerfeld-Watson Representations }

As in elementary Regge theory, it is necessary to introduce signature
before making Froissart-Gribov (F-G)  continuations of partial-wave amplitudes
and introducing Sommerfeld-Watson (S-W) representations. We 
define signatured hexagraph amplitudes
$$
\eqalign{M^{H,\til{\tau}}~& = ~ M^{H,(\tau_1,\tau_2,\tau_3)}\cr
& = ~{1 \over 8}~\biggl[ ~M^H ~+~ \tau_1 M^{{\cal T}_1 H} 
~+~ \tau_2 M^{{\cal T}_2 H} ~+~  \tau_3 M^{{\cal T}_3 H} 
~+~ \tau_1 \tau_2  M^{{\cal T}_1{\cal T}_2 H} \cr
&~~~~~~~~~~~ +~ \tau_2\tau_3  M^{{\cal T}_2{\cal T}_3 H} 
~+ ~\tau_3\tau_1  M^{{\cal T}_3{\cal T}_1 H} 
~+ ~ \tau_1\tau_2 \tau_3    M^{{\cal T}_1{\cal T}_2{\cal T}_3 H}~ \biggr] }
\auto\label{sig0}
$$
where $\tau_i = \pm 1 $ , and ${\cal T}_i H$ is the 
hexagaph obtained from the hexagraph $H$ 
by a twist about the ith horizontal line. The full amplitude, or rather the 
sum over hexagraph amplitudes, 
is recovered as a sum over signatured amplitudes, i.e.
$$
\sum_{\til{\tau}}~ M^{H,~\til{\tau}} = M^H + M^{{\cal T}_1 H} 
+ M^{{\cal T}_2 H} + M^{{\cal T}_3 H} + M^{{\cal T}_1{\cal T}_2 H} 
+ M^{{\cal T}_2{\cal T}_3 H} + M^{{\cal T}_3{\cal T}_1 H} 
+ M^{{\cal T}_1{\cal T}_2{\cal T}_3 H} 
\auto\label{sig01}
$$
For hexagraph amplitudes of the form $M^{{\cal C}_a}$ and $M^{{\cal C}_b}$,
(\ref{sig0}) is a simple generalisation of the analytic definition
of signature for elastic scattering amplitudes, where combinations of
amplitudes with right and left-hand cuts are formed. For $M^{{\cal C}_c}$
amplitudes it becomes the standard definition in the $s-s$ region.

In writing the initial dispersion relation (\ref{dis}) we are, of course, 
assuming a generalization of the usual crossing relation 
that there is a single analytic function that connects all the physical 
region amplitudes. When quark quantum numbers and helicities are involved, 
there are additional subtleties in the crossing relation. These subtleties
are resolved if we assume that our analytic definition of signatured
amplitudes in (\ref{sig0}) is equivalent to the following alternative
``group-theoretic'' definition of signature. 
Beginning with an N-point amplitude in a particular $s$-channel, we
form the positive (or negative) signatured amplitude, with respect to a
particular internal line of a Toller diagram, by adding (or subtracting) the
amplitude obtained by making a complete $CPT$ transformation on all external
particles connected (through the diagram) to one end of the internal line.
The fully signatured amplitude is formed by carrying out this procedure for
all internal lines of the Toller diagram. In this way, signature is
introduced at the amplitude level without introducing spectral components.
It is an operation defined directly on the external states and so
is often easier to implement. Although the 
equivalence of the two definitions has only been proven in the simplest 
cases, we have no reason to doubt that the equivalence is true in general. 
To understand the implications of signature for phases etc. it is, of 
course, essential to utilise the analytic formulation. 

In a $t$-channel the twisting process does not involve interchanging incoming 
and outgoing particles. Therefore, a signature twist becomes a $CP$ rather
than a $CPT$ transformation. The charge conjugation part of the transformation
will eventually be very important for our discussion of the anomaly. However,
if we ignore quantum numbers then a signature twist is
effectively a $t$-channel parity transformation of the final state relative 
to the initial state. In order to have three 
independent parity transformations, we must have a dependence on invariants
that involve directly the momenta at the central vertex, i.e. the $Q_i$. In
Fig.~5.4, only $s_{11'3}$ has this property. This is why only triplets of
the ${\cal C}_a$ kind produce three independent signatures. 

The S-W transform of (\ref{pw1}) is
obtained by converting the sums over $n_1, n_2, $ and $J_3$ to integrals. 
Initially this process is carried out with $z_1$ and $z_2$ small, although 
we will then use the representation to discuss large $z_1$ and $z_2$.
The conversion of sums over $n_1, n_2, $ to integrals effectively represents 
most of the asymptotic cuts as cuts in the $u_1$ and $u_2$ planes and therefore 
is most naturally carried out in the $s-s$ region where large $u_1$ and $u_2$
is part of the physical regions.

The treatment of that part of the expansion satisfying (\ref{pw4})
is straightforward. 
This contains $M^{{\cal C}_a}$ together with the corresponding contribution
from all the hexagraphs of Fig.~5.3. To 
illustrate the structure of the S-W transform we first omit 
the complications due to signature and (effectively) assume a F-G
continuation can be made for $M^{{\cal C}_a}$ alone. Because the
definition of $d^J_{n,0}$ changes non-analytically at $n=0$ we must make 
separate continuations for $n_1,n_2,n_1+n_2 ~
{\raisebox{1mm}{\centerunder{$\scriptscriptstyle >$}{$\scriptscriptstyle
<$}}} 0$. For $n_1, - n_2 , n_1 + n_2 \geq 0 $, we can write 
$$ 
\eqalign{M^{{\cal C}_a} =& ~{1 \over (2\pi)^3}~ 
\int_{> } ~{dn_1 ~(u_1)^{n_1}\over \sin \pi n_1}
\int_{< } ~{dn_2 ~(u_2)^{n_2} \over \sin\pi n_2} 
\cr
&\times 
\sum^{\infty}_{\centerunder{$\scriptstyle N_1=J_1-|n_1|= 
0$}{\raisebox{-5mm}{$\scriptstyle N_2=J_2 - |n_2|=0
$}}}
\int_{C_{N_1 + N_2}} 
{dJ_3 ~~d^{J_3}_{n_1+n_2,0} (z_3) \over \sin\pi(J_3-n_1+n_2)} 
d^{J_1}_{0,n_1}(z_1) 
d^{J_2}_{0,n_2}(z_2)\,
~a_{{\cal C}_a ,\til{J}, \til{n} ,\til{>}}
}
\auto\label{sw1}
$$
where each integration contour is asymptotically parallel to the imaginary 
axis and chosen to reproduce the partial-wave sum when closed in the
appropriate half-plane (because of the symmetry 
under $u_1 \to u_1^{-1}, u_2 \to u_2^{-1}$, $n_1 \leq 0, n_2 \geq 0$ 
gives an identical contribution). 
 The contour $C_{N_1 + N_2}$ imposes (\ref{pw4}) i.e.
$$
J_1 + J_2 ~=~ N_1 + n_1 + N_2 - n_2 \leq J_3
\auto\label{sw11}
$$
and so has the form shown in Fig.~5.5. The $>$ and $ <$ labels for the $n_i$
integrals indicate that they reproduce the associated positive/negative
helicity sums.
\begin{center}
\leavevmode
\epsfxsize=2.5in
\epsffile{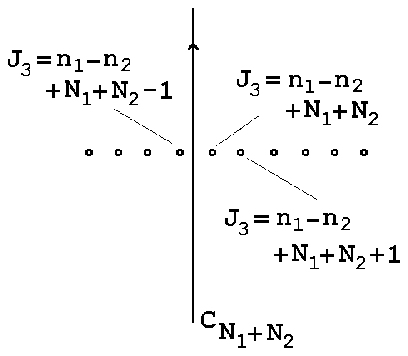}

Fig.~5.5 The Contour in the $J_3$ Plane.
\end{center}

For small $z_i,~i=1,2$, a twist in the $i$th channel corresponds to $u_i \to
~-u_i$ and $z_i \to ~-z_i$. For small $z_i$, and integer $j_i-|n_i|=N_i$, 
$$
d^{J_i}_{0,n_i}(-z_i) ~=~ (-1)^{J_i - n_i}  d^{J_i}_{0,n_i}(z_i)
\auto\label{sig00}
$$  
Consequently for $M^{{\cal C}_a}$ 
we can adapt (\ref{sw1}) to represent the sum over
amplitudes involving twists in the $1$ and $2$ channels 
by making the replacement
$$
(u_i)^{n_i}d^{J_i}_{0,n_i}(z_i) ~~\to ~~ (u_i)^{n_i}d^{J_1}_{0,n_i}(z_i) ~
+~ \tau_i ~ (-u_i)^{n_i} d^{J_i}_{0,n_i}(-z_i) 
~~~~~~~~i=1,2
\auto\label{sig1}
$$
A twist in the $3$ channel is more complicated since $d^{J_3}_{n_1+n_2,0} (z_3) 
$ depends on $n_1$ and $n_2$ as well as $J_3$. 
For amplitudes that already have specific
$\tau_1$ and $\tau_2$ signatures, we can introduce 
signature for $z_3 \to - z_3$ by writing
$$
d^{J_3}_{n_1+n_2,0} (z_3) 
~~ \to ~~d^{J_3}_{n_1+n_2,0} (z_3)~+~ 
~ \tau_3 \tau_1 \tau_2 (-1)^{N_1 +N_2} d^{J_3}_{n_1+n_2,0} (- z_3) 
\auto\label{sig2}
$$
Of course, we also write 
$$
a_{{\cal C}_a , \til{J}, \til{n},\til{>} }
~~\to ~~
a_{{\cal C}_a , \til{J}, \til{n},\til{>},\til{\tau} }
\auto\label{sig3}
$$
The construction of signatured F-G continuations 
$\raisebox{1mm}{$a_{{\cal C}_a , \til{J}, \til{n},\til{>},\til{\tau} }$}$ 
that are equal to 
the physical partial-waves at ``right-signature'' points (i.e. $J_i = $
even/odd for $\tau_i = +/-, ~ i=1,2,3~$)
is described in detail in \cite{arw1}. 

The S-W transform of that part of the expansion (\ref{pw1}) 
that satisfies (\ref{pw6}), and so contains both $M^{{\cal C}_b}$ and
$M^{{\cal C}_c}$, 
requires extra discussion. When (\ref{pw6}) is satisfied
at physical points, $n_1$ and $n_2$ will generally have opposite signs. (For 
leading helicities this must be the case, as we noted above.) 
Consider that part of 
(\ref{pw1}) with $n_1, -n_2, n_1+n_2  \geq 0 $. (\ref{pw6}) becomes 
$$
N_1+n_1 +N_2 -n_2 - J_3 ~=~ even ~integer ~\geq~0
\auto\label{sig03}
$$
Temporarily ignoring the full signature problem,
we can write 
$$ 
\eqalign{M^{{\cal C}_b} = &{1 \over (2\pi)^3}
\int dn_1 dn_2 (u_1)^{n_1} (u_2)^{n_2}  
\sum^{\infty}_{\centerunder{$\scriptstyle N_1,N_2  = 0
$}{\raisebox{-3mm}{$\scriptstyle N_1 + N_2 ~even $}}}
d^{N_1 + n_1}_{0,n_1}(z_1)~ d^{N_2-n_2}_{0,n_2}(z_2)\cr
& 
\int_{C_{J_3}} {dJ_3 ~~~~~~~~~~~~~~d^{J_3}_{n_1+n_2,0} (z_3)  
~~~~~~~~~~~~~~\over 
 \sin \pi (n_1+n_2)
\sin{\pi \over 2 }(J_3-n_1-n_2 \sin{\pi \over 2}(n_1-n_2 - J_3)} 
~~~a_{{\cal C}_b ,\til{J}, \til{n},\til{>}}
}
\auto\label{sw3}
$$
(The symmetry 
under $u_1 \to u_1^{-1}, u_2 \to u_2^{-1}$ implies that 
$n_1 \leq 0, n_2 \geq 0$ 
gives an identical contribution.) 
The integration contours are again asymptotically parallel to the 
imaginary axis and the contour $C_{J_3}$ is as shown in Fig.~5.6.
\begin{center}
\leavevmode
\epsfxsize=3in
\epsffile{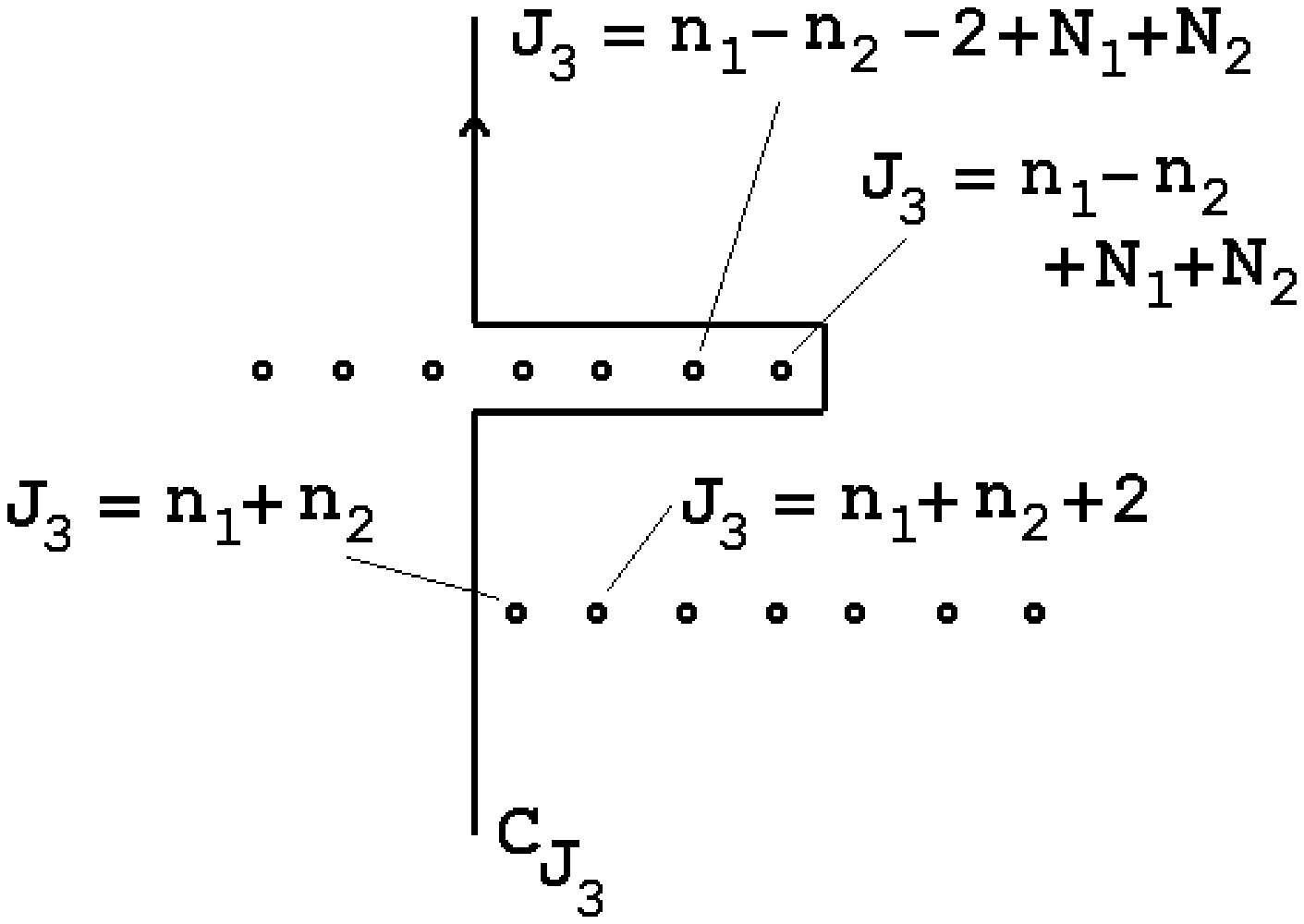}

Fig.~5.6 The $C_{J_3}$ Contour. 
\end{center}
This contour now imposes (\ref{sig03}). Poles at negative 
integer $n_2$ are produced when the poles 
at $J_3 - n_1 + n_2 -N_1 - N_2 = ~even~ integer~< 0$ collide with those
at $J_3 - n_1 - n_2 = ~even~ integer~ \geq 0~$.
As always the $~\raisebox{1mm}{$ \scriptstyle \til{\scriptstyle >}$}~ $
labels on $~\raisebox{1mm}{$a_{{\cal C}_b , \til{J},
\til{n},\til{>},\til{\tau} }$}~$ refer to distinct F-G continuations
made for distinct combinations of helicity signs. 

(\ref{sw3}) gives both even and odd values of $n_1$ and $n_2$.
Consequently (\ref{sig1}) can again be used to
introduce signature in the $1$ and $2$ channels. 
Since we must impose (\ref{pw60}), we need not add any further signature 
effects. However, we note that the arguments of two of the 
denominator sine factors in (\ref{sw3}) are already restricted
to even integer physical values. Therefore, each term in the signatured form 
of (\ref{sw3}) could be modified by a phase factor of the form $(-1)^E$ where
$E$ is (equivalently) either of these sine function arguments. In principle
the uniquenes of the F-G continuation resolves this anbiguity. We will
resolve it by determining the appropriate asymptotic phases of signatured
Regge pole amplitudes. We will find that this phase provides the crucial 
distinction between the contribution of the physical region discontinuities 
of the $M^{{\cal C}_b}$ (and $M^{{\cal C}_{b'}}$) 
kind and the unphysical region discontinuities 
of the $M^{{\cal C}_c}$ (and $M^{{\cal C}_{c'}}$) kind.

\subhead{5.8 Regge Behavior}

The Steinmann relations imply that for the hexagraph amplitudes with 
physical region asymptotic cuts, these cuts are completely represented by
signatured S-W integrals. Consequently, for $M^{{\cal C}_a}$ and 
$M^{{\cal C}_b}$ the sums over
$N_1=J_1 - n_1$ and $N_2 = J_2 - |n_2|$ should be uniformly convergent 
in the Regge asymptotic region with the limits and sums in the S-W 
representations liberally interchangeable. For the anomaly to appear in a
physical region in the configuration of Fig.~4.15(a) then 
it has to be due a physical
region singularity appearing in an amplitude that has cuts only in a
cross-channel or unphysical region.
As we elaborate further below, the anomaly can appear in this way in 
$M^{{\cal C}_c}$ (and $M^{{\cal C}_{c'}}$) amplitudes, where it has to produce
a divergence of the $N_1$ and $N_2$ sums. A-priori, it would appear that 
the anomaly could also appear in the 
$M^{{\cal C}_b}$
(and $M^{{\cal C}_{b'}}$) amplitudes, since they are so similar to the 
$M^{{\cal C}_c}$ amplitudes. However, we will give arguments below that this 
is not the case. 

For fixed $N_1$ and $N_2$, the integrals over $n_1$ and 
$n_2$ can be treated as integrals over either $n_1$ and $n_2$
or $J_1$ and $J_2$. Consequently asymptotic expansions can be obtained 
as either $z_1, z_2 \to \infty$ or as $u_1~(u_1^{-1}),u_2~(u_2^{-1}) 
\to \infty$ 
by pulling contours to the left (right) in the complex plane in the 
conventional manner. In this way, each of the triple-regge and helicity-pole 
limits defined above can be studied.
The replacement of the $d^J_{n,0} $ by second-type
representation functions proceeds in direct parallel with elementary regge 
theory. (This is a technical necessity to ensure that a genuine
asymptotic expansion is obtained but we will not describe it here. For our
purposes it is sufficient to assume that we simply pick up the leading
power behavior of the $d^J_{n,0}(z) $ of the form $z^J$, as contours.)

The most important point for studying limits via the S-W transform is that
(by analytically continuing $t$-channel unitarity equations in all complex 
angular momentum and helicity planes) it can be shown that the regge 
singularities of 
\newline $~\raisebox{0.5mm}{$a_{{\cal C}, \til{J}, \til{n}, \til{>}}$ } ~=~
 a_{{\cal C}, \til{>}}
(J_3,n_1,n_2,N_1,N_2,t_1.t_2,t_3)$ occur at fixed values of the $J_i$. 
In particular, Regge poles at $J_i = \alpha_i =
\alpha(t_i)$ occur in $a_{{\cal 
C},>,>}(J_3,n_1,n_2,N_1,N_2,t_1.t_2,t_3)$ (the F-G continuation made from
$n_1,n_2 >0$) at
$$
n_1 = \alpha_1 - N_1~, ~~~n_2 = \alpha_2 - N_2~, ~~~J_3 = \alpha_3
\auto\label{sw4}
$$
In $a_{{\cal 
C},>,<}(J_3,n_1,n_2,N_1,N_2,t_1.t_2,t_3)$ (the continuation from
$n_1,- n_2 >0$) the regge singularities occur at
$$
n_1 = \alpha_1 - N_1~, ~~~- n_2 = \alpha_2 - N_2~, ~~~J_3 = \alpha_3
\auto\label{sw41}
$$
etc. We will first study the contributions of regge poles to the S-W integral 
and then discuss the (minor) differences when the regge poles are replaced by
regge cuts.

In the triple-regge limit, regge poles give contributions
to each of the terms in the double sums in (\ref{sw1}) and (\ref{sw3}). 
We initially omit the
denominator sine factors since they are modified by the introduction of 
signature. We can then write the 
triple regge pole contribution to $M^{{\cal 
C}_a}$ via (\ref{sw1}) as 
$$
\eqalign{M^{{\cal C}_a}
\centerunder{$\large\sim$}{\raisebox{-4mm} 
{\centerunder{$z_1,z_2,$}{\raisebox{-4mm} 
{$ z_3, \rightarrow\infty$}}}}
~~&z_1^{\alpha_1}z_2^{\alpha_2}z_3^{\alpha_3}\sum^{\infty}_{N_1, N_2=0}
~\biggl[~u_1^{\alpha_1 - N_1}u_2^{\alpha_2 - 
N_2}\beta^{a}_{\alpha_1, \alpha_2, \alpha_3, N_1, N_2}~+ \cr
&~
u_1^{-\alpha_1 + N_1}u_2^{\alpha_2 - 
N_2}\beta^{a}_{-\alpha_1, \alpha_2, \alpha_3, N_1, N_2}
~+~ u_1^{\alpha_1 - N_1}u_2^{-\alpha_2 + N_2}\beta^{a}_{\alpha_1, 
-\alpha_2, \alpha_3, N_1, N_2}\cr
&~~~
~~~~~~~~~+ u_1^{-\alpha_1 + N_1}u_2^{-\alpha_2 +  N_2}\beta^{a}_{-\alpha_1,
-\alpha_2, \alpha_3, N_1, N_2}~\biggr]}
\auto\label{trp}
$$
where $\beta^{a}_{\alpha_1, \alpha_2, \alpha_3, N_1, N_2}$ is the Regge-pole 
residue of the ``non-flip'' F-G amplitude 
\newline $a_{{\cal C},>,>}(J_3,n_1,n_2,N_1,N_2,t_1.t_2,t_3)$ 
at $J_i=\alpha_i,~ i=1,2,3$ and $n_i=J_i - N_i,~i=1,2$ and 
$\beta^{a}_{-\alpha_1, \alpha_2, \alpha_3, N_1, N_2}$
is the corresponding residue of the ``helicity-flip'' F-G amplitude
$a_{{\cal C},<,>}(J_3,n_1,n_2,N_1,N_2,t_1.t_2,t_3)$.
Because of the symmetry under
$u_1 \to 1/u_1, u_2 \to 1/u_2$, the first and last sums in 
(\ref{trp}) can be identified, as can the second and third. When the 
hexagraph containing $M^{{\cal C}_a}$ 
is part of a larger hexagraph this symmetry is, in general, not present.

The contribution of Regge poles to $M^{{\cal C}_b}$ (and $M^{{\cal C}_b}$), 
in the triple-Regge 
limit, has less structure than (\ref{trp}). From (\ref{sw3}) we obtain
$$
\eqalign{M^{{\cal C}_b}
\centerunder{$\large\sim$}{\raisebox{-4mm} 
{\centerunder{$z_1,z_2,$}{\raisebox{-4mm} 
{$ z_3, \rightarrow\infty$}}}}
~~z_1^{\alpha_1}z_2^{\alpha_2}z_3^{\alpha_3}\sum_{N_1+ N_2 even }
~\biggl[u_1^{-\alpha_1 + N_1}&u_2^{\alpha_2 - 
N_2}\beta^{b}_{-\alpha_1, \alpha_2, \alpha_3, N_1, N_2}\cr
& +~ u_1^{\alpha_1 - N_1}u_2^{-\alpha_2 + N_2}\beta^{a}_{\alpha_1, 
-\alpha_2, \alpha_3, N_1, N_2}\biggr]}
\auto\label{trpb}
$$
Again the symmetry under $u_1 \to 1/u_1, u_2 \to 1/u_2$ 
implies that (in this case) the two terms in (\ref{trpb}) can be identified. 
$M^{{\cal C}_c}$ has an identical contribution but with, of course,
$\beta^{b}_{\pm\alpha_1, \pm \alpha_2, \alpha_3, N_1, N_2}
~\to ~ \beta^{c}_{\pm\alpha_1,\pm \alpha_2, \alpha_3, N_1, N_2}$.

To obtain the behavior of the full amplitude in the 
triple-regge limit we add $M^{{\cal C}_a}$, $M^{{\cal C}_b}$ and 
$M^{{\cal C}_c}$ $M^{{\cal C}_{b'}}$ and 
$M^{{\cal C}_{c'}}$, together with the analagous contributions corresponding to
the additional hexagraphs illustrated in Fig.~5.2. These contributions will
have the same form as (\ref{trp}) and (\ref{trpb}) but with the indices
$1,2$ and $3$ cyclically rotated. Finally, the twisted graphs also have to be
added by incorporating signature factors. Before discussing signature in
detail, it will be useful to first discuss the contribution of regge poles in
helicity-pole limits.

The non-flip helicity-pole limit (\ref{hp1}) picks out
only the first term of the first (and identical third) sum in (\ref{trp}) i.e. 
$$
M^{{\cal C}_a} ~~
\centerunder{$\large\sim$}{\raisebox{-4mm} 
{\centerunder{$u_1,u_2,$}{\raisebox{-4mm} 
{$ z_3, \rightarrow\infty$}}}}
 (z_1u_1)^{\alpha_1}\, (z_2u_2)^{\alpha_2}\, z_3^{\alpha_3}\, 
\beta^a_{\alpha_1,\alpha_2,\alpha_3,0,0}
\auto\label{hp1r}
$$
There is no triple-regge contribution from $M^{{\cal C}_b}$ or $M^{{\cal C}_c}$
in this limit. In the helicity-flip limit each of $M^{{\cal C}_a}$, 
$M^{{\cal C}_b}$, and $M^{{\cal C}_c}$ give contributions, i.e.
$$
M^{{\cal C}_{a,b,c}}~~ 
\centerunder{$\large\sim$}{\raisebox{-4mm} 
{\centerunder{$u_1,1/u_2,$}{\raisebox{-4mm} 
{$ z_3, \rightarrow\infty$}}}}
 (z_1u_1)^{\alpha_1}\, (z_2u_2^{-1})^{\alpha_2} z_1^{\alpha_3} 
\beta^{a,b,c}_{\alpha_1,-\alpha_2,\alpha_3,0,0}
\auto 
$$

We see that distinct leading helicity amplitudes, 
i.e. non-flip and flip, contribute in
the distinct helicity-pole limits while the complete series of 
both amplitudes contribute in the
full triple-regge limit. This explains why we refer 
to (\ref{hp1}) and (\ref{hp2}) 
respectively as non-flip and helicity-flip limits. 
Note that in both limits the dependence on 
both $z_1$ and $z_2$ is determined by the $u_1$ and $u_2$ dependence. This is 
necessary for the amplitudes to be directly expressible in terms of 
invariants, as we see in the next sub-section.

\subhead{5.9 Asymptotic Analytic Structure}

We can now discuss how the cuts of $M^{{\cal C}_a}$, $M^{{\cal C}_b}$, 
and $M^{{\cal C}_c}$ are 
represented asymptotically in triple-regge formulae. Again we discuss regge 
poles in detail. We will then illustrate how the discussion generalises to
regge cuts. Similarly to the rewriting of (\ref{pw2}) in the form (\ref{pw3}),
we can use (\ref{ap1}) - (\ref{ap5}) to write 
$$
\eqalign{  \beta^a_{\alpha_1,\alpha_2,\alpha_3,0,0}
~(z_1u_1)^{\alpha_1} (z_2u_2)^{\alpha_2} z_3^{\alpha_3}~ 
&= ~\beta^a_{\alpha_1,\alpha_2,\alpha_3,0,0}~
(z_1z_3u_1)^{\alpha_1}(z_1z_3u_2)^{\alpha_2}~
(z_3)^{\alpha_3-\alpha_1-\alpha_2}\cr
& \sim ~\beta^a_{\alpha_1,\alpha_2,\alpha_3,0,0}~
(s_{13})^{\alpha_1}(s_{2'3'})^{\alpha_2}(s_{11'3})^{ \alpha_3 - 
\alpha_1 - \alpha_2} }
\auto\label{hp1i} 
$$
showing how (\ref{hp1r}) represents the cuts of $M^{{\cal C}_a}$ 
in both the non-flip helicity-pole limit and the full
triple-Regge limit. The non-leading helicity terms in (\ref{trp}) that 
contribute in the triple-regge limit are represented in terms of invariants
by writing
$$
u_1^{-N_1}~u_2^{-N_2}~\sim~ \biggl({s_{133'}~s_{311'}\over s_{13} }
\biggr)^{N_1}~\biggl({s_{2'33'}~s_{3'22'} \over s_{23}}\biggr)^{N_2}
\auto\label{n1n2}
$$
The result is a power series expansion in terms of the invariants
$s_{133'},s_{311'},s_{2'33'},s_{3'22'}$ in which the Steinmann relations 
determine there are no singularities. Therefore this series is convergent
and the cut structure is fully represented by 
(\ref{hp1i}). Similarly to (\ref{hp1i}) we can write 
$$
\eqalign{  \beta^a_{\alpha_1,-\alpha_2,\alpha_3,0,0}~
(z_1u_1)^{\alpha_1} (z_2u_2^{-1})^{\alpha_2} z_3^{\alpha_3}~ 
&= ~\beta^a_{\alpha_1,-\alpha_2,\alpha_3,0,0}~
(z_1z_3u_1)^{\alpha_1}\biggl({z_2z_3 \over u_2}\biggr)^{\alpha_2}~
(z_3)^{(\alpha_3-\alpha_1-\alpha_2)}\cr
& \sim ~\beta^a_{\alpha_1,-\alpha_2,\alpha_3,0,0} ~
(s_{13})^{\alpha_1}(s_{2'3'})^{\alpha_2}(s_{11'3})^{ \alpha_3 - 
\alpha_1 - \alpha_2}}
\auto\label{hp2i} 
$$
to see the cuts of  $M^{{\cal C}_a}$ also represented in the 
helicity-flip limit and in helicity-flip contributions to the 
triple-Regge limit. 

For the helicity-flip contribution from $M^{{\cal C}_b}$ 
we utilise (\ref{ap6}) and write 
$$
\eqalign{ & \beta^b_{\alpha_1,-\alpha_2,\alpha_3,0,0} ~
(z_1u_1)^{\alpha_1} (z_2u_2^{-1})^{\alpha_2} z_3^{\alpha_3} \cr
& ~~~~~~~ = \beta^b_{\alpha_1,-\alpha_2,\alpha_3,0,0} ~
(z_1z_3u_1)^{(\alpha_1+\alpha_3-\alpha_2)/2}
\biggl({z_2z_3 \over u_2}\biggr)^{(\alpha_2+\alpha_3-\alpha_1)/2}
\biggl({z_1z_2 u_1 \over u_2}\biggr)^{(\alpha_1+\alpha_2-\alpha_3)/2}\cr
& ~~~~~~~ \sim ~ \beta^b_{\alpha_1,-\alpha_2,\alpha_3,0,0}~
(s_{13})^{(\alpha_1+\alpha_3-\alpha_2)/2}
(s_{2'3'})^{(\alpha_2+\alpha_3-\alpha_1)/2}
(s_{123})^{(\alpha_1+\alpha_2
-\alpha_3)/2}}
\auto\label{hp2i1}
$$
showing how the cuts of $M^{{\cal C}_b}$ are represented in the 
helicity-flip limit and in helicity-flip contributions to the 
triple-Regge limit. For $M^{{\cal C}_c}$ we write, in close analogy,
$$
\eqalign{ & \beta^c_{\alpha_1,-\alpha_2,\alpha_3,0,0} ~
(z_1u_1)^{\alpha_1} (z_2u_2^{-1})^{\alpha_2} z_3^{\alpha_3} \cr
& ~~~~~~~ \sim ~ \beta^c_{\alpha_1,-\alpha_2,\alpha_3,0,0}~
(s_{13})^{(\alpha_1+\alpha_3-\alpha_2)/2}
(s_{2'3'})^{(\alpha_2+\alpha_3-\alpha_1)/2}
(s_{1'2})^{(\alpha_1+\alpha_2
-\alpha_3)/2}}
\auto\label{hp2i2}
$$

To add signature factors to (\ref{hp1i}) note that 
for $J_i- |n_i| =N_i = 0$
$$
{(u_i)^{n_i}d^{J_1}_{0,n_i}(z_i) ~
+~ \tau_i ~ (-u_i)^{n_i} d^{J_i}_{0,n_i}(-z_i) \over
sin \pi n_i}
~~\centerunder{$\sim $}{\raisebox{-5mm}{$u_i,z_i \to \infty$}}
~~ |u_iz_i|^{J_i} \biggl[{1 + \tau_i e^{i\pi J_i} \over 
sin \pi J_i} \biggr]
\auto\label{sig571}
$$
and 
$$
{d^{J_3}_{n_1+n_2,0} (z_3)+ 
 \tau_3 \tau_1 \tau_2 d^{J_3}_{n_1+n_2,0} (- z_3) 
\over sin \pi (J_3 - J_1 - J_2)}
~\centerunder{$\longrightarrow$}{\raisebox{-5mm}{$z_3 \to \infty$}}
~ |z_3|^{J_3 } \biggl[ {1 + \tau_1 \tau_2 \tau_3 
e^{i\pi (J_3 - J_1 - J_2)} \over 
sin \pi (J_3 - J_1 - J_2)} \biggr]
\auto\label{sig572}
$$
Therefore the signatured form of the S-W representation (\ref{sw3})
will give (\ref{hp1i}) multiplied by a factor 
$$
\biggl[{1 + \tau_1 e^{i\pi \alpha_1} \over 
sin \pi \alpha_1} \biggr]
\biggl[{1 + \tau_2 e^{i\pi \alpha_2} \over 
sin \pi \alpha_2} \biggr]
\biggl[ {1 + \tau_1 \tau_2 \tau_3 
e^{i\pi (\alpha_3 - \alpha_1 - \alpha_2)} \over 
sin \pi (\alpha_3 - \alpha_1 - \alpha_2)} \biggr]
\auto\label{sig5720}
$$
giving 
$$
\eqalign{ ~~~~& \beta^a_{\alpha_1,\alpha_2,\alpha_3,0,0}~
\biggl[{(s_{13})^{\alpha_1} + \tau_1 (- s_{13})^{\alpha_1}  \over 
sin \pi \alpha_1 } \biggr]
\biggl[{(s_{2'3'})^{\alpha_2} + \tau_2 (- s_{2'3'})^{\alpha_2} \over 
sin \pi \alpha_2} \biggr] \cr
& ~~~~~~~~~~~~~~~~~~~~~~~~~~~~~~~~~ \times~ 
\biggl[ { (s_{11'3})^{ \alpha_3 - 
\alpha_1 - \alpha_2}  + \tau_1 \tau_2 \tau_3 
(- s_{11'3})^{ \alpha_3 - \alpha_1 - \alpha_2} \over 
sin \pi (\alpha_3 - \alpha_1 - \alpha_2)} \biggr] }
\auto\label{sig573}
$$
This expression now represents the leading helicity (non-flip)
triple-regge contribution of the
sum of amplitudes corresponding to the eight hexagraphs appearing in 
Fig.~5.3. 

Each hexagraph can be identified with a term in (\ref{sig573}), with 
the phase appropriately representing the cut structure.
It is therefore straightforward to take discontinuities in (\ref{sig573}) and to
recover a single hexagraph amplitude (\ref{hp1i}), e.g. 
$$
\eqalign{ [Disc]_{s_{13}} ~~= ~~
2~ & \beta^a_{\alpha_1,\alpha_2,\alpha_3,0,0}~
(s_{13})^{\alpha_1} 
\biggl[{(s_{2'3'})^{\alpha_2} + \tau_2 (- s_{2'3'})^{\alpha_2} \over 
sin \pi \alpha_2} \biggr] \cr
& ~~~~~~~~~~~~  \times~ 
\biggl[ { (s_{11'3})^{ \alpha_3 - 
\alpha_1 - \alpha_2}  + \tau_1 \tau_2 \tau_3 
(- s_{11'3})^{ \alpha_3 - \alpha_1 - \alpha_2} \over 
sin \pi (\alpha_3 - \alpha_1 - \alpha_2)} \biggr] }
\auto\label{sig574}
$$
\medskip
$$
\eqalign{ [Disc]_{s_{13}} ~[Disc]_{s_{2'3'}} ~~=& ~~
4~\beta^a_{\alpha_1,\alpha_2,\alpha_3,0,0}~
(s_{13})^{\alpha_1} (s_{2'3'})^{\alpha_2} \cr
&\times~ \biggl[ { (s_{11'3})^{ \alpha_3 - 
\alpha_1 - \alpha_2}  + \tau_1 \tau_2 \tau_3 
(- s_{11'3})^{ \alpha_3 - \alpha_1 - \alpha_2} \over 
sin \pi (\alpha_3 - \alpha_1 - \alpha_2)} \biggr] }
\auto\label{sig575}
$$
\medskip
$$
[Disc]_{s_{13}} ~[Disc]_{s_{2'3'}} [Disc]_{s_{11'3}} 
~~= ~~ 8~\beta^a_{\alpha_1,\alpha_2,\alpha_3,0,0}~
(s_{13})^{\alpha_1} (s_{2'3'})^{\alpha_2} (s_{11'3})^{ \alpha_3 - 
\alpha_1 - \alpha_2}  
\auto\label{sig576}
$$
Note that for a reggeized gluon with $\alpha_i = 1 + O(g^2)$ 
$$
sin\pi \alpha_2 ~\sim ~ sin\pi(\alpha_3 -\alpha_1 - \alpha_2)
~ \sim ~ O(g^2)
\auto\label{5076}
$$
and since $\tau_i = -1~\forall~i$, 
each discontinuity reduces the amplitude by $O(g^2)$. With two reggeon 
states in each channel $\tau_i = -1~\forall~i$. In this case 
the leading terms in each of the square brackets in (\ref{sig5720}) cancel and
taking discontinuities does not introduce extra powers of $g^2$.
 
Moving on to the contributions to $M^{{\cal C}_b}$ obtained from the
signatured form of (\ref{sw3}). As we noted earlier, for the signature
constraint (\ref{pw60}) to also emerge from the hexagraph definition of
signature, then it has to be that the sum of the 
amplitudes $M^{{\cal C}_b}$ and $M^{{\cal
C}_{b'}}$ is equal in the two physical regions in which they appear, 
as must also be the amplitudes $M^{{\cal C}_c}$ and $M^{{\cal C}_{c'}}$.
As we emphasized, the regge amplitudes we discuss can not distinguish
the contribution of asymptotically equivalent cuts. In the following, 
therefore, we identify $M^{{\cal C}_b}$ with the sum of $M^{{\cal C}_b}$ and 
$M^{{\cal C}_{b'}}$. 

Since we only have two signature factors
to add, (\ref{sig5720}) is replaced by 
$$
{[1 + \tau_1 e^{i\pi \alpha_1} + \tau_2 e^{i\pi \alpha_2}
+ \tau_1 \tau_2 e^{i\pi (\alpha_1 + \alpha_2}) ] \over 
[sin \pi (\alpha_1 - \alpha_2)
sin {\pi \over 2} (\alpha_2 + \alpha_3 - \alpha_1)
sin {\pi \over 2} (\alpha_1 + \alpha_2 - \alpha_3)]} 
\auto\label{sig57210}
$$
After adding the term corresponding to (\ref{sw3}) from $n_1 + n_2 < 0$, we 
obtain the factor
$$
{[1 + \tau_1 e^{i\pi \alpha_1} + \tau_2 e^{i\pi \alpha_2}
+ \tau_1 \tau_2 e^{i\pi (\alpha_1 + \alpha_2}) ] \over 
[sin {\pi \over 2} (\alpha_1 + \alpha_3 - \alpha_2)
sin {\pi \over 2} (\alpha_2 + \alpha_3 - \alpha_1)
sin {\pi \over 2} (\alpha_1 + \alpha_2 - \alpha_3)]} 
\auto\label{sig5721}
$$
The interpretation of the phases in 
terms of cut structure is now more subtle. 
The $\tau_1$ twist of the hexagraph containing 
$M^{{\cal C}_b}$ sends $s_{123} \to - s_{123}$ in addition to $s_{13} \to
- s_{13}$. The corresponding triple-regge behavior should therefore be
$$
\eqalign{ (- s_{13})^{(\alpha_1+\alpha_3-\alpha_2)/2}&
(s_{2'3'})^{(\alpha_2+\alpha_3-\alpha_1)/2}
(- s_{123})^{(\alpha_1+\alpha_2 -\alpha_3)/2} \cr
&~=~ e^{i\pi \alpha_1} (s_{13})^{(\alpha_1+\alpha_3-\alpha_2)/2}
(s_{2'3'})^{(\alpha_2+\alpha_3-\alpha_1)/2}
(s_{123})^{(\alpha_1+\alpha_2 -\alpha_3)/2} }
\auto\label{57600}
$$
which is the phase-factor corresponding to the $\tau_1$ term in 
(\ref{sig5721}). This phase is due to the contribution of two cuts
rather than the conventional single cut, as 
was the case for $M^{{\cal C}_a}$. 
The $\tau_2$ twist gives the analagous result and explains the $\tau_2$ term in 
(\ref{sig5721}). Because of the 
signature constraint (\ref{pw60}) we have $\tau_1 \tau_2\equiv \tau_3$. We 
can obtain the corresponding phase for the $\tau_1 \tau_2$ term 
if we multiply by 
$e^{i\pi (\alpha_3 - \alpha_1 -\alpha_2)} $. The presence or absence 
of a phase of this kind is the ambiguity in (\ref{sw3}) that we discussed at 
the end of sub-section 5.5. Invoking this phase, we can write 
the analogue of (\ref{sig573}) as 
$$
\eqalign{& \beta^b_{\alpha_1,- \alpha_2,\alpha_3,0,0}~
\biggl[~(s_{13})^{(\alpha_1+\alpha_3-\alpha_2)/2}
(s_{2'3'})^{(\alpha_2+\alpha_3-\alpha_1)/2}
(s_{123})^{(\alpha_1+\alpha_2 -\alpha_3)/2} \cr
& ~~~~~+~\tau_1 (- s_{13})^{(\alpha_1+\alpha_3-\alpha_2)/2}
(s_{2'3'})^{(\alpha_2+\alpha_3-\alpha_1)/2}
(- s_{123})^{(\alpha_1+\alpha_2 -\alpha_3)/2} \cr
& ~~~~~~~~~~+~ \tau_2 (s_{13})^{(\alpha_1+\alpha_3-\alpha_2)/2}
(- s_{2'3'})^{(\alpha_2+\alpha_3-\alpha_1)/2}
(- s_{123})^{(\alpha_1+\alpha_2 -\alpha_3)/2} \cr
& ~~~~~~~~~~~~~~~+~\tau_3 
(-s_{13})^{(\alpha_1+\alpha_3-\alpha_2)/2}
(-s_{2'3'})^{(\alpha_2+\alpha_3-\alpha_1)/2}
(s_{123})^{(\alpha_1+\alpha_2 -\alpha_3)/2}~\biggr] \cr 
& ~~~~~~~~~~~~~~~~~~~\bigg/ ~ 
[sin {\pi \over 2} (\alpha_1 + \alpha_3 - \alpha_2)
sin {\pi \over 2} (\alpha_2 + \alpha_3 - \alpha_1)
sin {\pi \over 2} (\alpha_1 + \alpha_2 - \alpha_3)]
}
\auto\label{57601}
$$
which is now the sum of the 
leading helicity-flip triple-regge contributions of cuts of
the form of $M^{{\cal C}_b}$ in the hexagraphs of Fig.~5.3. (Note
that we can assign the $\tau_i$ factors in many different ways using the
relations $\tau_1\tau_2\tau_3=1$ and $\tau_i^2=1,~ i=1,2,3$). 
If we assume that 
$\beta^b_{\alpha_1,- \alpha_2,\alpha_3,0,0}$ is real then the phases of the
four terms in (\ref{57601}) naturally represent the four possible cut 
structures. There are only four terms because of the equalities leading to the
signature constraint.

Taking discontinuities in (\ref{57601}) (bearing in mind asymptotic 
equivalence) 
$$
\eqalign{ [Disc]_{s_{13}} ~[Disc]_{s_{2'3'}} ~~=& ~~
2~\beta^b_{\alpha_1,-\alpha_2,\alpha_3,0,0}~
(s_{13})^{(\alpha_1+\alpha_3 -\alpha_2)/2 } 
(s_{2'3'})^{(\alpha_2+\alpha_3 -\alpha_1)/2 } \cr
&~~~~~~~~~~ \times~  { (s_{123})^{ (\alpha_1+\alpha_2 -\alpha_3)/2 }  
\over 
sin {\pi \over 2} (\alpha_1 + \alpha_1 - \alpha_3)} }
\auto\label{sig5751}
$$
and the triple discontinuity is 
$$
\eqalign{[Disc]_{s_{13}} ~[Disc]_{s_{2'3'}} [Disc]_{s_{123}} 
~~= ~~&8~ \beta^b_{\alpha_1,-\alpha_2,\alpha_3,0,0}~
(s_{13})^{(\alpha_1+\alpha_3 -\alpha_2)/2 } 
(s_{2'3'})^{(\alpha_2+\alpha_3 -\alpha_1)/2 } ~~~~~~ \cr
& \times ~ (s_{123})^{ (\alpha_1 + \alpha_2 -\alpha_3)/2 }  }
\auto\label{sig5761}
$$
Note that because 
$$
sin{\pi \over 2}(\alpha_i +\alpha_j - \alpha_k)
~=~sin{\pi \over 2}(1 + O(g^2)) ~=~ 1 + O(g^4)
\auto\label{sig57610}
$$
taking discontinuities, in lowest-order, simply introduces
factors of 2 as in (\ref{211dc}). 
This  simplicity holds for the leading helicity amplitude and (because 
the azimuthal angle sums are convergent) also for 
the full triple-regge amplitude. Using (\ref{212}) we see that in lowest-order 
each of the amplitudes in (\ref{57601}) is pure imaginary.

Finally we come to the unphysical triplets of the form $M^{{\cal C}_c}$.
The regge behavior obtained from $M^{{\cal C}_c}$ has to have 
essentially the same form as that obtained from $M^{{\cal C}_b}$. 
However, with the appropriate choice of the phase ambiguity in (\ref{sw3}) 
we can obtain, instead of (\ref{57601}), the triple-regge amplitude
$$
\eqalign{& \beta^c_{\alpha_1,- \alpha_2,\alpha_3,0,0}~
\biggl[~(s_{13})^{(\alpha_1+\alpha_3-\alpha_2)/2}
(s_{2'3'})^{(\alpha_2+\alpha_3-\alpha_1)/2}
(-s_{12})^{(\alpha_1+\alpha_2 -\alpha_3)/2} \cr
& ~~~~~+~\tau_1 (-s_{13})^{(\alpha_1+\alpha_3-\alpha_2)/2}
(s_{2'3'})^{(\alpha_2+\alpha_3-\alpha_1)/2}
( s_{12})^{(\alpha_1+\alpha_2 -\alpha_3)/2} \cr
& ~~~~~~~~~~+~ \tau_2 (s_{13})^{(\alpha_1+\alpha_3-\alpha_2)/2}
(- s_{2'3'})^{(\alpha_2+\alpha_3-\alpha_1)/2}
( s_{12})^{(\alpha_1+\alpha_2 -\alpha_3)/2} \cr
& ~~~~~~~~~~~~~~~+~\tau_3 
(-s_{13})^{(\alpha_1+\alpha_3-\alpha_2)/2}
(-s_{2'3'})^{(\alpha_2+\alpha_3-\alpha_1)/2}
(- s_{12})^{(\alpha_1+\alpha_2 -\alpha_3)/2}~\biggr]  \cr
& ~~~~~~~~~~~~~~~~~~~~\bigg/ ~ 
[sin {\pi \over 2} (\alpha_1 + \alpha_3 - \alpha_2)
sin {\pi \over 2} (\alpha_2 + \alpha_3 - \alpha_1)
sin {\pi \over 2} (\alpha_1 + \alpha_2 - \alpha_3)]
}
\auto\label{57602}
$$
Again we have only four terms because of the signature constraint. 

Note that both (\ref{57601}) and (\ref{57602}) are symmetric with respect to
$\alpha_1, \alpha_2$ and $\alpha_3$ and that more generally
the breaking of cyclic symmetry by our choice of hexagraphs has been restored. 
Full expansions of the form of 
(\ref{trpb}) still reflect the hexagraphs used in their derivation. 
However, this is no more than a choice of variables to expand in. 
Also, neither (\ref{57601}) nor (\ref{57602}) has any $t$-channel poles
when any of the $\alpha_i$ pass through integer values. Indeed, any 
relationship between analytic structure and the hexagraph formalism 
used has essentially been lost. The asymptotic structure of
the $M^{{\cal C}_b}$ and $M^{{\cal C}_c}$ amplitudes has no preference for 
the $t$-channel in which the partial-wave analysis is carried out.
Ultimately this goes back to the constraint (\ref{pw6}) which, in fact, is
satisfied symetrically. Note, however, that the denominator in both 
(\ref{57601}) and (\ref{57602}) does appear to produce unphysical poles in the
$\alpha_i$. The $M^{{\cal C}_b}$ and 
$M^{{\cal C}_c}$ amplitudes have to combine to cancel these poles. This is a
consistency condition which clearly requires the 
presence of the unphysical triple discontinuities in the dispersion relation.

\subhead{5.10 Regge Cuts}

(\ref{sig573}) and (\ref{57601}) contain the contribution of Regge poles only. 
To replace a 
Regge pole by the Regge cut corresponding to a two-reggeon state is 
straightforward in principle but in practise can be quite complicated. 
However, for odd-signature
reggeized gluons with $\alpha \sim 1$ it is relatively simple to describe
the first-order approximation. 
For example, replacing the Regge pole in the $1$ channel in (\ref{sig573}) 
by an even signature two-reggeon state gives,  
$$
\eqalign{ \biggl[{(s_{13})^{\alpha_1} + (- s_{13})^{\alpha_1}  \over 
sin \pi \alpha_1 } \biggr] ~ & \longrightarrow~
\int { d^2k \over 
sin \pi \alpha (k^2)  sin \pi \alpha ((Q_1 -k)^2)  } \cr
&~~~~~~~ \times 
\biggl[{(s_{13})^{(\alpha(k^2) + \alpha(Q_1-k)^2 -1)} 
+ (- s_{13})^{(\alpha(k^2) + \alpha(Q_1-k)^2 -1)} \over
sin \pi (\alpha(k^2) + \alpha(Q_1-k)^2 -1)}  \biggr] ~~~~ \cr
&\sim ~~~ \int { d^2k \over 
(k^2)  (Q_1 -k)^2  }~~ \bigl[~s_{13}~ \bigr] ~\biggl[ 1 + O(g^2) \biggr]\cr
&\sim ~~~ J_1(Q_1^2) ~~ \bigl[~s_{13}~ \bigr] ~\biggl[ 1 + O(g^2) \biggr]}
\auto\label{sig5731}
$$
Similarly 
$$
\eqalign{ ~~~~ [Disc]_{s_{13}} ~ & \longrightarrow~
~\int { d^2k \over 
sin \pi \alpha (k^2)  sin \pi \alpha ((Q_1 -k)^2)  } 
(s_{13})^{(\alpha(k^2) + \alpha(Q_1-k)^2 -1)} \cr
&\sim ~~~ J_1(Q_1^2) ~~ \bigl[~s_{13}~ \bigr] ~\biggl[ 1 + O(g^2) \biggr]
}
\auto\label{sig57310}
$$
Analagous changes occur if we replace any of the other Regge pole 
contributions in the foregoing by Regge cuts. The isolated power behavior 
corresponding to a regge pole is replace by a continuum integral of the power
behavior involved together with a corresponding replacement of signature
factors. Apart from this, 
all the above discussion of hexagraph contributions,
analytic structure of triple-regge, helicity-flip, and helicity non-flip,
amplitudes goes through in complete parallel for amplitudes containing general 
multi-reggeon states in each $t$-channel.
 
To give another specific example that is directly relevant for our 
discussion of anomaly amplitudes, we consider the first term in (\ref{57602}).
If each of the regge poles is replaced by a two-reggeon state we obtain
$$
\eqalign{ &~~\prod_i~\int ~~{ d^2k_i \over 
sin \pi \alpha (k_i^2)  sin \pi \alpha ((Q_i -k_i)^2)  } \cr
&\beta^c_{[\alpha (k_1^2)+\alpha ((Q_1 -k_1)^2) -1], 
-[\alpha (k_2^2)+\alpha ((Q_2 -k_2)^2) -1], 
[\alpha (k_3^2)+\alpha ((Q_3 -k_3)^2) -1],0,0}(k_1,k_2,k_3,Q_1,Q_2,Q_3)\cr 
&\biggl[ ~(s_{13})^{[\alpha (k_1^2)+\alpha ((Q_1 -k_1)^2) + 
\alpha (k_3^2)+\alpha ((Q_3 -k_3)^2) -
\alpha (k_2^2)-\alpha ((Q_2 -k_2)^2) -1]/2} \cr
& ~~~~~~~~ (s_{2'3'})^{[\alpha (k_3^2)+\alpha ((Q_3 -k_3)^2) + 
\alpha (k_2^2)+\alpha ((Q_2 -k_2)^2) -
\alpha (k_1^2)-\alpha ((Q_1 -k_1)^2) -1]/2} \cr
& ~~~~~~~~~~~~(-s_{12})^{[\alpha (k_1^2)+\alpha ((Q_1 -k_1)^2) + 
\alpha (k_2^2)+\alpha ((Q_2 -k_2)^2) -
\alpha (k_3^2)-\alpha ((Q_3 -k_3)^2) -1]/2} \biggr]~ \bigg/ \cr
&\biggl[[sin {\pi \over 2} [\alpha (k_1^2)+\alpha ((Q_1 -k_1)^2) + 
\alpha (k_3^2)+\alpha ((Q_3 -k_3)^2) -
\alpha (k_2^2)+\alpha ((Q_2 -k_2)^2)] \cr
&~~ sin {\pi \over 2} [\alpha (k_3^2)+\alpha ((Q_3 -k_3)^2) + 
\alpha (k_2^2)+\alpha ((Q_2 -k_2)^2) -
\alpha (k_1^2)+\alpha ((Q_1 -k_1)^2) ] \cr
& ~~ sin {\pi \over 2} [\alpha (k_1^2)+\alpha ((Q_1 -k_1)^2) + 
\alpha (k_2^2)+\alpha ((Q_2 -k_2)^2) -
\alpha (k_3^2)+\alpha ((Q_3 -k_3)^2) ] \biggr]\cr
&\centerunder{$\sim$}{\raisebox{-6mm}{$ g^2 \to 0$}}~
(s_{13})^{1/2}(s_{2'3'})^{1/2}(s_{1'2})^{1/2}
~\prod_i~\int ~~{ d^2k_i \over 
k_i^2  (Q_i -k_i)^2  }~
\beta^c_{1, -1, 1,0,0}(k_1,k_2,k_3,Q_1,Q_2,Q_3)}
\auto\label{57602c}
$$
The last line can be identified directly with the general reggeon diagram
amplitude (\ref{211}) so that $\beta^c_{1, -1, 1,0,0}(k_1,k_2,k_3,Q_1,Q_2,Q_3)$
is identified with the reggeon vertex $R(k_1,k_2,k_3,Q_1,Q_2,Q_3)$. More
specifically, the last line of (\ref{57602c}) can also be identified with
reggeon diagram amplitudes such as (\ref{578}) discussed in the last Section.
From (\ref{sig57310}) and (\ref{57602c}) it is clear that the simple 
leading order properties of discontinuities that follows from
(\ref{sig57610}) hold for regge cut as well 
as regge pole amplitudes.

In general, as we have already emphasized, we expect amplitudes containing
regge cuts (i.e. multi-reggeon states) in any channel to have the 
non-planarity properties necessary to produce simultaneous right and left hand 
cuts in integrated invariants. This will lead to closely related
right and left hand cuts in external invariants.
Comparing (\ref{57601}) and (\ref{57602}) we see
that for fixed $s_{13}$ and $s_{2'3'}$ the two contributions provide right
and left-hand cuts in the $s_{1'2} \sim z_1z_2$ plane. We anticipate, and in 
the next Section will find, that amplitudes containing regge cuts in each 
channel will have closely related cuts of this kind and so if they contribute 
to $M^{{\cal C}_b}$ they will typically contribute also 
to $M^{{\cal C}_c}$.
Indeed, in many respects the $M^{{\cal C}_c}$ amplitudes provide the additional
four amplitudes that would need to be added to the $M^{{\cal C}_b}$ amplitudes
to obtain a complete set of signatured amplitudes with no signature constraint.
However, the analytic relationship between the the asymptotic cuts and the
angular invariants in which signature properties are necessarily determined
prevents such a relationship. Also we will see that the anomaly can consistently
appear in the $M^{{\cal C}_c}$ amplitudes but not the 
$M^{{\cal C}_b}$ amplitudes.

\subhead{5.11 Dimensions of Reggeon Interactions}

Next we note a crucial difference between (\ref{hp2i}) and 
(\ref{hp2i1}) that is vital for the appearance of the triangle anomaly.
First we set 
$$
\alpha_1~= ~\alpha_2~= ~\alpha_3~= ~1
\auto\label{aas1}  
$$
corresponding to the contribution of (multi-)gluon reggeon states. 
We then compare the momentum dimension of (\ref{hp2i}) and 
(\ref{hp2i1}). For (\ref{hp2i}) we obtain
$$
(s_{13})^{\alpha_1}(s_{2'3'})^{\alpha_2}(s_{11'3})^{ \alpha_3 - 
\alpha_1 - \alpha_2} ~\equiv~ [s]^{1+1 -1} ~=~ [s]
\auto\label{aas2}
$$
while for (\ref{hp2i1}) we obtain
$$
(s_{31})^{(\alpha_1+\alpha_3-\alpha_2)/2}
(s_{23})^{(\alpha_2+\alpha_3-\alpha_1)/2}
(s_{12})^{(\alpha_1+\alpha_2
-\alpha_3)/2} ~\equiv~ [s]^{{1 \over 2} + {1 \over 2} + {1 \over 2}}
~=~[s]^{{3 \over 2}}
\auto\label{aas3}
$$
Since the contribution of any multi-reggeon state always carries the same 
transverse dimension dimension 
$$
\int { d^2 k_{1}d^2 k_{2}  
~ \cdots \delta^2 (Q -  k_{1} -  k_{2} - \cdots)
\over k_{ 1}^2  k_{2}^2 ~\cdots} ~~~\equiv~~[Q]^{-2} 
\auto\label{aas30}
$$
the difference in dimensions of (\ref{aas2}) and (\ref{aas3}) has to be 
compensated by a difference in the transverse momentum dimension of the 
accompanying reggeon vertex. If we anticipate that (as is the case)
the dimension of the vertex accompanying (\ref{aas2}) is the normal $[Q]^2$
for a reggeon vertex in QCD, then the vertex accompanying (\ref{aas3}) will
have the ``anomalous dimension'' of $[Q]$. This anomalous dimension 
allows the reggeon interaction vertices generated by 
$M^{{\cal C}_b}$ or $M^{{\cal C}_c}$ to potentially
contain the anomaly of the four-dimensional triangle diagram - which is
linear in it's momentum dimension and is independent of any other scale. 

\subhead{5.12 Multi-Regge Amplitudes and the Anomaly}

Note that the amplitude for the process given by Fig.~4.15(a),
that contains the anomaly, has no initial ($s_{13}$) or final ($s_{2'3'}$)
state discontinuities. If we suppose that external $G_i$ couplings can be 
chosen such that the anomaly in an amplitude of this kind does not cancel 
then the amplitude must be reproduced by
a triple-regge amplitude with the anomaly in the six-reggeon vertex. In QCD
we will find that a quantum number (color parity) prevents the anomaly 
from appearing in the triple-regge amplitude unless it effectively appears
already in the $G_i$ couplings, so that they violate color parity.
If it does appear as a physical region singularity in the triple-regge
amplitude, as in Fig.~4.15(a), we know from the last Section (and will see
explicitly in the next Section) that it has to appear in an $M^{{\cal C}_c}$
and/or an $M^{{\cal C}_b}$ amplitude. It, therefore, has to appear
in the regge cut analogue of the last term of (\ref{57602}) and/or the last 
term  of (\ref{57601}) since these are the only terms without
($s_{13}$) or ($s_{2'3'}$) discontinuities. But, 
since the anomaly has to appear in a reggeon vertex, if it appears in 
any of the terms in 
(\ref{57602}) or (\ref{57601}) then it must appear in all of them.
However, the multi-regge behavior in 
the first term of (\ref{57601}) already represents 
the maximum set of physical region cuts allowed by the 
Steinmann relations and so can not contain the anomaly.
Consequently it can not appear in any of the terms in (\ref{57601}). 
We can also argue that it should appear in (\ref{57602}) as follows.

We can draw the quark loop of Fig.~4.10 as in Fig.~5.7(a) with the attached
gluon lines labeled by the index for the corresponding external momentum 
entering or exiting the associated external coupling. 
The double discontinuity in $s_{13}$ and $s_{2'3'}$ puts the hatched lines
on-shell as illustrated in Fig.~5.7(a). The anomaly puts the remaining
unhatched lines on-shell as discussed in Section 4.
\begin{center}
\leavevmode
\epsfxsize=3.5in
\epsffile{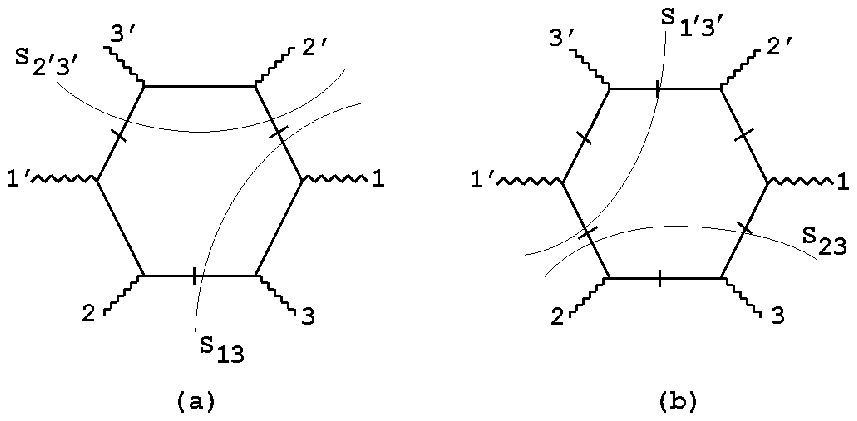}

Fig.~5.7 The Quadruple Discontinuity.
\end{center}
As illustrated in Fig.~5.7(b) the remaining unhatched 
lines are also put 
on-shell by taking the double discontinuity in the asymptotically equivalent
set of cuts $s_{1'3'}$ and $s_{23}$. The anomaly therefore occurs in
combination with the asymptotic cuts of the triple-regge amplitude
when there is a quadruple discontinuity in two sets of asymptotically
equivalent cuts, in particular $s_{13}$ and $s_{2'3'}$  together
with $s_{1'3'}$ and $s_{23}$. (As we discuss in the next Section, the third
discontinuity involved in triple-regge behavior is produced by the 
addition of a further gluon.) Either of the two sets of cuts
can produce the anomaly while the other produces the asymptotic regge hehavior
and, in fact, the two possibilities correspond to
the two anomaly contributions of Fig.~4.8(a) and (b) discussed in Sections 2
and 4 that are related by a parity transformation. Note that the chirality
violating quark line is different in the two cases.

It is straightforward to determine, by comparing the asymptotic relations
(\ref{ap1})-(\ref{ap6}) with the light-cone formulae (\ref{npl01}) and 
(\ref{npl1}), that in the anomaly configuration $t_1=t_2$ and $cos~ \omega_1
=cos~ \omega_2$ so that, asymptotically, $s_{13}=s_{23}=s_{1'3'} =s_{2'3'}$.
This allows the quadruple discontinuity to appear in the
asymptotic region. Because the Steinmann relations should be valid
(asymptotically) in a physical region, the quadruple discontinuity can not
approach the asymptotic region (as a function of the non-asymptotic variables)
from within a physical region. It must do so from an unphysical region where,
just because it is unphysical, it can contain a chirality transition. Indeed,
from Fig.~5.7 it is clear that if all quark lines are on-shell then this can
only be achieved in a region where there are also discontinuities in $s_{12'}$
and $s_{21'}$. This is the unphysical region where the triple discontinuity
${\cal C}_c$ occurs. In this region, the anomaly clearly can appear as a
``physical region'' singularity in combination
with a triple discontinuity. If it then appears as a singularity in a 
reggeon interaction vertex, it can consistently appear in all four terms 
of (\ref{57602}), since all four represent unphysical triple discontinuities.
The last term can contain the amplitude of Fig.~4.15(a) and,
as will be come clear from the next Section, the initial calculation of 
Fig.~4.8(a) can be regarded as calculating the double discontinuity present in
the first term of (\ref{57602}).

\newpage

\mainhead{6. Multiple Discontinuities} 

In this Section we evaluate multiple discontinuities and look for the 
anomaly in reggeon interactions via the multi-regge 
formalism of the last Section. 
The writing of an asymptotic 
dispersion relation, without subtractions, 
depends\cite{arw1,sw} on the feature that all asymptotic behavior originates 
from multi-regge singularities and our analysis implicitly assumes that all
asymptotic contributions of a Feynman graph can be assigned to multi-regge
amplitudes of some form.

\subhead{6.1 The Simplest Diagrams}

From the previos Section we know that the anomaly can only appear in 
amplitudes of the $M^{{\cal C}_c}$ form with triple
discontinuities corresponding to tree diagrams of the form illustrated 
in Fig.~2.9(c). To count all triple 
discontinuities, and the diagrams that contribute to them, we 
first consider the initial and final state double discontinuities 
that are uniquely associated with each hexagraph. However, we then find that,
in the triple-regge limit, the lowest-order graphs do not have a non-trivial
third discontinuity that can be obtained by putting further propagators 
on-shell. Additional discontinuities of this kind, 
associated with triple-regge
behavior, appear only as additional gluons are added and the reggeization
effects appear as in (\ref{sig57310}) and (\ref{57602c}). 

An additional discontinuity can be trivially taken by using the equivalent of
(\ref{211dc}) but to strictly justify this again requires calculating the same
reggeization effects. To carry out a complete study of how such triple 
discontinuities appear, therefore, we would need to go to at least the next
order of perturbation theory. In the following we will do something in between.
We know from the discussion of the previous Section that we should evaluate 
triple discontinuities of the form of Fig.~2.9(c). Having evaluated a double
discontinuity, we will check diagrammatically that the appropriate additional
discontinuity does indeed appear as additional gluons are added. We then appeal
to (\ref{sig5751})-(\ref{sig57610}), as applied to (\ref{57602c}),
and extract the reggeon interaction directly from the double discontinuity -
in effect simply using (\ref{211dc}).

A double discontinuity requires a minimum of 
two gluons exchanged in each $t$-channel. To obtain the 
double discontinuity in $s_{13}$ and $s_{2'3'}$ that is 
associated with the first hexagraph of Fig.~5.2
we consider the diagrams of Fig.~6.1. 
The desired double discontinuity is obtained 
by putting the hatched quark lines on-shell. 
If we ignore gluon self-interactions, we can argue that 
these diagrams are a complete set as follows. 
The initial scattering process producing the $s_{13}$ intermediate state 
is necessarily the production of a quark-antiquark pair
and without loss of generality we can draw this process as in 
the bottom part of Fig.~6.1(a), provided we don't distinguish a quark 
direction on the exchanged quark line. Similarly the $s_{2'3'}$
intermediate state is associated with the reverse of this production process. 
The quark loop obtained by joining the amplitudes for these initial and 
final scatterings is either planar, as in the first six diagrams of 
Fig.~6.1, or it has a twist 
in it, as in the second six diagrams. The six diagrams of each kind are 
obtained by attaching the two gluons that do not participate 
in either the 
initial or final scattering process, in all possible ways. 
\begin{center}

\leavevmode
\epsfxsize=5.3in
\epsffile{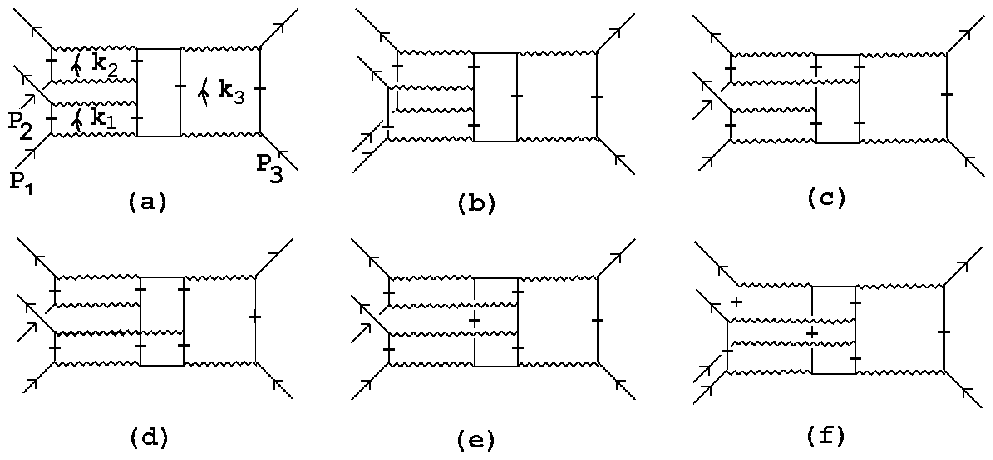}

\leavevmode
\epsfxsize=5.3in
\epsffile{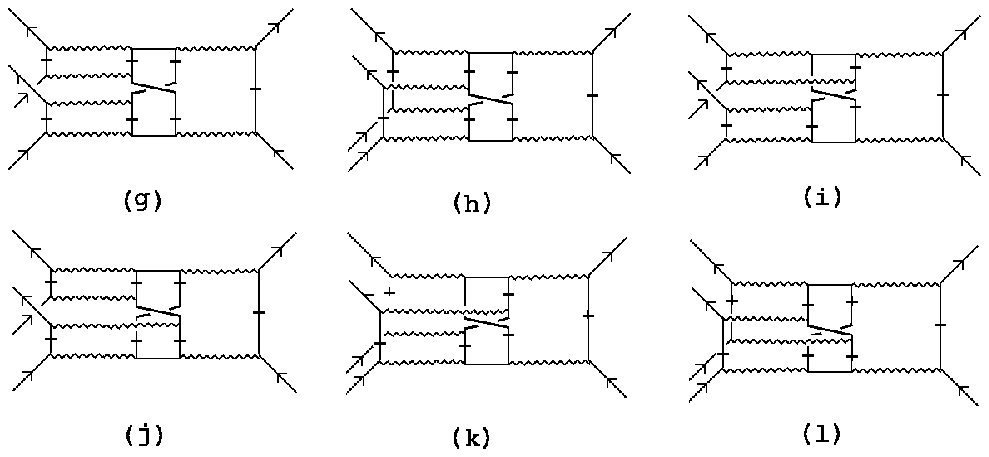}

Fig.~6.1 Diagrams with Two Gluons in each $t$-channel.

\end{center}
Apart from the need to sum over the direction of the quark line around the 
loop, the diagrams of Fig.~6.1  are all of the lowest-order diagrams with 
both an $s_{13}$ and an $s_{2'3'}$ discontinuity. 
We evaluate diagrams in the full 
triple Regge limit (\ref{np3}) in which 
the $P_i$ become lightlike in distinct directions and the $Q_i$ have the 
general form given in (\ref{np3}). 
In each case, the double discontinuity 
provides a sufficient number of $\delta$-functions to perform all 
longitudinal integrations. 

\subhead{6.2 The Diagram of Fig.~6.1(a).}

We have already discussed this diagram at length in sub-section 4.1. Indeed
the hatched lines of Fig.~6.1(a), that are placed on-shell to obtain the
double discontinuity, are the same as those of Fig.~4.2. Our previous analysis
is, therefore, sufficient to determine that the anomaly is not present.
There is, however, an important point concerning further discontinuities
that we referred to above and applies to our analysis of all the remaining
diagrams.

A-priori, there is an additional $s_{11'3}$ discontinuity which we can take
by putting the only unhatched vertical line on shell. If
we repeat our evaluation of Fig.~4.2 but instead use 
co-ordinates for the $k_1$ and $k_2$ integrations in which
$\underline{n}_{1^+}$ and $\underline{n}_{2^+}$ 
are the basic light-like momenta, the longitudinal 
integrations will lead to the $\gamma$-matrix couplings 
shown in Fig.~6.2 
\begin{center}
\leavevmode
\epsfxsize=1.1in
\epsffile{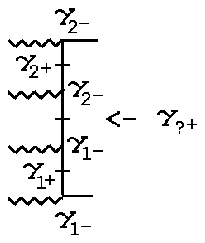}

Fig.~6.2 $\gamma$-Matrix Couplings for Fig.~6.1(a)
\end{center}
If the middle quark line is to put be put on-shell and give a leading 
contribution, then it must be 
helicity-conserving with respect to both the upper and lower on-shell 
states. However, this is clearly not possible since both options give a 
product of $\gamma$-matrices that is zero.

As we will see below, this last point applies generally to all the diagrams 
we discuss. The triple-regge behavior
we are looking for is inconsistent with taking a discontinuity through a
remaining uncut quark line. For the diagram under
discussion we note that while we can not introduce an $s_{11'3}$ discontinuity 
without cutting the forbidden quark line we can 
introduce an unphysical $s_{1'2}$ discontinuity by 
adding an extra gluon, as illustrated in Fig.~6.3.
\begin{center}
\leavevmode
\epsfxsize=2in
\epsffile{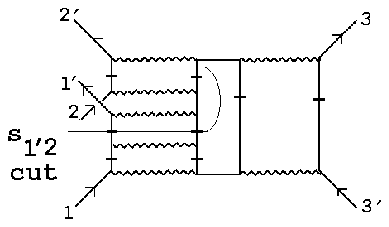}

Fig.~6.3 An $s_{1'2}$ Discontinuity Introduced by Adding an Extra Gluon. 
\end{center}
(The lines placed on-shell by the additional discontinuity are indicated by 
the hatches on the thin line showing how the discontinuity is taken.)
In this way we can replace 
a single real gluon in the original diagram by the one-loop contribution to 
the reggeization of this gluon. We must, of course, remember that 
a cut gluon of this kind has necessarily to be considered as reggeized when we
extract reggeon interactions.

In general we expect that 
the sum total of higher-order triple discontinuity contributions that produce 
reggeization effects will simply determine that the lower-order diagram is
obtained as a generalised ``real part''. The reggeon vertex obtained from the
original diagram, with no additional discontinuity taken, will then be 
obtained. 
The existence of the $s_{1'2}$ discontinuity implies that 
the reggeon interaction
obtained from Fig.g~6.1(a) will appear in the triple regge amplitude 
associated with the triplet ${\cal C}_c$ discussed in the last Section.
However, since only one set of cuts appears, we expect the arguments of Section
4 that the six-reggeon interaction computed from Fig.~6.1(a) will be zero
when all-orders reggeization effects are included will be valid.

\subhead{6.3 Isolating the Anomaly}

In all of the remaining diagrams of Fig.~6.1, one or more of the $k_i$ loop
momenta flow through more than one line of the internal quark loop.
Consequently, as we already saw in Section 4, the reduction 
of the $k_i$ integrations to two dimensions is not as straightforward as it 
was for Fig.~6.1(a). The internal quark loop and the remaining two-dimensional
$k_i$ integrations are not, in general, coupled only by an effective
point-coupling and the reggeon vertices generated are very 
complicated. As we have made clear already, 
we will not attempt to obtain complete expressions 
for the vertices generated by 
the remaining diagrams in Fig.~6.1. Rather we will concentrate on isolating 
contributions that might contain the anomaly. 

Our search for the anomaly will, as in Section 4,  
be based on the discussion of Appendix A.
We will look for effective vector-like point-couplings for the three
vertices of a quark triangle diagram with an odd number of axial couplings. 
We will also look for the appropriate flow of a light-like momentum through
the reggeon vertex. We will assume that the anomaly, if present in a diagram, 
can be found using any of the
possible sets of 
light-cone variables discussed in Appendix B, provided we consider all 
choices for assigning particular quark propagators to particular 
longitudinal $k_i$ integrations. As will become clear, the appropriate
choice of variables will often enable us to see immediately whether a local 
coupling occurs or whether only non-local
couplings arise. 

\subhead{6.4 The Diagram of Fig.~6.1(b).}

At first sight this diagram has an 
$s_{123}$ discontinuity obtained by cutting the remaining uncut 
vertical line in Fig.~6.1(b) However, as illustrated in Fig.~6.4, 
the helicity conservation problem again arises.
\begin{center}
\leavevmode
\epsfxsize=1.1in
\epsffile{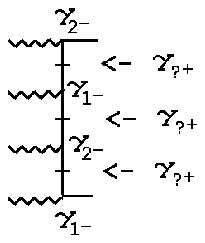}

Fig.~6.4 $\gamma$-Matrix Couplings for Fig.~6.1(b)
\end{center}
There are two on-shell scatterings for which it is impossible to choose
helicities such that both give the leading behavior. Instead we can 
introduce an $s_{123}$ discontinuity by 
adding an extra gluon, as illustrated in Fig.~6.5(a).
\begin{center}
\leavevmode
\epsfxsize=5in
\epsffile{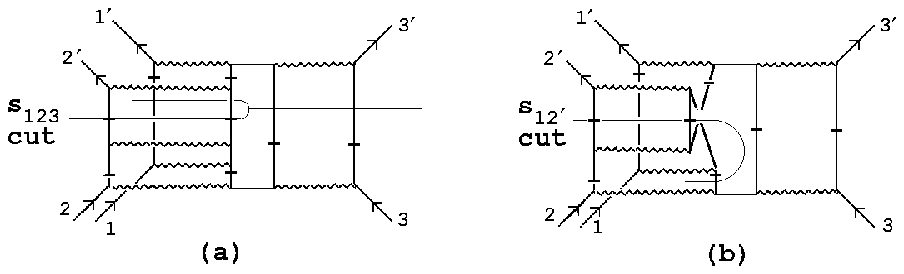}

Fig.~6.5 Adding a Gluon to Fig.~6.1(b) to Give Additional Discontinuities.
\end{center}
Alternatively, adding a gluon line as in Fig.~6.5(b) gives
an unphysical discontinuity in $s_{12'}$. (As before, the lines placed
on-shell by the additional discontinuity are indicated by the hatches
place directly on the thin line showing how the discontinuity is taken.) 
The two discontinuities are closely related, as anticipated in the 
previous Section. Again using 
the additional gluon loop to provide the reggeization of the gluon,
the appropriate reggeon vertex is that given
by the original diagram of Fig.~6.1(b) (apart from a normalization factor 
that we are not attempting to determine anyway).

The momentum flow through the 
internal quark loop of Fig.~6.1(b) 
and the $\gamma$ matrices involved are shown in 
Fig.~6.6.
As in our discussion of diagrams in Section 4, we use the light-cone
co-ordinates ($k_{i1^-},k_{i2^-}, \underline{\tilde{k}}_{i\perp}$)
introduced in Appendix B to perform the $k_1$ and $k_2$ integrations and to
evaluate the $\gamma$-matrix trace associated with the quark loop. For the
$k_3$ integration, the choice of co-ordinates is not critical. For
simplicity, we choose 
conventional light-cone co-ordinates ($k_{3^+},k_{3^-},k_{3 \perp}$).
Our evaluation of the integral $I_i$ of Fig.~4.3 can be repeated to 
perform the $k_{11^-},k_{22^-}$ and $k_{3^-}$ integrations 
using the $\delta$-function associated with the correponding 
external quark line. 
\begin{center}
\leavevmode
\epsfxsize=3in
\epsffile{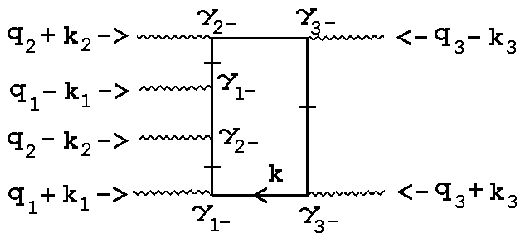}

Fig.~6.6 The Quark Loop in Fig.~6.1(b).
\end{center}

To perform the remaining longitudinal integrations we first route the $k_i$
momenta along the shortest path through the quark loop. This 
matches a unique on-shell (hatched-line) propagator in the loop with each $k_i$.
As before, to look for a potential $\gamma$-matrix point-coupling, we look for 
that momentum factor within the numerator of the 
on-shell quark that is multiplied by the same momentum 
that is scaling the longitudinal momentum 
integrated over via the $\delta$-function.
In particular the $k_{12^-}$ integration has the form
$$
\eqalign{ \int d k_{12^-} &~\delta\biggl((k_1 +k + Q_1)^2 -m^2\biggr)
~\gamma_{1^-}~\biggl((k_1 +k + Q_1)\cdot \gamma -m \biggr) 
 ~\gamma_{2^-}\times ~~\cdots  \cr
&= ~\int d k_{12^-} ~\delta\biggl( (k_{11^-} +k_{1^-} + Q_{11^-})k_{12^-}
 + \cdots \biggr) \cr
& ~~~~~~~~~ \times \gamma_{1^-} ~
\biggl( (k_{11^-} +k_{1^-} + Q_{11^-}) \cdot 
\gamma_{2^-} + \cdots \biggr) ~
\gamma_{2^-}~ \times~ \cdots  \cr
&= ~~ \gamma_{1^-} \gamma_{2^-}^2 ~~ + ~~\cdots \cr
&= ~~0~~ + ~~\cdots
}
\auto\label{541}   
$$
In this case the potential point-coupling from the 
($k_{11^-} +k_{1^-} + Q_{11^-}$)$\gamma_{2^-}$ term in the quark numerator
is eliminated by one of the adjacent $\gamma$-matrices (c.f. our evaluation 
of Fig.~4.8(a) in sub-section 4.3). Since the 
$k_{21^-}$ integration has the same structure as the
$k_{12^-}$ integration, performing each of these integrations will produce 
couplings of the form 
(\ref{rlc8}) rather than the point couplings 
necessary to produce the anomaly. In any alternative momentum flow and 
assignment of $\delta$-functions, it is straightforward to show that 
either the $k_{12^-}$ or the $k_{21^-}$
integration gives an analagous result to (\ref{541}), i.e. only non-local 
couplings remain. (Note that our choice of light-cone co-ordinates has 
enabled us to reach this conclusion rather simply).

\subhead{6.5 The Diagrams of Fig.~6.1(c) and (d).}
		
Because of the number of lines put on-shell by taking the $s_{13}$ and 
$s_{2'3'}$ discontinuities, there are not three quark lines off-shell
in either of these diagrams. As a result 
there is no possibility to generate the anomaly divergence.

\subhead{6.6 The Diagram of Fig.~6.1(e).}

This diagram is similar to that of Fig.~4.8(a) (which can be identified
with Fig.~6.1(f) discussed next) and can be analysed similarly. In fact, as 
we discussed in Section 2 and discuss further below, the reggeon vertices
obtained from Figs.~6.1(e) and (f) are related by reggeon Ward identities and
so must have similar properties. We route
the $k_i$ momenta through the (unique) shortest path combination
and again use the light-cone
co-ordinates ($k_{i1^-},k_{i2^-}, \underline{\tilde{k}}_{i\perp}$)
together with 
conventional light-cone co-ordinates ($k_{3^+},k_{3^-},k_{3 \perp}$).
Integrating the longitudinal momenta and keeping local couplings
produces the $\gamma$ matrix assignment of Fig.~6.7(a). Comparing with
Fig.~4.13 and the following analysis we see that the three $\gamma_5$ couplings 
generated are identical to those generated by Fig.~4.8(a) and Fig.~6.1(f).
A momentum configuration for Fig.~6.1(e) that parallels Fig.~4.15(a) is 
shown in Fig.~6.7(b). This configuration has already appeared in Fig.~2.6(b).
\begin{center}
\leavevmode
\epsfxsize=4.5in
\epsffile{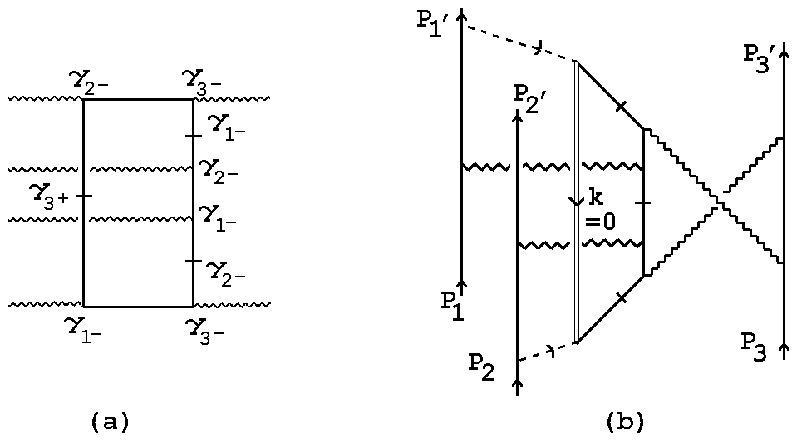}

Fig.~6.7 (a) $\gamma$-Matrices and (b) Anomaly Momentum Configuration for 
Fig.~6.1(e).
\end{center}

As in Fig.~4.15(a), the scattering process containing
the anomaly takes place in a part of the physical region where the
original discontinuities used to evaluate the reggeon interaction vertex are
no longer present. This is consistent with the discussion at the end of the
last Section, provided the unphysical $s_{1'2}$ discontinuity is present so 
that the anomaly is associated with an unphysical triple discontinuity.
The helicity conflict again prevents a 
further discontinuity being taken by cutting the remaining vertical quark line.
Considering discontinuities obtained by cutting gluon lines we find that the 
unphysical $s_{1'2}$ discontinuity is indeed the only one that
can be taken. We conclude that Fig.~6.1(e) does generate the anomaly,
just as the reggeon
Ward identity relationship that we discuss below requires. 

A-priori, we might suspect that the contribution of Fig.~6.1(e) 
will not persist 
if all-orders reggeization effects are included.
The lack of additional discontinuities 
can be traced to the essential
planarity of the coupling to the $t_3$-channel gluon exchanges.
As discussed in Section 4, this would be expected to allow
a contour closing that will give zero as higher-order
reggeization effects are added. However, as we discuss briefly in the next
Section, the anomaly produces both ultra-violet and infra-red effects. If it
is not canceled in the sum of all diagrams it's ultra-violet effects
could prevent such contour closing arguments. Alternatively, if the contour
closing arguments can be carried
through it would imply that all properties of the anomaly 
and it's relationship to reggeon Ward identity zeroes would be contained in 
the maximally non-planar diagram.

\subhead{6.7 The Diagram of Fig.~6.1(f).}

This diagram has already appeared extensively in early Sections, as the 
maximally non-planar diagram of Figs.~2.2 and 2.4, and in Fig.~4.8(a)
with the hatched lines put on-shell just as in Fig.~6.1(f). The extensive
discussion in Section 4 showed that the the anomaly is present although,
very importantly, the momentum configuration in which it appears occurs
in a part of the physical region distinct from that in which the 
discontinuities are evaluated. In the previous Section we 
concluded that this is resolved by associating the
anomaly with an unphysical triple discontinuity.
The helicity mismatching encountered above again implies that to take 
additional discontinuities we must cut through gluon lines. In Fig.~6.8 we have
shown that both an $s_{123}$ discontinuity and $s_{1'2}$ discontinuity 
can be obtained. 
\begin{center}
\leavevmode
\epsfxsize=4.5in
\epsffile{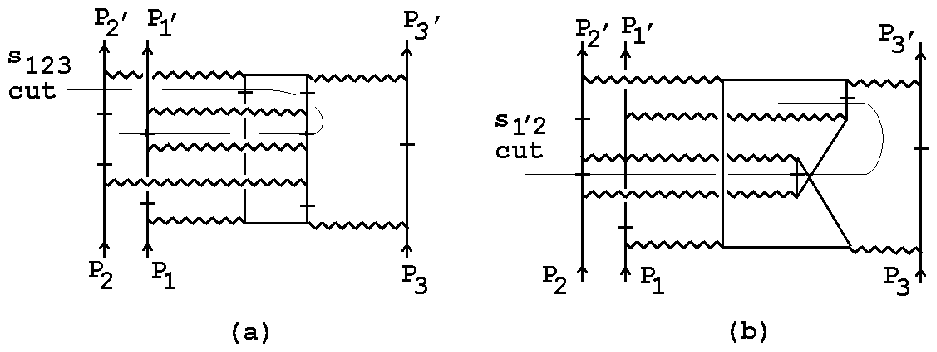}

Fig.~6.8 Further Discontinuities of Fig.~6.1(f).
\end{center}
(The hatched lines directly on the thin line again indicate 
the on-shell particles producing the new discontinuity).
This implies the diagram will continue to provide a six-reggeon interaction
containing the anomaly as reggeization effects are added. 
According to the argument of the last Section, the anomaly 
has to go into the reggeon interaction vertex associated with
the unphysical $s_{1'2}$ discontinuity. 

Moving on to the twisted diagrams of Fig.~6.1, we will find that 
two of these diagrams contain the appropriate local couplings, but can not 
satisfy all the constraints on the momentum flow. 
These two diagrams are also related to Fig.~6.1(f) via reggeon Ward 
identities.

\subhead{6.8 The Diagrams of Fig.~6.1(g) - (j).}

The diagrams of Fig.~6.1(g) and (h) also appear in Fig.~4.6, except that an 
extra line is now on-shell. In Section 4 we argued that such diagrams do not 
contain the anomaly. With the extra line on-shell we again have only two 
quark lines off-shell and so clearly there is no triangle anomaly.

\subhead{6.9 The Diagrams of Fig.~6.1(k) and (l).}
 
The diagrams of Fig.~6.1(k) and (l) both have $s_{123}$ and unphysical
$s_{1'2}$ discontinuities that can be taken through a gluon line. 
Fig.~6.1(l) is simply obtained from Fig.~6.1(k) by
time reversal of the scattering process and so has analagous properties.
Therefore our discussion below of Fig.~6.1(k) will immediately extend to
Fig.~6.1(l). 

We can repeat much of the discussion of Fig.~6.1(e) and (f) for Fig.~6.1(k).
We do this briefly as follows. For reasons that will soon become apparent
we reverse the sign of $k_2$ and obtain the internal quark loop 
contribution shown in Fig.~6.9.
\begin{center}
\leavevmode
\epsfxsize=3in
\epsffile{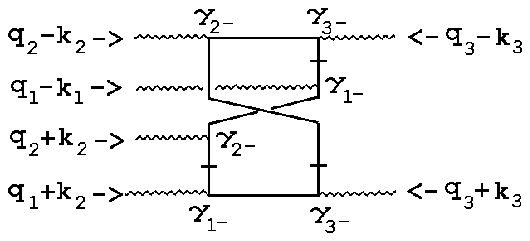}

Fig.~6.9 The Quark Loop in Fig.~6.1(k).
\end{center}
If we take the shortest routes for
each of the $k_i$ momenta then we find that, in parallel
with our discussion of Fig.~6.1(b), neither the $k_{12^-}$ nor the $k_{21^-}$ 
integrations give local couplings.
The only $\delta$-function assignment giving local couplings at all vertices 
is that shown in Fig.~6.10(a), with the corresponding momentum flow shown in 
Fig.~6.10(b).
\begin{center}
\leavevmode
\epsfxsize=3.4in
\epsffile{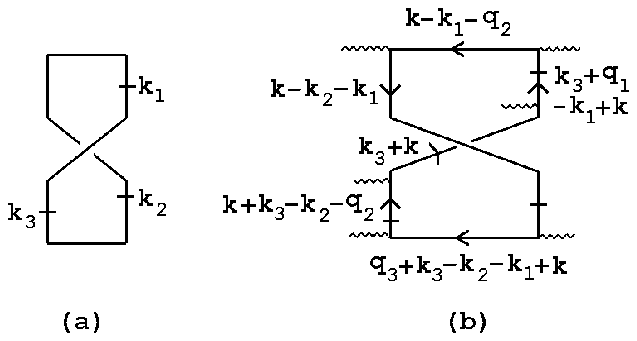}

Fig.~6.10 Another (a) $\delta$-function Assignment and (b) Momentum Flow
for Fig.~6.9.
\end{center}

The calculation of local couplings proceeds as usual.
The couplings generated differ from those of 
Fig.~4.13 only in that 
$$
\hat{\gamma}_{31}~ = ~\gamma_{3^-}\gamma_{2^-}\gamma_{1^-}
\to~
\hat{\gamma}_{13 } ~=~\gamma_{1^-}\gamma_{2^-}\gamma_{3^-} 
= ~\gamma_{-,+,-}~-~ i~
\gamma_{-,-,-} ~\gamma_5 
\auto\label{589}
$$
and so the three $\gamma_5$ couplings needed for
the anomaly are again present. The momentum flow 
and couplings in the corresponding triangle diagram are shown in
Fig.~6.11. 
\begin{center}
\leavevmode
\epsfxsize=2.2in
\epsffile{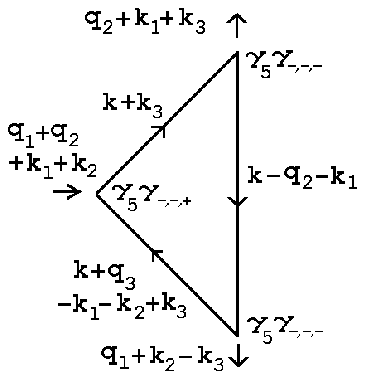}

Fig.~6.11 The Triangle Diagram Generated by Fig.~6.1(k). 

\end{center}
At this point we note that Fig.~6.11 is identical to Fig.~4.14, apart from a 
shift of the internal momentum
$$
k~ ~\to ~~ k-q_2 - k_1
\auto\label{shift}
$$
We would then expect the anomaly to appear in the limit (\ref{5801}),
with $(q_1+q_2+k_1+k_2)=0$ and $ (q_2+k_1 +k_3)$ light-like in the limiting
configuration. However,
if the shift (\ref{shift}) is made, $k_1$ is routed along a different
path
and the $\delta$-function assignment of Fig.~6.9(a) can no longer be made.
Therefore, the momentum configurations of Fig.~6.11 must be kept.
Since the anomaly has to be generated when the vertical momentum line of 
Fig.~6.10 carries stricyly zero momentum, in the limiting configuration we 
must also have 
$$
k=q_2 + k_1
\auto\label{shift0}
$$
and must combine this with
the mass-shell $\delta$-constraints 
determining $k_{12^-}, k_{21^-}$ and $k_{33^+}$ that replace
(\ref{5851})-(\ref{5853}). Imposing (\ref{shift0}), these constraints give
\beqa (k_3 + q_1 + k - k_1)^2  ~=~(k_3-q_3)^2~&=&~0 \label{5951} \\
(k + k_1 + k_2)^2~=~(q_2+k_2+2k_1)^2 ~=~(k_1-q_1)^2~&=&~0 \label{5952} \\
(k +k_3 - q_2 -  k_2)^2~=(k_3+k_1-q_2)^2~&=&~0 \label{5953}
\eeqa
From (\ref{5951}) we see immediately that the anomaly divergence associated
with Fig.~6.1(k) can only coincide with that of Fig.~6.1(f) at $q_3-k_3)^2 =0$.

\subhead{6.10 Reggeon Ward Identities}

It is not an accident that the diagram of Fig.~6.1(k) contains the same
$\gamma$-matrix structure as that of Fig.~6.1(f). In fact a reggeon Ward
identity determines that it has to contribute equally (and, when color factors
are appropriate, with opposite sign) at a zero momentum point, such as 
$(q_3-k_3)^2 =0$.
Consider the two sets of 
amplitudes forming $s_{13}$ discontinuities as in Fig.~6.12, 
\begin{center}
\leavevmode
\epsfxsize=3.7in
\epsffile{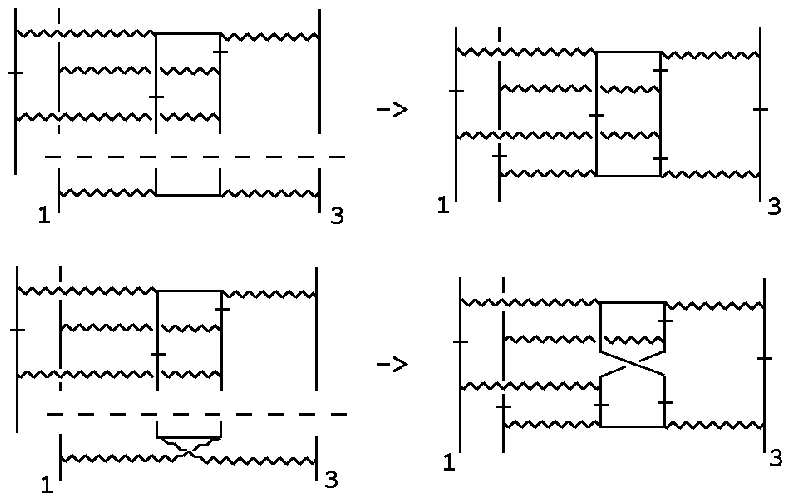}

Fig.~6.12 Forming $s_{13}$ Discontinuities
\end{center}                             
The upper set gives Fig.~6.1(f) while the lower set gives Fig.~6.1(k). The two
lower production amplitudes that distinguish the diagrams are related by the
reggeon Ward identity illustrated in Fig.~C6. Therefore, when the central
quark/antiquark pair carries zero color, the two diagrams must cancel at the
zero momentum point. Fig.~6.1(l) is
similarly related to Fig.~6.1(f) via final state amplitudes satisfying a
reggeon Ward identity. 

As we noted above, Fig.~6.1(e) is also related to Fig.~6.1(f) by a reggeon
Ward identity. In this case the unphysical $s_{1'2}$ discontinuity 
has to be considered.
The triple gluon diagram of Fig.~C6 (the third diagram) can not contribute
when the quark exchange exchange in the first two diagrams involves a zero 
momentum chirality transition, as is the case in the anomaly divergence.
Therefore, when the quark/antiquark pair (involved in the $s_{1'2}$
discontinuity) carries octet color the anomaly contributions in Fig.~6.1(e)
and (f) will not cancel and there will be no triple gluon contribution. The
reggeon Ward identity will necessarily be violated when the light-cone momenta
corresponding to the anomaly are present. That a reggeon Ward identity could
fail for a quark loop in which all lines are are on-shell was emphasized 
in \cite{arw98}.

The reggeon Ward identity is, however,
sufficient to ensure that if the anomaly in maximally non-planar diagrams 
cancels then so must the contribution of all diagrams having the form of
Fig.~6.1(e). Note that for the contribution of the maximally non-planar
diagram of Fig.~2.4 to other hexagraphs (such as that associated with the
scattering processes of Fig.~4.9), the Feynman diagram corresponding to
Fig.~6.1(k) actually plays the role of Fig.~6.1(e).
We conclude, therefore, that we can focus only on Fig.~6.1(f) in Fig.~6.1.
To discuss whether the anomaly cancels we have to consider only
the sum of the double discontinuities of the form of Fig.~6.1(f), each of 
which is associated with a separate hexagraph. In Section 4, we have already 
discussed the kinematical symmetries that will produce a cancelation. We
enlarge this with a discussion of color factors in the next Section. 

\subhead{6.11 Feynman Diagrams Versus Multiple Discontinuities}

At the end of the previous Section we explained how it is that the anomaly
can occur in physical region momentum configurations where the discontinuities
associated with the propagators that are put on-shell are no longer present.
To be clear we would like to reiterate the logic that we are employing.
As we outlined in Section 2, in principle we can study Feynman diagrams 
directly and look for propagators that are placed on-shell (or close to 
on-shell) by the triple-regge limit. In part this is what we did in Section 4.
The very large number of diagrams, as well as their complexity, makes it 
essentially impossible to apply this procedure to all diagrams. 
We have instead proceeded by using the multi-regge 
theory of the previous Section which tells us that the anomaly could appear in 
specific multi-regge amplitudes which have the discontinuities that 
we have calculated directly. We then extract the reggeon vertices calculated 
from the discontinuities and insert them back into the multi-regge formulae.
In this way we obtain 
amplitudes that describe triple-regge scattering away from the discontinuities.
If the anomaly divergence then occurs in a physical region within the
multi-regge formula, for consistency it should occur in a corresponding way in 
some Feynman diagram. This is what we demonstrate when we show space-time
scattering diagrams such as those of Fig.~4.15 and Fig.~6.7(b). 
We have emphasized, however, that the discontinuity structure given by the
multi-regge amplitudes is not the same as is found in individual diagrams.
Consequently, a multiple discontinuity and real anomaly configuration that 
appear in the same Feynman graph do not generally appear in the same
multi-regge amplitude. 

\newpage

\mainhead{7. Color Factors, Cancelations and Divergences}

We have narrowed down a discussion of the cancelation of the anomaly,
at lowest-order,
to contributions from double (or triple) discontinuities occurring in Feynman 
diagrams of the maximally non-planar type. In Section 4
we already 
discussed the kinematical symmetries that can produce a cancelation. As a
result we need give only a minimal discussion of 
the role played by color quantum numbers and signature properties.

When signatured amplitudes are formed the two-reggeon state appears only
in even signature channels. The reggeon interactions
containing the anomaly that we have discussed couple two reggeized gluons in
each $t_i$-channel and 
so all three channels have $\tau_i=1$. Therefore the signature 
rule of Section 5 is immediately satisfied. 
To obtain an 
amplitude for which all signatures are positive we add the 
contributions from all eight of the hexagraphs in Fig.~5.3. This requires 
that we add the contributions of the twisted diagrams of the form of 
Fig.~4.16 to the untwisted contributions of Figs.~4.8(a) and (b).

To begin our discussion of color factors we first consider
the external coupling of two gluons (or reggeons)
to a scattering quark. The color factor that appears can be written 
as shown diagrammatically
in Fig.~7.1.
\begin{center}
\leavevmode
\epsfxsize=4.5in
\epsffile{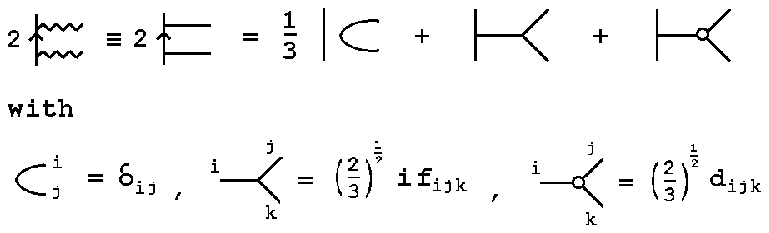}

Fig.~7.1 Color Factors for Two Gluons Coupling to Two Quarks
\end{center}
$f_{ijk}$ and $d_{ijk}$ are the usual antisymmetric and symmetric 
tensors for SU(3) color. In lowest order, the $G_h$ have no momentum 
dependence and so, in the even signature amplitude, only
the symmetric $\delta_{ij}$ and
$d_{ijk}$ couplings survive. Therefore, 
the (two-gluon) two-reggeon state has to be in either 
a color zero state,  or a ``symmetric octet'' ($8_s$) state. At 
this order it is obvious that a single scattering quark does not couple to
an ``anti-symmetric octet'' ($8_a$) two-reggeon state.

It will be important to discuss the color parity of 
reggeon states. Color charge conjugation on gluon fields is 
defined by the transformation of gluon color matrices 
$$
A^i_{ab}~\to ~~ -~A^i_{ba}
\auto\label{cpm}
$$
For SU(3) we can choose $A^i \sim \lambda^i$ so that 
$$
A^i ~\to ~ - A^i ~~~ i=1,3,4,6,8~, ~~~~~~~ A^i ~\to ~  A^i ~~~ i=2,5,7~,
\auto\label{cpm1}
$$  
For a trace of gluon matrices the color charge 
conjugation reverses the trace order. In particular, in a space-time
path-ordered integral of gluon fields 
it reverses the direction of the path integration. For gauge-invariant 
states involving such integrals there may be an inter-relation between 
color parity and space-time symmetry properties.
 
We consider the minus sign in (\ref{cpm}) as defining the negative color
parity of the gluon. The odd-signature reggeized gluon then has a color
parity equal to it's signature. 
Color-zero combinations of color matrices also 
have a definite color parity, e.g.
$$
\eqalign{\delta_{ij}~A^iA^j ~&\to~ \delta_{ij}~A^iA^j~, ~~~~
f_{ijk}~A^iA^jA^k ~\to ~
f_{ijk}~A^iA^jA^k ~,\cr
&d_{ijk}~A^iA^jA^k ~\to ~-~ d_{ijk}~A^iA^jA^k }
\auto\label{cpm2}
$$
i.e. the $d$-tensor provides a ``color parity violating'' coupling
for gluon fields. Ultimately our main interest is in color zero multi-reggeon 
states and these can immediately be assigned a color parity. Also since 
$$
\eqalign{f_{ijk}~A^jA^k~/~ A^i ~&\to ~f_{ijk}~A^jA^k~/~ A^i ~,\cr
d_{ijk}~A^jA^k~/~ A^i ~&\to ~- ~d_{ijk}~A^jA^k~/~ A^i }
\auto\label{cpm3}
$$
we can assign negative and positive color parities, respectively,
to the $8_a$ and $8_s$ states discussed above. 
We can also assign color 
parities to multi-reggeon states with color factors containing combinations
of $f$- and $d$- tensors. Any reggeon state, and in particular 
an even-signature $8_a$ two-reggeon state, has ``anomalous color
parity'' if it has a color parity not equal to it's signature. We will argue
below that, in general, anomalous color parity reggeon states do not couple 
to a scattering quark.  

Color charge conjugation invariance implies color charge parity conservation
and so, after summing over quark directions, the quark loop color factor
must contain an even number of $d$-tensors. 
Given the color structure of the external couplings, the possible color
couplings for the $\Gamma_6$ reggeon interaction extracted from the 
lowest-order diagrams are those shown in Fig.~7.2
\begin{center}
\leavevmode
\epsfxsize=3.2in
\epsffile{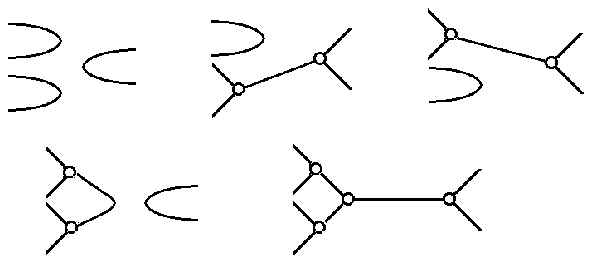}

Fig.~7.2 Color Factors for the Lowest-Order Reggeon Interaction
\end{center}
In lowest-order, therefore, both the color factors and the remaining
$k_i$-integrations are symmetric with respect to the two reggeons in each of
the $t_i$ channels. From Section 4 we know that this
implies the anomaly is canceled. 

In higher-orders, helicity conserving 
couplings $G_h(q_i,k_i)$ that appear within multiple discontinuities, need
not be symmetric under $k_i \leftrightarrow -k_i$. Therefore, 
the $8_a$ two-reggeon state could appear in such 
discontinuities. However, as explained in Section 5, a positive 
signatured amplitude can be obtained either by adding hexagraph amplitudes
or by adding full 
amplitudes related by a $CPT$ transformation applied selectively to 
external states.
For an external (left-handed) scattering quark $q_L$,
the second procedure gives directly that the full signatured 
coupling is as shown in Fig.~7.3.
Because of helicity conservation, the two vertices in Fig.~7.3 are also 
related by a $CP$ transformation. Therefore, since $CP$ is conserved, 
their equality in lowest-order
must extend to all orders. Consequently, the two-reggeon coupling remains
symmetric to all orders and the $8_a$ state does not couple. 
\begin{center}
\leavevmode
\epsfxsize=3in
\epsffile{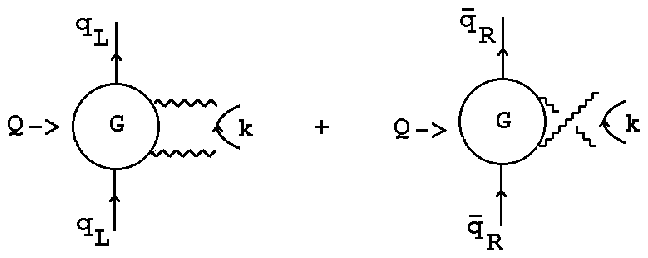}

Fig.~7.3 The Signatured Two-Reggeon Coupling to a Quark
\end{center}

More generally, even signature implies that the external ``state'' 
formed by the initial and final scattering particles is even under
$CPT$. Therefore, the internal two-reggeon state must similarly be even.
Since the reggeon state lies entirely in the transverse plane, it is 
independent of the $T$ transformation. Therefore, it must be even 
under $CP$. (The same conclusion could be reached by working in the 
$t$-channel.) The antisymmetry in the $k_i$ integrations required for the 
anomaly is equivalent to requiring $P= -1$ for the two-reggeon state, which
must, therefore carry anomalous color parity, i.e. $C= -1$. 

A-priori, the necessary parity antisymmetry for the two-reggeon state 
could appear if there is helicity 
non-conservation. 
If we consider scattering gluons then helicity-flip vertices coupling
a reggeized gluon do appear in next-to-leading order\cite{fl1}. However, 
parity conservation, applied when the reggeized gluon goes on-shell, 
implies there is a change of sign when the gluon helicity is reversed. This
determines that the ``anomalous color parity'' 
$8_a$ two-reggeon state again decouples in all orders. More generally, we
anticipate that no reggeon states with anomalous color parity 
couple to scattering quarks (or gluons).  
By appealing to 
instanton interactions we could introduce hypothetical external
couplings that are helicity non-conserving and that violate $CP$ conservation.
However, our belief is that the anomaly will ultimately force a choice of 
scattering states in order to satisfy unitarity. Therefore, we wish to first
determine whether there is a level at which the anomaly does cause a problem if
we use the quark and gluon states of perturbation theory.

The parity asymmetry needed to couple the anomalous color parity 
two-reggeon state can be obtained if we add an extra particle (or particles)
to the initial or final state as in Fig.~7.4. 
\begin{center}
\leavevmode
\epsfxsize=0.8in
\epsffile{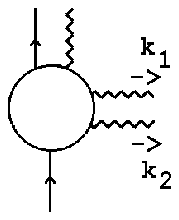}

Fig.~7.4 A Two-Reggeon Coupling with an Additional Final State Gluon.
\end{center}
This coupling can be directly studied in the two-to-four amplitude\cite{sw}
where the novel signature properties produced by an imbalance between
discontinuities is well-known. Nevertheless, even if all three external
couplings have the required asymmetry between initial and final states, a
triple-regge amplitude containing the anomaly still can not exist, because
of the conservation of color parity. An equivalent way of stating this is to 
say that for the anomaly to appear in the coupling of three $8_a$ 
two-reggeon states, a $d$-tensor coupling is required that violates
color parity conservation. Unless the external couplings for two-reggeon 
states violate color parity conservation (or, equivalently, an analogue of the
anomaly appears in the external couplings) overall color parity conservation
will force the cancelation of the triple-regge anomaly amplitude. 

We can outline how we anticipate the anomaly does appear in amplitudes as 
follows, although more explicit verification is clearly required.
In a ditriple regge limit reggeon diagrams, of the form illustrated in
Fig.~7.5, containing two anomaly vertices can appear. A single $d$-coupling 
can be present for each anomaly vertex while the full
amplitude conserves color parity. 
\begin{center}
\leavevmode
\epsfxsize=2in
\epsffile{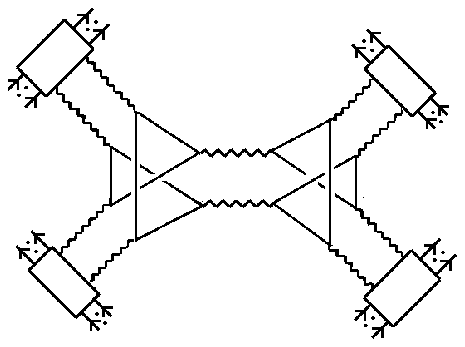}

Fig.~7.5 A Ditriple-Regge Limit Amplitude. 
\end{center}
It is then important to note that the external coupling will have reggeon 
Ward identity zeroes\cite{arw98} (which follow from gauge invariance). 
For example, the coupling of Fig.~7.4 has a zero when either $k_1$ or $k_2
\to 0$. The anomaly divergence occurs at just such points. If the
corresponding zeroes are present in all four of the external couplings
of Fig.~7.5, the linear divergence of the anomaly will always be compensated
by at least two Ward identity (linear) zeroes and this will be sufficient to
prevent an infra-red divergence of the full amplitude. (The logarithmic 
divergences due to zero mass gluon propagators do not 
affect this argument.)

We anticipate that all reggeon states coupling to anomaly 
vertices will have anomalous color parity to compensate for the 
antisymmetric parity properties of the anomaly. 
For example, reggeon 
interactions containing the anomaly will appear in any amplitude of the form of 
Fig.~7.6
\begin{center}
\leavevmode
\epsfxsize=2in
\epsffile{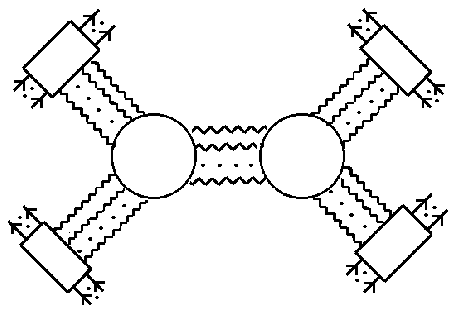}

Fig.~7.6 A General Ditriple-Regge Reggeon Amplitude 
\end{center}
provided that there is sufficient imbalance between the initial and final 
states that anomalous color parity reggeon states appear and provided 
that the signature conservation rule is satisfied. 
Reggeon Ward identity zeroes will continue to
prevent the occurrence of divergences in full amplitudes. 

In general multi-regge limits, reggeon diagrams 
containing any number of pairs of 
anomalous vertices will similarly appear\cite{arw98}. 
Even though infra-red divergences will not appear, the
ulltraviolet presence of the anomaly (that must accompany it's infra-red 
appearance) most likely still causes problems. We expect the 
large momentum region of the triangle graph to give behavior
of the form 
of (\ref{05847}) but with $l \sim p_1,p_2,p_3 ~\to \infty$ 
When embedded in diagrams such as that of Fig.~7.6 we expect this behavior
to produce a powerlike enhancement of the asymptotic behavior
that ultimately conflicts
with unitarity.  In \cite{arw98} we proposed to avoid this conflict 
by introducing large (but finite) mass fermion Pauli-Villars
regulators at finite (but small) physical
quark mass.  If the ``physical'' reggeon S-Matrix is obtained, as we
anticipate, by taking the quark mass to zero and extracting infra-red
divergent contributions from anomaly 
amplitudes, the regulator fermions will not appear. To produce 
infra-red divergent amplitudes, however, we have to introduce 
external reggeon couplings that produce a reggeon condensate. This is 
essentially equivalent to introducing the anomaly directly in external 
couplings. This is the program, mentioned in the Introduction, 
that is outlined at length in 
\cite{arw98} and that we plan to return to in succeeding papers.

Essentially the correct 
phenomenon is outlined in \cite{arw98}. However, there are some differences.
In particular, because of the signature conservation for anomalous 
amplitudes, there is no triple anomalous odderon vertex, as we assumed.
Instead, the anomaly divergence occurs within 
the primary momentum carrying interactions of a reggeon diagram and not just 
in accompanying vertices as was suggested in \cite{arw98}. This is possible 
because, as we now understand, the anomaly divergence occurs when only some 
of the interacting reggeized gluons carry zero transverse momentum. As we noted 
above, a very important consequence of the signature rule is that it
promises to explain the even signature of the pomeron - a property that
previously we had not clearly seen the origin of. We will not describe the
infra-red divergence phenomenon any further in this paper, but simply
 note that in Fig.~2.8 we have already portrayed the general phenomenon that
we expect to occur.

\vspace{0.5in}

\noindent {\Large \bf Acknowledgements}

I am grateful to both Jochen Bartels and Lev Lipatov
for very useful discussions on the contents of this paper.

\newpage

\renewcommand{\theequation}{A.\arabic{equation}}
\setcounter{equation}{0}
\vskip 1cm \noindent
\noindent {\large\bf Appendix A. ~ The Infra-Red Triangle Anomaly and}
\newline \centerline{\large\bf Chirality Violation}
\vskip 3mm \noindent

It has been known\cite{dz} for a long time 
that the triangle anomaly is not only an ultra-violet phenomena but is also
manifest in the infra-red region when the 
fermions involved are massless. This was elaborated in detail by Coleman and
Grossman\cite{cg} in the context of establishing 't Hooft's anomaly matching
condition for confining theories. Closely related 
results were also obtained in \cite{bfsy}.

In the body of the paper we use the infra-red properties of the anomaly
to establish it's presence in particular reggeized gluon interactions.
The Coleman and Grossman analysis establishes that the vertex function for
three axial vector currents has a singularity when the quark fields involved
are massless. As we describe below, the maximal divergence is obtained  when
all the spacelike momenta flowing through the vertex are scaled uniformly to
zero while a finite light-like momentum remains. The presence of the
light-like momentum is a crucial ingredient. 

In the notation of Fig.~A1, $J^a$ is the axial current and 
\begin{center}
\leavevmode
\epsfxsize=2in
\epsffile{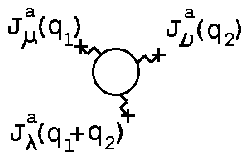}

Fig.~A1 The Three-point Function 
\end{center}
the three current vertex function $\Gamma^{\mu\nu\lambda}$ 
can be decomposed in terms of invariant amplitudes as follows
$$
\Gamma^{\mu\nu\lambda}({\hbox{\q}}_1,{\hbox{\q}}_2)~
= ~A~ \epsilon^{\nu\lambda\alpha\beta} 
 {\hbox{\q}}_{1\alpha}{\hbox{\q}}_{2\beta} {\hbox{\q}}_1^{\mu}
~+~ \cdots ~+~B~ \epsilon^{\mu\nu\lambda\alpha} {\hbox{\q}}_{1\alpha} 
 ~+ ~ \cdots ~  
\auto\label{a1}
$$
The omitted terms are obtained from those shown explicitly by appropriate
permutations. The crucial result from \cite{cg} is 
that the anomaly equation 
$$
{\hbox{\q}}_{1\mu}~\Gamma^{\mu\nu\lambda}({\hbox{\q}}_1,{\hbox{\q}}_2)~= 
~\tilde{A}~ \epsilon^{\nu\lambda\alpha\beta} 
 {\hbox{\q}}_{1\alpha}{\hbox{\q}}_{2\beta}
\auto\label{a01}
$$
implies that, when ${\hbox{\q}}_1^2 ~\sim {\hbox{\q}}_2^2 ~\sim 
({\hbox{\q}}_1 +
{\hbox{\q}}_2)^2 ~ \sim {\hbox{\q}}^2 ~\to 0$ the invariant amplitude $A$ 
has a pole at ${\hbox{\q}}_1^2 = 0$ with the
coefficient $\tilde{A}$ given by the anomaly. Therefore, as 
${\hbox{\q}}^2 \to 0$ we
have 
$$
\Gamma^{\mu\nu\lambda}({\hbox{\q}}_1,{\hbox{\q}}_2)~= ~ \tilde{A} ~ 
\epsilon^{\nu\lambda\alpha\beta} ~ {
{\hbox{\q}}_{1\alpha}{\hbox{\q}}_{2\beta}{\hbox{\q}}_{1\mu} \over 
{\hbox{\q}}_1^2 }
~+~ \cdots   
\auto\label{a2}
$$
The ultra-violet anomaly appears also in the 
vertex function for one axial current and two vector currents and in 
Ref.~\cite{bfsy} it is shown how the corresponding Ward identities similarly
imply the presence of the divergence (\ref{a2}) when the quarks involved are 
massless. We also refer to this result in our discussion of reggeon
vertices. 

If the chiral symmetry associated with the axial current $J^a_{\mu}$ is 
spontaneously-broken by a quark condensate, the pole at 
${\hbox{\q}}_1^2 = 0$ is 
associated with the corresponding Goldstone boson. In our case,
the anomaly equation (\ref{a01}) for the $U(1)$ current 
is invalidated by non-perturbative,
non-trivial topological, gluon field 
configurations - instantons in particular\footnote{In 't Hooft's 
solution\cite{gth2}
of the U(1) problem, instantons produce a quark interaction (an $\eta'$ mass
term) that moves the ``perturbative'' $\eta'$ pole away from 
${\hbox{\q}}_1^2 = 0$. In our regge limit analysis, it is not clear how
such an interaction could contribute.}. However, our
initial purpose is to first discover a ``perturbative'' contribution of the
anomaly within reggeon diagrams and only later determine it's dynamical 
significance. In this case we can use a divergence of the form (\ref{a2}) 
as a signal of the anomaly. 

If we simply take all components of ${\hbox{\q}}_1$ and 
${\hbox{\q}}_2$ to scale with ${\hbox{\q}}$ then 
(\ref{a2}) gives the (dimensional) result
$$
\Gamma^{\mu\nu\lambda} ~
\centerunder{$\sim$}{\raisebox{-4mm}{${\hbox{\q}} \to 0$}}
~~{\hbox{\q}} 
\auto\label{a20}
$$
We obtain more singular behavior as follows. First, 
choose $  \mu$  to be a light-cone index ``$+$'' and choose 
$$
{\hbox{\q}}_1^{\mu} ~=~{\hbox{\q}}_1^+~
=~{\hbox{\q}}_{1-}~=~\hbox{\p} ~\st{\to}~0~,~~~{\hbox{\q}}_1^-~
=~{\hbox{\q}}_{1+}~=~0
\auto\label{200}
$$
Choosing ${\hbox{\q}}_2$, and all spacelike momenta flowing 
through the diagram, to be 
$O({\hbox{\q}})$ and to lie in a spacelike 
plane orthogonal to the space component of ${\hbox{\q}}_1^+$,
we obtain from (\ref{a2}) 
$$
\Gamma^{++\lambda}({\hbox{\q}}_1,{\hbox{\q}}_2)~~
\centerunder{$\sim$}{\raisebox{-4mm}{${\hbox{\q}}^2 \to 0$}} 
~~ \tilde{A} ~ \epsilon^{+ \lambda - \beta}~ 
{{\hbox{\q}}_{2\beta}~ {\hbox{\q}}_{1-} 
{\hbox{\q}}_{1}^+ \over {\hbox{\q}}^2}  ~
\sim~\tilde{A} ~ {{\hbox{\q}}_{2}^{\beta}~ {{\hbox{\p}}^2 \over 
{\hbox{\q}}^2}  
~\sim ~ \tilde{A} ~ {{\hbox{\p}}^2 \over {\hbox{\q}}}  }~~~
\auto\label{a3}
$$
Note that if we leave the spacelike momenta unchanged but instead choose
$$
{\hbox{\q}}_1^{\mu} ~=~{\hbox{\q}}_1^-~
=~{\hbox{\q}}_{1+}~= ~ \hbox{\p} ~\st{\to}~0~,~~~{\hbox{\q}}_1^+~
=~{\hbox{\q}}_{1-}~=~0
\auto\label{201}
$$
then we obtain
$$
\Gamma^{--\lambda}({\hbox{\q}}_1,{\hbox{\q}}_2)~~
\centerunder{$\sim$}{\raisebox{-4mm}{${\hbox{\q}}^2 \to 0$}} 
~~ \tilde{A} ~ \epsilon^{- \lambda + \beta}~ 
{{\hbox{\q}}_{2\beta}~ {\hbox{\q}}_{1+} 
{\hbox{\q}}_{1}^- \over {\hbox{\q}}^2}  ~
~\sim ~ - \tilde{A} ~ { \hbox{\p}^2 \over {\hbox{\q}}}  
\auto\label{a30}
$$
The change of sign compared to (\ref{a3}) has very important consequences 
for our discussion of the cancellation of the anomaly in Sections 4 and 7. 
Clearly we could equally well have changed the sign of ${\hbox{\q}}_{2\beta}$
while keeping the same light-cone space component. In either case there is
a form of parity transformation involved and the antisymmetry of the
anomaly is a direct consequence of the chirality violation discussed below.
Note that the structure of the anomaly divergence involves each of the four
dimensions of Minkowski space in distinct roles. This is, in part, why a
triple-Regge limit which fully utilises all four dimensions is necessary to
see the anomaly appear.

The infra-red behavior (\ref{a3}) arises directly from a combination of normal
thresholds and the Landau triangle singularity (or anomalous threshold) 
in the quark triangle diagram shown in Fig.~A2, i.e.
$$
\Gamma^{\mu \nu \lambda}({\hbox{\q}}_1,{\hbox{\q}}_2) 
= i\int {  d^4 k~ Tr \{ \gamma_5
\gamma^{\mu} ~\st{k}  )~ \gamma_5 \gamma^{\nu}~ (\st{{\hbox{\q}}_2} 
+ \st{k})~ 
\gamma_5 \gamma^{\lambda}~ (-\st{{\hbox{\q}}}_1 + \st{k} ) \} 
\over  k^2  ({\hbox{\q}}_2 + k)^2 
 (k - {\hbox{\q}}_1)^2 }
\auto\label{a4}
$$
The triangle diagram singularity can be thought of as due to a 
space-time scattering as indicated by the arrows in Fig.~A2. 
${\hbox{\q}}_2$ is a 
spacelike momentum transfered by the $J^{a\nu}$ current 
and ${\hbox{\q}}_1$ has the 
light-like component necessary to produce an initial pair of massless 
particles. 
\begin{center}
\leavevmode
\epsfxsize=2in
\epsffile{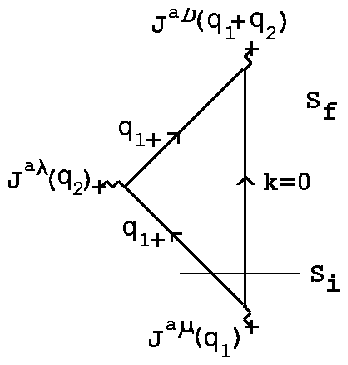}

Fig.~A2 The Triangle Diagram
\end{center}
Therefore, the vertices where the lightlike momenta enters and leaves are 
respectively associated with the production and annihilation of a pair of 
massless fermions. 

When $q_1^2 \sim (q_1+q_2)^2 \to 0$ all momenta become parallel and the
thresholds at $q_1^2 = 0$ and $ (q_1+q_2)^2 = 0$ can enhance the
triangle singularity. However, when the helicities of the fermions are 
determined\cite{cg} it is found that the situation is symmetric in that in 
both intermediate states (the produced and annihilated states) there is a net 
fermion chirality, i.e. a fermion/antifermion state with the same sign
center-of-mass helicities (opposite sign spin components). 
The axial-vector coupling implies that the two possible alignments for the 
helicities involved give contributions that add rather than cancel, as they
would do for a vector coupling (i.e. for a vector coupling intermediate
states with non-zero chirality are not present). Since the spacelike current
$J^{a \lambda}$ flips the helicity of the fermion that it scatters, the
unscattered fermion must also flip it's helicity. This is only possible if this
fermion carries strictly zero momentum so that it's helicity is undefined (as is
indeed the case\cite{cg}). The finite light-like momentum is carried by the
scattered fermion. That the unscattered fermion carries zero momentum implies 
that both propagator poles are involved in producing (\ref{a3}), thus allowing 
the chirality transition. 

In effect the coincindence of both propagator poles for a zero momentum 
fermion and the resulting chirality transition is the essence of the infra-red 
occurrence of the anomaly. It is the chirality transition that produces the
pseudotensorial asymmetry with respect to light-cone components discussed
above. It is also the ``chirality violation'' that we refer to often in the
main text. Clearly the alignment of helicities producing this violation has to
be an asymptotic effect of the multi-regge limit, which is not present at
finite momentum. The antisymmetry in going from (\ref{a3}) to (\ref{a30}) is
the feature that we expect to lead to cancellation of the anomaly, unless
there is some background asymmetry accompanying the reggeon interaction.

Coleman and Grossman also argued for the infra-red equivalent of the 
``non-renormalization'' theorem that holds for the ultra-violet 
manifestation of the anomaly. They argued that Feynman diagrams with a
Landau singularity and helicity structure other than that of the
triangle diagram with a chirality transition, can not reproduce the behaviour
(\ref{a3}). In our case the reggeon vertices we obtain will not contain the
full Lorentz tensor amplitude (\ref{a4}) but rather will contain only
particular light-cone related momenta and $\gamma$-matrix components. We
will show, however, that we do have all the necessary components to produce
the infra-red divergence (\ref{a3}). The argument of Coleman and Grossman 
then determines that the infra-red divergence we find can not be canceled 
by the contribution of other diagrams to the reggeon vertices we discuss.

It will also be important for our analysis to discuss the momenta
$k$ involved in generating the pole at ${\hbox{\q}}^2 = 0$ in (\ref{a4}). The 
numerator in (\ref{a4}) gives directly the numerator in (\ref{a3}) and so we 
can write
$$
\Gamma^{++\lambda}({\hbox{\q}}_1,{\hbox{\q}}_2)~~
\centerunder{$\sim$}{\raisebox{-4mm}{${\hbox{\q}}^2 \to 0$}} 
~~ \tilde{A} ~ \epsilon^{+\lambda-\beta}~ 
{\hbox{\q}}_{2 \beta}~ {{\hbox{\q}}_+}^2 \int {  d^4 k~ 
\over  k^4 (k^2 - {\hbox{\q}}_+k_-) }
\auto\label{a5}
$$
Superficially this integral depends on ${\hbox{\q}}_+$ 
and so might be expected to be
$O(1/{\hbox{\q}}{\hbox{\q}}_+)$. However, it is straightforward 
to make the scaling
$$
k_+~\to ~ \Lambda~k_+~,~~~~ k_-~\to~ {\Lambda}^{-1}~k_-
\autolabel{a6}
$$
so that 
$$
\int {  d^4 k~ 
\over  k^4 (k^2 - {\hbox{\q}}_+k_-)}~=~\int {  dk_+dk_-d^2 k_{\perp} 
\over  k^4 (k^2 - {\hbox{\q}}_+k_-)}~\to~ \int {  d^4 k~ 
\over  k^4 (k^2 - {\hbox{\q}}_+k_- /\Lambda)}
\auto\label{a7}
$$
showing that the integral is independent of ${\hbox{\q}}_+$. (In the limit 
$\Lambda \to \infty$ the ${\hbox{\q}}_+$ dependence can be 
scaled out of the integral altogether, the only trace being the location of 
the integration contour.) Therefore we can write
$$
\int {  d^4 k~ 
\over  k^4 (k^2 - {\hbox{\q}}_+k_- )} ~~\sim ~~ \int {  d^4 k~ 
\over  k^6 } ~~\sim~~ {1 \over {\hbox{\q}}^2}
\auto\label{a8}
$$
and take all components of $k$ to be $O({\hbox{\q}})$.
 
\renewcommand{\theequation}{B.\arabic{equation}}
\setcounter{equation}{0}
\vskip 1cm \noindent
\noindent {\large\bf Appendix B. ~ Light Cone Kinematics }
\vskip 3mm \noindent

Regge limits are conventionally related to light-cone momenta by writing a
general 4-momentum $p^{\mu} = (p_0,p_1,p_2,p_3) $ in the form
$$
p^{\mu} ~= ~\hbox{${1 \over 2 }$}
~p_{1^+}~ \underline{n}_{1^+} ~+~ 
\hbox{${1 \over 2 }$}~p_{1^-} ~ \underline{n}_{1^-}  ~+ ~ 
\underline{p}_{1\perp}
\auto
\label{lcd1}
$$
where $\underline{n}_{1^+} = (1,1,0,0)$ 
and $\underline{n}_{1^-} = (1,-1,0,0)$ are 
and $\underline{p}_{1\perp}$ is a two-dimensional 
``transverse momentum'' orthogonal 
to both $\underline{n}_{1^+}$ and $\underline{n}_{1^-}$. It is simple
to determine that
$$
p_{1^+}~ = ~p_0 +p_1~, ~~p_{1^-}~=~ p_0 - p_1~, ~~ 
\underline{p}_{1\perp}~=~ (p_2, p_3)
\auto
\label{lcd2}
$$
We regard 
$\underline{n}_{1^+}$ as euclidean vectors and form Minkoski space products 
by introducing 
$$
p_{\mu} ~= ~\hbox{${1 \over 2 }$}
~p_{1^+}~ \underline{n}_{1^-} ~+~ 
\hbox{${1 \over 2 }$}~p_{1^+} ~ \underline{n}_{1^-}  ~- ~ 
\underline{p}_{1\perp}
\auto
\label{lcd11}
$$
The euclidean product $p^{\mu} p_{\mu}$ then, as usual, gives the Minkowski 
product.
Clearly we can similarly define $p_{2^+},p_{2^-}, \underline{p}_{2\perp}$ and 
$p_{3^+},p_{3^-}, \underline{p}_{3\perp}$ by, respectively, projecting on 
vectors $\underline{n}_{2^+} = (1,0,1,0)$ 
and $\underline{n}_{2^-} = (1,0,-1,,0)$ or vectors 
$\underline{n}_{3^+} = (1,0,0,1)$ 
and $\underline{n}_{3^-} = (1,0,0,-1,)$.

In this paper we make use of alternative, but formally 
parallel, decompositions of the form
$$
p^{\mu}  ~= ~p_{2^-}~ \underline{n}_{1^+} ~+~ p_{1^-} ~ 
\underline{n}_{2^+}  ~+ ~ 
\underline{p}_{12+}
\auto
\label{lcd3}
$$
where $\underline{p}_{12+}$ is 
now a two-dimensional vector orthogonal 
to both $\underline{n}_{1^+}$ and $\underline{n}_{2^+}$. This determines that
$$
\underline{p}_{12+}=~ 
p_{12-}~\underline{n}_{12+} 
+ p_3~ \underline{n}_3 ~,~ ~~~~~p_{12-} = 
p_1 + p_2 - p_0
\auto
\label{lcd4}
$$
where $\underline{n}_{12+} = (1,1,1,0)$ and $\underline{n}_{3} = (0,0,0,1)$
are again euclidean vectors. 
We can also write
$$
p_{\mu}  ~= ~p_{2^-}~ \underline{n}_{1^-} ~+~ p_{1^-} ~ 
\underline{n}_{2^-}  ~+ ~ 
p_{12-}\underline{n}_{12-} - p_3 \underline{n}_3
\auto
\label{lcd5}
$$
where 
$\underline{n}_{1^-} = (1,-1,0,0)$, $\underline{n}_{2^-} = (1,0,-1,0)$.
and $\underline{n}_{12-} = (1,-1,-1,0)$. $p^{\mu} p_{\mu}$ is, of course,  
again the Minkowski product and 
if $\underline{q}$ is a second four-momentum
$$
p\cdot q~=~p^{\mu}q_{\mu}~ = ~p_{1^-}q_{2^-} ~+~ 
p_{2^-}q_{1^-}
 ~-~ p_{12-} q_{12-} ~-~ p_3 q_3
\auto
\label{lcd6}
$$

The analagous decomposition to (\ref{lcd3}) for 
$\gamma$-matrices is 
$$
\eqalign{\gamma^{\mu}  ~&= ~\gamma_{2^-}~ \underline{n}_{1^+} ~+~ 
\gamma_{1^-} ~ \underline{n}_{2^+}  ~+ ~ 
\underline{\gamma}_{12+} \cr 
&= ~\gamma_{2^-}~ \underline{n}_{1^+} ~+~ 
\gamma_{1^-} ~ \underline{n}_{2^+}  ~+ ~ 
\gamma_{12-}~\underline{n}_{12-}~+ \gamma_3~\underline{n}_3 }
\auto
\label{lcd7}
$$
where 
$$
\eqalign{  \gamma_{1^-} &= \gamma_0 -\gamma_1~, 
~~\gamma_{2^-}= \gamma_0 - \gamma_2~, \cr
\gamma_{12-} &=\gamma_1 + \gamma_2 - \gamma_0 }
\auto
\label{lcd8}
$$
Similarly
$$
\gamma_{\mu}  ~= ~\gamma_{2^-}~ \underline{n}_{1^-} ~+~ 
\gamma_{1^-} ~ \underline{n}_{2^-}  ~+ ~ 
\gamma_{12-} \underline{n}_{12-} - \gamma_3 \underline{n}_3 
\auto\label{lcd80}
$$
The $\gamma$-matices introduced in this way then satisfy
$$
\eqalign{&\gamma_{1^-}^2 ~= ~\gamma_{2^-}^2 ~=~ 0 ~, 
~~~~~~\gamma_{12-}^2 ~= ~\gamma_3^2 ~=~ -1~,\cr 
&\gamma_{1^-}\gamma_{2^-} ~+~ \gamma_{2^-}\gamma_{1^-}~=~ 2 ~, ~~~~~
 \gamma_{3}\gamma_{1^-} ~+~ \gamma_{1^-}\gamma_{3} ~= ~0~, \cr
& \gamma_{3}\gamma_{2^-} ~+~ \gamma_{2^-}\gamma_{3} ~=~ 
\gamma_{12-}\gamma_{1^-} ~+~ \gamma_{1^-}\gamma_{12-}~= ~0~, \cr 
&\gamma_{12-}\gamma_{2^-}~ +~ \gamma_{2^-}\gamma_{12-} ~= ~ 
\gamma_{12-}\gamma_{3} ~+ ~ \gamma_{3}\gamma_{12-} ~= ~ 0 ~. }
\auto
\label{lcd9}
$$
Clearly all the usual algebraic properties of both four-momenta and
$\gamma$-matrices in terms of conventional light-cone coordinates are the
same in the ``new light-cone coordinates''. 

For our discussion of the anomaly it is useful to note 
that the $\epsilon$-tensor can also 
be expressed in the new co-ordinates, i.e. we can write
$$
\epsilon^{\mu\nu\gamma\delta} P_{\mu}Q_{\nu}R_{\gamma}S_{\delta}
~=~p_{2^-}q_{1^-}r_{12-}s_3 ~- ~p_{1^-}~q_{2^-}~r_{12-}~s_3 ~ 
+ ~~\cdots
\label{lcd10}
$$
where there is a term corresponding to each permutation of 
($2^-,1^-,12-,3$), with the sign determined by the usual antisymmetry 
property of the $\epsilon$-tensor. 

Finally we note that we can use any two (non-parallel) 
light-cone momenta and introduce appropriate ``light-cone co-ordinates''.
In particular we can obviously choose $n_{1^+}$ and $ n_{3^+}$, or $n_{2^+}$ and
$ n_{3^+}$, instead of $n_{1^+}$ and $ n_{2^+}$ , and trivially repeat all 
of the above discussion.

\newpage

\renewcommand{\theequation}{C.\arabic{equation}}
\setcounter{equation}{0}
\vskip 1cm \noindent
\noindent {\large\bf Appendix C. ~Regge Limit Calculations }
\vskip 3mm \noindent

In this Appendix we discuss some simple Regge limit 
calculations using the light-cone variables introduced in the previous 
Appendix. We consider first two quarks scattering via single gluon exchange 
as illustrated in Fig.~C1.
\begin{center}
\leavevmode
\epsfxsize=2in
\epsffile{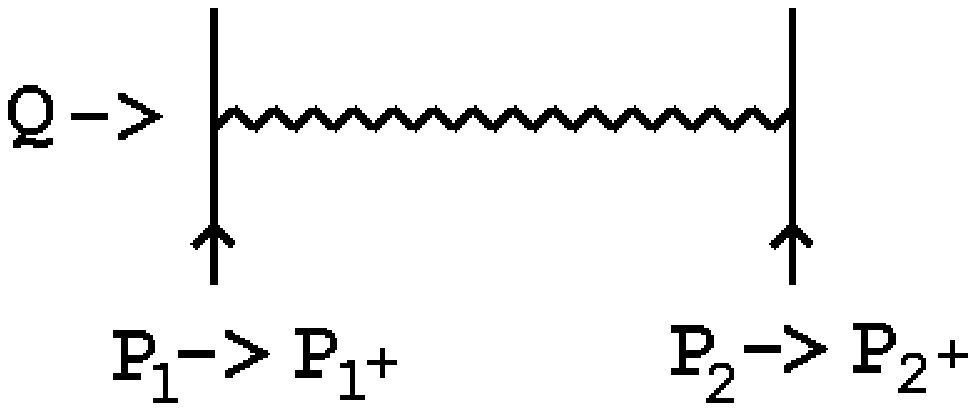}

Fig.~C1 Single Gluon Exchange
\end{center}
We consider the Regge limit in which 
$$
\eqalign{P_1 & \to ~ P_{1^+} = ~p_{12^-} ~ \underline{n}_{1^+}~, 
~~~~p_{12^-}   
\to \infty \cr
 P_2 &\to ~P_{2^+}  = ~p_{21^-}~\underline{n}_{2^+}~,~~~~p_{21^-}   
\to \infty \cr
Q &\to ~\underline{Q}_{12+} }
\auto\label{1rlc}
$$
This is, perhaps, a counter-intuitive way to discuss high-energy forward 
scattering. Nevertheless, we can proceed in complete parallel with 
conventional calculations. 

The spinor $\psi(P)$ for an on-shell quark satisfies 
$$
 \eqalign{ m~\psi(P)&~= ~(~p_{2^-}\gamma_{1^-} + p_{1^-}\gamma_{2^-} -  
\underline{p}_{12+} \cdot \underline{\gamma}_{12+}) 
~ \psi(P ) \cr 
& \centerunder{$\longrightarrow$}{\raisebox{-5mm}{$P \to P_{1^+}$}}
~~ p_{2^-} ~\gamma_{1^-} ~\psi(p) } 
\auto
\label{rlc0}
$$
Therefore the vertex for such a fast quark to couple to a single gluon carrying 
momentum transfer $\underline{Q}_{12+}$ is given by 
$$
\eqalign{
{p_{2^-} \gamma_{1^-} \over m }~  \gamma_{\mu}~ {(p_{2^-} \gamma_{1^-} -
\underline{Q}_{12+} \cdot \underline{\gamma}_{12+}) \over m }~  
~& = ~ {p_{2^-}  \gamma_{1^-} \over m  }~\biggl(
{p_{2^-} \over m }~\delta_{2^-,\mu}~ - ~ 
{\gamma_{\mu}~ 
\underline{Q}_{12+} \cdot \underline{\gamma}_{12+} \over m }~ \biggr)\cr
&= ~{p_{2^-} \over m }
~\delta_{2^-,~\mu} \bigl(1~+~O(1/p_{2^-}) \bigr) }
\auto\label{rlc01}
$$ 
where we have used the formulae of Appendix B and have reused (\ref{rlc0}) 
to obtain the last equality. Using this result for the $P_1$ vertex 
and the  analagous result for the $ P_{2}$ vertex,
we obtain the familiar result for the full amplitude
$$
A(s,t) ~\centerunder{$\sim$}{\raisebox{-4mm}{$s \to \infty$}}
 ~ {p_{12^-}~ g^{2^-,1^-} ~p_{21^-} \over Q^2} ~\sim ~ {s \over t}
\auto\label{rlc011}
$$

Moving on to the two-gluon exchange diagram illustrated in Fig.~C2,

\begin{center}
\leavevmode
\epsfxsize=2.5in
\epsffile{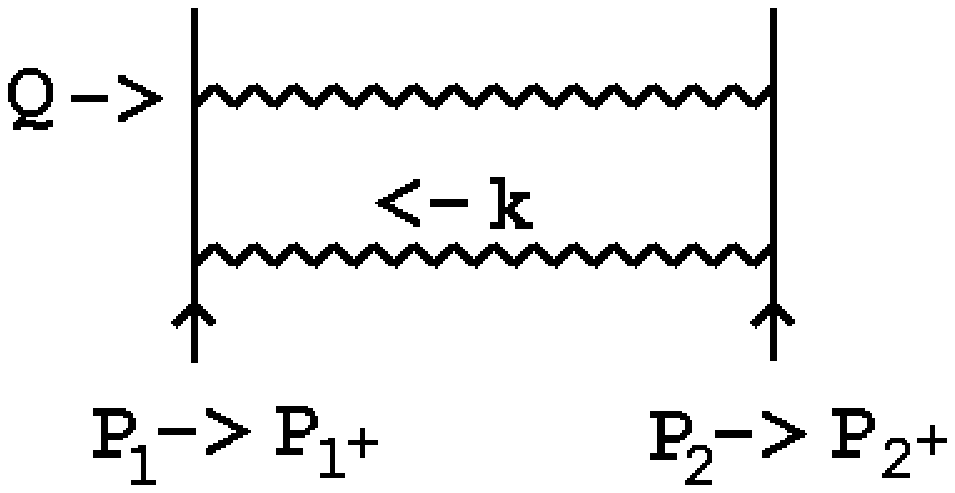}

Fig.~C2 Two Gluon Exchange 

\end{center}
we calculate the imaginary part by first writing 
$$
\int~ d^4k ~= ~ \int ~dk_{1^-} dk_{2^-} d^2 \underline{k}_{12+}
\auto
\label{rlc1}
$$
Then, for the internal quark propagator along which $P_1$ flows, we write 
$$
\eqalign{
{ \gamma \cdot (P + k) +m \over (P+k)^2-m^2 }
\quad \centerunder{$\large\sim$}{\raisebox{-4mm}{$p_{2^-}
\rightarrow\infty$}}\quad &
{ \gamma_{1^-}p_{2^-}  +\cdots 
\over 2p_{2^-} k_{1^-} - \underline{k}_{12+}^2 
-m^2}\cr
\equiv ~~~~ &{\gamma_{1^-}+ ~ 0(1/p_{2^-}) 
\over \left[ k_{1^-} -  (\underline{k}_{12+}^2 
- m^2) /  p_{2^-}  \right]}
}
\auto\label{rlc2}
$$
Putting this 
quark on-shell by performing the $k_{1-}$ integration, 
the vertex for two gluons to couple to the fast quark is then 
$$
{p_{12^-} \gamma_{1^-} \over m }~ ( \gamma_{\mu}~\gamma_{1^-}~ \gamma_{\nu})
~ {p_{12^-} \gamma_{1^-}
\over m }~  
~= ~{p_{12^-} \over m }
~\delta_{2^-,\mu}~\delta_{2^-,\nu} 
\auto\label{rlc21}
$$
where we have again used (\ref{rlc0}). The essential feature here, is that 
the infinite momentum limit leads to the exchange of gluons that will couple 
to a second scattering quark with a $\gamma_{1^-}~$-coupling only. Note that 
this feature would be the same if we had used conventional light-cone
co-ordinates (or, in fact, any other light-cone co-ordinates).

Using the analagous result for $P_2 \to P_{2^+}$, 
to perform the $k_{2^-}$ integration, the
kinematic part of the full result, is 
$$
p_{12^-}\delta_{2^-,\mu}\delta_{2^-,\nu} g^{\mu\alpha}g^{\nu\beta}
\delta_{\alpha,1^-}\delta_{\beta, 1^-} 
{p'}_{21^-} \int { d^2 \underline{k}_{12+} \over
\underline{k}_{12+}^2 (\underline{k}_{12+} +  
\underline{Q}_{12+})^2 }  = 
s\int { d^2 \underline{k}_{12+} \over
\underline{k}_{12+}^2 (\underline{k}_{12+} +  
\underline{Q}_{12+})^2 }  
\auto
\label{rlc3}
$$
showing that the familiar tranverse momentum integral is simply replaced by 
an integral over the new ``transverse momentum'' $\underline{k}_{12+}$. Since 
$$
\int { d^2 \underline{k}_{12+} \over
\underline{k}_{12+}^2 (\underline{k}_{12+} +  
\underline{Q}_{12+})^2 }  ~ = ~J_1(Q^2) ~ = 
\int { d^2 \underline{k} \over
\underline{k}^2 (\underline{k} +  
\underline{Q})^2 }  
\auto\label{J1}
$$
this is a relatively
trivial modification. Nevertheless is important for the
arguments made in the body of the paper that the same result is clearly 
obtained whatever light-cone co-ordinates are used.

It is also interesting to calculate the Regge limit of Fig.~B2 keeping 
$P_2$ finite. In this case the choice of ``light-cone co-ordinates'' is not 
determined by the large momenta in the problem, since there is only one. We
can equally well use the conventional choice  (\ref{lcd11}), or the novel
co-ordinates utilised above. In either case we can arrive rapidly at the
correct answer by arguing as follows. We again use (\ref{rlc2}) to perform 
one longitudinal momentum integration ($k_{1^-}$). The two exchanged gluons
then couple to the $P_2$ quark via 
$$
\eqalign{
\gamma_{1^-}~ { \gamma \cdot (P_2 -k) \over (P_2 - k)^2 - m^2 } ~ \gamma_{1^-}
& = ~\gamma_{1^-}~{\gamma_{1^+} (P_2 - k)_{1^-} ~+ 
~\cdots \over k_{1^+}(P_2 - k)_{1^-} 
 ~+ ~\cdots }~\gamma_{1^-}~=~{ \gamma_{1^-} 
\over (k_{1^+}~+ ~\cdots)} \cr
or ~~~~~~~~~~~~~~~~~~~~~~~~~~~~ &~ \cr
& = ~\gamma_{1-}~
{\gamma_{2^-} (P_2 - k)_{1^-} ~+ ~\cdots \over k_{2^-}(P_2 - k)_{1^-} 
 ~+ ~\cdots }~\gamma_{1^-}~=~{ \gamma_{1^-} 
\over (k_{2^-}~+~ \cdots)}
} 
\auto\label{2gv}
$$
in either case, we use this last pole to carry out a second longitudinal
momentum integration ($k_{1^+}$ or $k_{2^-}$) and obtain the 
corresponding two-dimensional transverse 
integral. (Whether $k_{1^+}$ or $k_{2^-}$ is used, the exchanged 
gluon propagators become independent of this variable as $P_1 \to P_{1+}$.) 
We then use the Dirac equation, as in (\ref{rlc0}), to 
write either 
$$
\gamma_{1^-} ~= ~\gamma_{1^-} ~{ \gamma \cdot p \over m} ~
=~ p_{1^+} /m ~+~\cdots
\auto
$$ 
or 
$$
\gamma_{1^-} ~= ~\gamma_{1^-}~{ \gamma \cdot p \over m}  ~
=~ p_{1^-} /m ~+~\cdots
\auto
$$ 
and argue that only the first term, shown explicitly, is capable of forming 
a Lorentz invariant with the momentum of the 
fast quark. The result is then either the conventional transverse 
momentum integral or (\ref{rlc3}). We conclude that when a fast quark 
scatters off a quark carrying finite momentum we can calculate using any 
light-cone co-ordinates. The result will be the same, but will be expressed 
in terms of transverse momenta that depend on the co-ordinates chosen.

We consider next some double-Regge and triple-regge amplitudes. 
The main results are not used directly in the text but they are instructive
and some of the intermediated results are used. We briefly discuss 
the kinematics of single particle (gluon) production first. We can parallel
our elastic scattering discussion using the notation of Fig.~C3.
\begin{center}
\leavevmode
\epsfxsize=3in
\epsffile{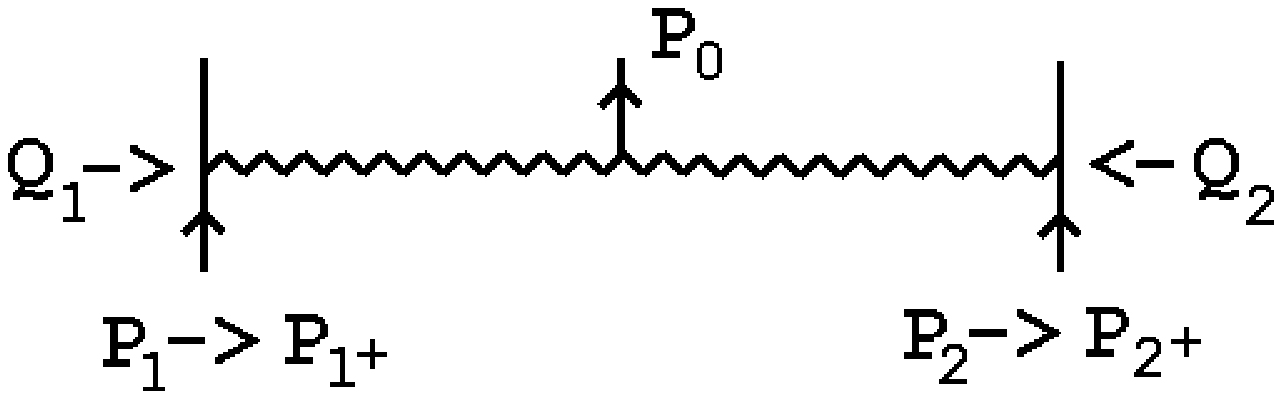}

Fig.~C3  Double Regge Kinematics

\end{center}
We take $P_1 \to P_{1^+}$ and $P_2 \to P_{2^+}$ as before and also 
$$
\eqalign{&
Q_1 \to (q_{11^-},q_{12^-},q_{112-},q_{13}) 
~\equiv (q,0,\tilde{q},q_{13}) \cr
& Q_2 \to (q_{21^-},q_{22^-},q_{212-},,q_{23}) ~\equiv 
(0,q,-\tilde{q},q_{23})}
\auto
\label{rlc4}
$$
with $P_0 = Q_1 + Q_2$.  In this notation we have 
six independent variables, $p_{2^-},p_{1^-}',q,\tilde{q},q_{13}$ and $q_{23}$. 
The necessary 
reduction to five variables is achieved by putting $P_0$ on mass-shell. This 
determines $q$ in terms of $q_{13}$ and $q_{23}$. 

Consider now the double-regge 
amplitude shown in Fig.~C4
for producing a quark-antiquark pair via gluon exchange.
\begin{center}
\leavevmode
\epsfxsize=3in
\epsffile{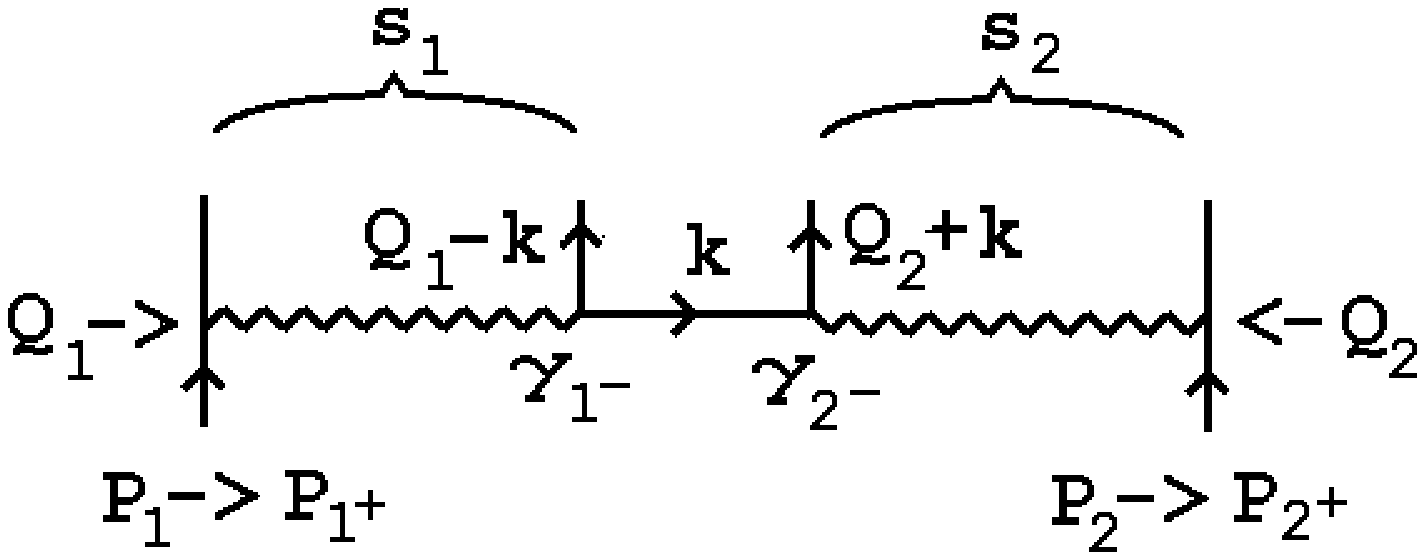}

Fig.~C4  Quark-antiquark Production in the Double-Regge Limit

\end{center}
We define $\underline{k}$ to be the
four-momentum flowing along the exchanged quark propagator 
and use the same notation for $Q_1$ and $Q_2$ as in (\ref{rlc4}), except that 
we take $q_{11-} \neq ~q_{22-}$. We can then fix 
both of $q_{11-}$ and $q_{22-}$ by putting both produced particles on shell. 

By applying (\ref{rlc0}) to the fast particles we determine that, as
illustrated, the gluons 
couple to the quark-antiquark pair via $\gamma_{1-}$ and $\gamma_{2-}$ 
couplings. This implies that only the transverse part of the exchanged
quark propagator contributes, i.e.
$$
\gamma_{1^-}~{\underline{k} \cdot \underline{\gamma} -m 
\over \underline{k}^2 -m^2} ~\gamma_{2^-}
~~~=~~ \gamma_{1^-} ~{ - ~\underline{k}-{12+} \cdot 
\underline{\gamma}{12+} -m \over k^2 -m^2} ~\gamma_{2^-}
\auto
\label{rlc5}
$$
The full amplitude for Fig.~C4 is then 
$$
A(p_{12^-},p_{21^-},\tilde{q},q_{13},q_{23},\underline{k}_{12+},
k_{1^-}k_{2^-})
=  {p_{12^-}p_{21^-}~\gamma_{1^-} ~(- \underline{k}_{12+} \cdot 
\underline{\gamma}_{12+} -m)~\gamma_{2^-}
 \over m^2 (\tilde{q}^2 + q_{13}^2)
(\tilde{q}^2 + q_{23}^2)(k^2 -m^2) } 
\auto\label{rlc51}
$$
As must be the case, the amplitude is a function of 
eight independent variables. 

To extract an amplitude expressed in terms of invariants consider, in
particular, the case in which the produced quark and antiquark spin
dependence contributes similarly to 
(\ref{rlc0}), i.e. we write
$$
 \eqalign{ m~\bar{\psi}(Q_1-k)~&= ~~ k_{1^-} ~\gamma_{2^-} 
~\bar{\psi}(Q_1-k) ~+ ~\cdots \cr
m~\psi(Q_2 + k)~&= ~~ k_{2^-} ~\gamma_{1^-} ~\psi(Q_2 +k) ~+ ~\cdots 
} 
\auto
\label{rlc6}
$$
and keep only the spinor components shown explicitly. 
In this case the production amplitude of Fig.~C4 has the simple form
$$
{s~s' \over m^2 Q_1^2 Q_2^2 }~~ {- ~\underline{k}_{12+} \cdot 
\underline{\gamma}_{12+} -m \over (k^2 - m^2)} 
\auto\label{rlc7}
$$

Note that with the polarizations of the 
produced pair given by (\ref{rlc6}), the diagram of Fig.~C5 does not contribute.
\begin{center}
\leavevmode
\epsfxsize=2.5in
\epsffile{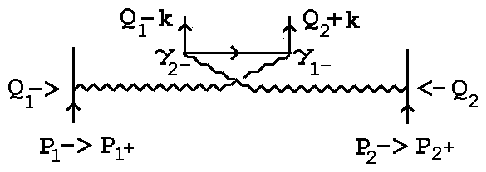}

Fig.~C5  An Alternative Gluon Coupling
\end{center}

A reggeon Ward identity requires that when all diagrams are summed
the central reggeon amplitude (contained in the square brackets of 
(\ref{rlc7})) should 
vanish when either $Q_1$ or $Q_2$ vanish. This is achieved by 
adding the three diagrams of Fig.~C6.
\begin{center}
\leavevmode
\epsfxsize=4.5in
\epsffile{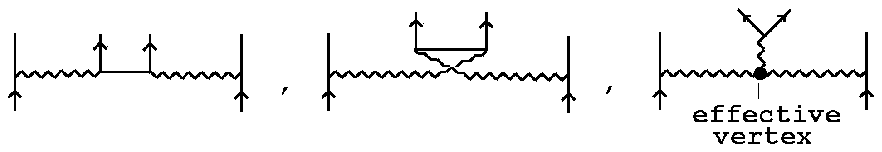}

Fig.~C6  Diagrams Required for the Reggeon Ward Identity. 
\end{center}
The third diagram involves an effective regge limit vertex\cite{fl} rather
than the gauge coupling. The quark/antiquark state can be written as a sum of 
symmetric and antisymmetric combinations that, when color factors are 
introduced, respectively carry zero and octet color. The third diagram appears
only in the color octet channel. 
For the special polarizations given by (\ref{rlc6}) it 
directly cancels the first when $Q_1$ or $Q_2 \to 0$.

Consider next the diagram of Fig.~C7 in which an additional gluon is 
exchanged in the $Q^2$ channel.
\begin{center}
\leavevmode
\epsfxsize=2.6in
\epsffile{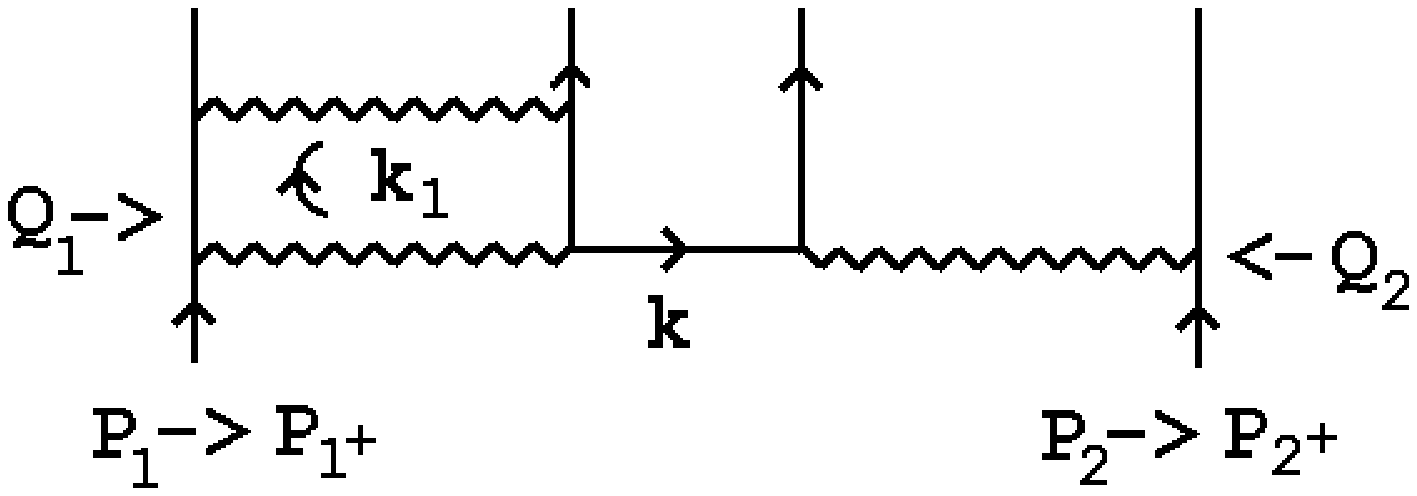}

Fig.~C7  An Additional Gluon Exchanged
\end{center}
We can calculate the discontinuity in $s_1$, or simply carry out two 
longitudinal integrations, by 
repeating the analysis that we applied to Fig.~C2, we obtain 
$$
\eqalign{ A(p_{12^-},p_{21^-},\tilde{q},q_{13}&,q_{23},
\underline{\tilde{k}}_{\perp},
k_{1^-}k_{2^-}) \cr
&=  
{p_{12^-}p_{21^-}' \over m^2  Q_2^2 } \int { d^2 \underline{k}_{112+} 
\over 
\underline{k}_{112+}^2 (\underline{k}_{112+} -  
\underline{Q}_{12+})^2} {\gamma_{1^-} (- \underline{k}_{12+} \cdot 
\underline{\gamma}_{12+} -m)\gamma_{2^-} \over (k^2 -m^2)} 
\cr &=  {p_{12^-}p_{21^-} \over m^2  Q_2^2 } J_1(Q_1^2)
~ {\gamma_{1^-} (- \underline{k}_{12+} \cdot 
\underline{\gamma}_{12+} -m)\gamma_{2^-} \over (k^2 -m^2)} 
}
\auto\label{rlc71}
$$
Comparing with (\ref{rlc51}), we see that the additional gluon has simply 
replaced one gluon transverse momentum propagator by a transverse momentum 
integral. The integral also has a $\gamma_{1^-}$ ``point-coupling'' to the 
central vertex. The pointlike nature of this coupling is, of course, 
essential if Fig.~C7 is to be added to Fig.~C3 and the $J_1(Q_1^2)$ is to 
produce the reggeization of the gluon in the $Q_1^2$ channel. However, there 
will also be a pointlike coupling when the quantum numbers 
in the $Q^2$ channel are such that each of the two 
gluons involved in the loop integral in Fig.~C7 reggeize separately
and the two reggeon cut appears. 

If the additional gluon is attached to the outgoing quark as in 
Fig.~C8 (rather than to the antiquark as in Fig.~C7) then we 
no longer obtain a point-coupling for the two-gluon exchange in the $Q_1$
channel.
\begin{center}
\leavevmode
\epsfxsize=2.6in
\epsffile{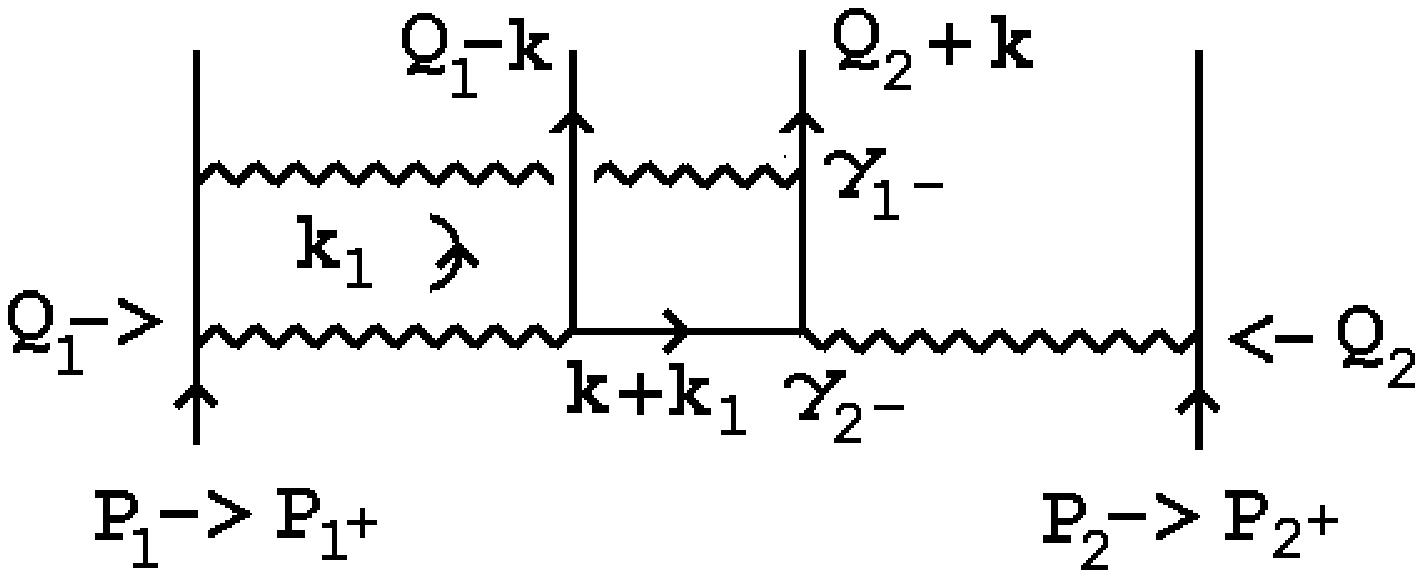}

Fig.~C8  An Additional Gluon Exchange Giving no Point Coupling.
\end{center}
The contribution of the on mass-shell hatched quark line and the 
adjacent $\gamma$-couplings to the $k_1$ integral is now
$$
\eqalign{\int &d k_{12^-} ~\delta\biggl((k_1 +k + Q_2)^2 -m^2\biggr)
\gamma_{2^-}~\biggl((k_1 +k + Q_2)\cdot \gamma -m \biggr) 
 ~\gamma_{1^-} \cr
~~=&~ \int d k_{12^-} ~\delta\biggl( k_{12^-}(k_1 +k + Q_2)_{1^-} ~
 \cdots \biggr) \gamma_{2^-} ~
\biggl(- ~(\underline{k}_{112+} + \underline{k}_{12+} 
+ \underline{Q}_{212+} )\cdot 
\underline{\gamma}_{12+} -m\biggr) 
~\gamma_{1^-} \cr
&~~~=~~~{\gamma_{2^-} ~
\biggl(- ~(\underline{k}_{112+} + \underline{k}_{12+} 
+ \underline{Q}_{212+} )\cdot 
\underline{\gamma}_{12+} -m\biggr) 
~\gamma_{1^-}
\over (k_1 +k + Q_2)_{1^-} }
}
\auto
\label{rlc8}
$$
We do not obtain a point-like coupling 
because (unlike in (\ref{2gv}), for example)
the argument of the $\delta$-function contains an integrated
longitudinal momentum multiplied by a momentum factor that does not 
multiply a $\gamma$-matrix appearing in the numerator of the propagator.
The relevant part of the propagator numerator is eliminated by the surrounding
$\gamma$ matrices.

Finally we move on to the process that is of central interest in the main 
body of the paper. This is the triple Regge scattering illustrated in 
Fig.~C9. 
\begin{center}
\leavevmode
\epsfxsize=3in
\epsffile{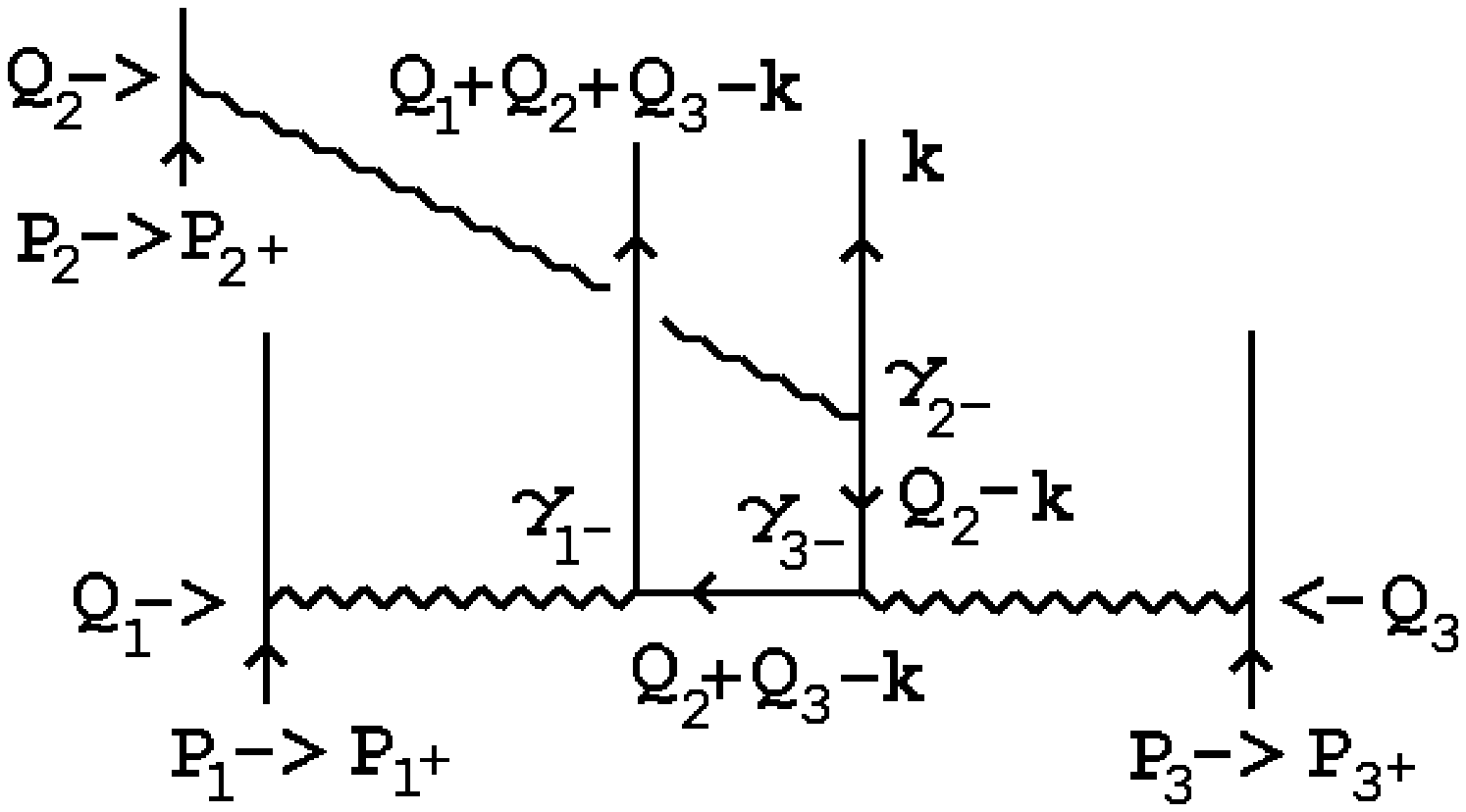}

Fig.~C9  A Triple Regge Amplitude. 
\end{center}
A triple-Regge limit can be defined as $P_i ~\to P_{i+},~ i=1,2,3$ with 
the $Q_i$ kept finite. We will not give a complete description of the
quark-antiquark intermediate state in this limit since it will not be needed 
in the body of the paper. The important point for our purposes is that 
to directly obtain the triple discontinuities studied in Sections 5 and 6, the
amplitude in Fig.~C9 should be combined with another amplitude of the same form
and all three of the $Q_i$ integrated over. We do not do this in Section 6
but instead discuss only double discontinuities explicitly. However, we can 
make the following comment on the direct construction of triple 
discontinuities.

The reduction of 
$Q_i$ integrations to two-dimensional integrals is achieved by using all the 
longitudinal integrations to put on-shell all quark lines involved in the 
multiple discontinuity. In particular, the internal quark propagator 
in Fig.~C9 carrying momentum $Q_2 - k$ should be placed on-shell. If we use the 
$Q_2$ integration to put this line on-shell, and also use the
$1^-$,$2^-$, ... co-ordinates of Appendix B, the combination of 
the $\gamma_{2^-}$ and $\gamma_{3^-}$ factors with the on-shell propagator 
produces the effective coupling
$$
\eqalign{ ~\int &~dQ_{21^-} 
~\delta\biggl(Q_{21^-}(Q_{22^-} - k_{2^-}) ~- ~\cdots\biggr)~
  \gamma_{2^-}~ (\st{Q}_2 - \st{k})~ \gamma_{3^-} \cr 
~=&~ \int ~dQ_{22^-}
~\delta\biggl(Q_{22^-}(Q_{21^-} - k_{1^-}) ~- ~\cdots\biggr)
  ~\gamma_{2^-}~ \biggl(\gamma_{1^-} (Q_{21^-} - k_{1^-}) ~- ~\cdots\biggr)
~ \gamma_{3^-} \cr 
&~~~~~~~~ = ~\gamma_{2^-} \gamma_{1^-} 
 \gamma_{3^-} ~+ ~\cdots }
\auto\label{rlc9}
$$
Using the identity
$$
\gamma_{\alpha}\gamma_{\beta}\gamma_{\lambda}~=~
g_{\alpha\beta} \gamma_{\lambda} ~+~ g_{\beta\lambda} \gamma_{\alpha} 
~-~ g_{\alpha\lambda} \gamma_{\beta} + i \epsilon_{\mu\alpha\beta\gamma}
\gamma^{\mu} \gamma_5
\auto\label{3ga}
$$
we obtain
$$
~\gamma_{2^-} \gamma_{1^-} 
 \gamma_{3^-} ~= ~ \gamma_0 + \gamma_1 -  \gamma_2 -  \gamma_3 
~+~ i \gamma_5 ~(\gamma_0 + \gamma_1 +  \gamma_2 +  \gamma_3)
\auto\label{3ga1}
$$
showing that the coupling (\ref{rlc9}) contains the  
effective $\gamma_5$-coupling shown in Fig.~C10

\begin{center}
\leavevmode
\epsfxsize=2.5in
\epsffile{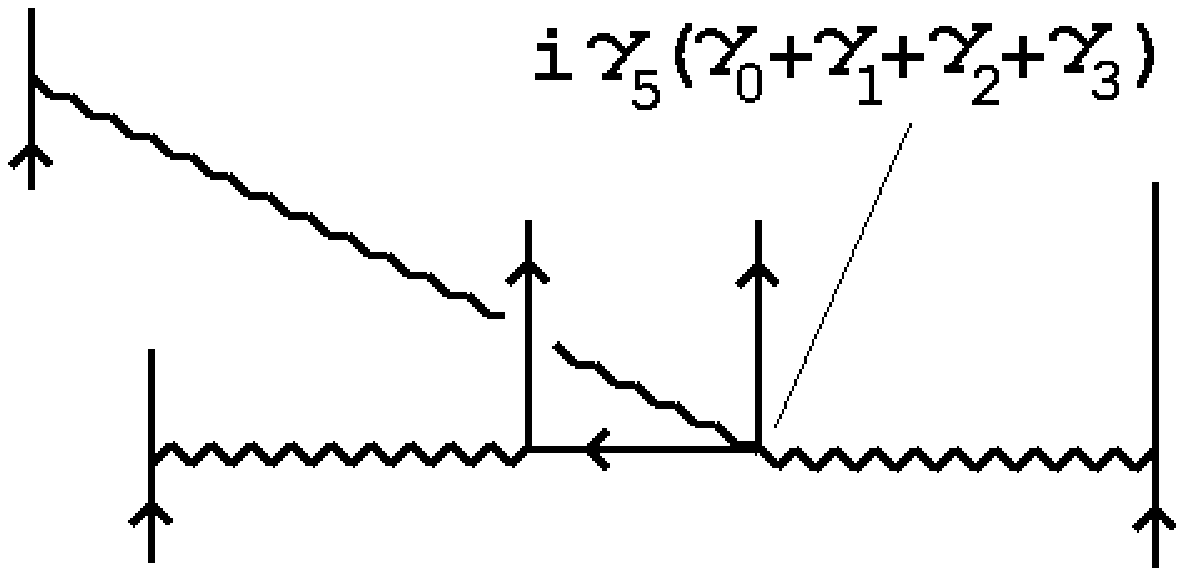}

Fig.~C10  The $\gamma_5$-coupling Generated by Fig.~C9

\end{center}
This illustrates how the triple-Regge limit introduces sufficient 
orthogonality for the large momenta to produce an effective 
$\gamma_5$-coupling. 

\newpage

\renewcommand{\theequation}{D.\arabic{equation}}
\setcounter{equation}{0}
\vskip 1cm \noindent
\noindent {\large\bf Appendix D. Angular Variables }
\vskip 3mm \noindent

To introduce angular variables for a six-particle amplitude it is necessary 
to define a set of six standard Lorentz frames ${\cal F}_1, {\cal F}_2, 
{\cal F}_3, \tilde{{\cal F}}_1, \tilde{{\cal F}}_2,  \tilde{{\cal F}}_3$.
These frames are associated with the vertices of the Toller diagram,
as indicated in Fig.~D1, by requiring that the momenta meeting at a vertex
take a standard form. For each internal 
vertex there are three frames, in each of 
which one of the momenta lies either along the $t$-axis or the $z$-axis. 
As we will see, once the standard frames are defined,
the angular variables parametrize ``little group'' Lorentz
transformations between the frames. 

\begin{center}
\leavevmode
\epsfxsize=3.5in
\epsffile{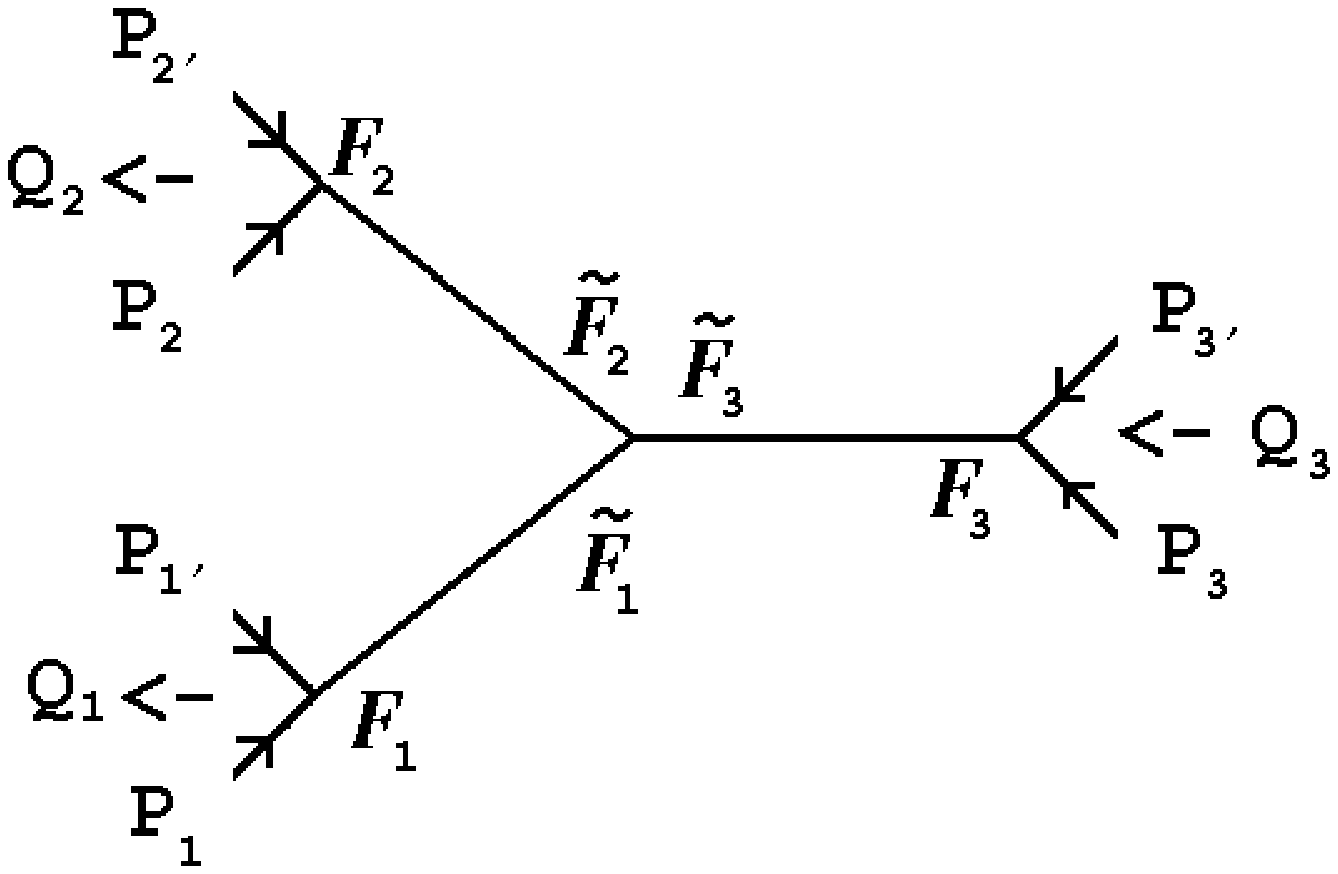}

Fig.~D1 Special Frames
\end{center}

Not surprisingly, the definition of the 
standard frames, together with the little groups involved (and their 
parametrization) depend on the physical region discussed. Since the 
multi-regge theory we develop in Section 5 effectively moves backwards and 
forwards between various 
$t$ and $s$-channels we need to determine how the variables introduced in 
different channels are analytically related. For this purpose we explicitly
calculate below, expressions for invariants in terms of angular
variables in each of the channels we discuss. We 
take the mass of all external particles to be $m$. We can then distinguish
the three $t$-channels and four $s$-channels that we study as follows.
In the $t_i$-channel ($i=1,2,3$), 
$|Q_i| \geq |Q_j| + |Q_k|$ ($i\neq j\neq k$) with 
$Q_j^2, Q_k^2 \geq 4m^2$. In the $s$-shannels the $t_i = Q_i^2$ are all
negative. The four channels are that in which 
the particles with momenta $P_1, P_2$ and $P_3$ scatter, with final 
momenta $P_1', P_2'$ and $P_3'$ respectively, and those in which one of the 
$P_i'$ is exchanged with the corresponding $P_i$. 

In Fig.~D2 we have shown (topographically) the three $t_i$-channels and one of
the $s$-channels. In this figure, we have also indicated that a single 
$s$-channel breaks up into four distinct sub-regions. There are three 
``$s-t$'' sub-regions in which one of the transverse momenta has longer length
than the sum of the other two. In these regions the plane containing the
$Q_i$ must have a timelike component. In the ``$s-s$'' sub-region the $Q_i$
satisfy euclidean inequalities and can be taken to have only spacelike
components. We will discuss how the variables introduced in all regions are
related by analytic continuation. 
\begin{center}

\leavevmode
\epsfxsize=5.5in
\epsffile{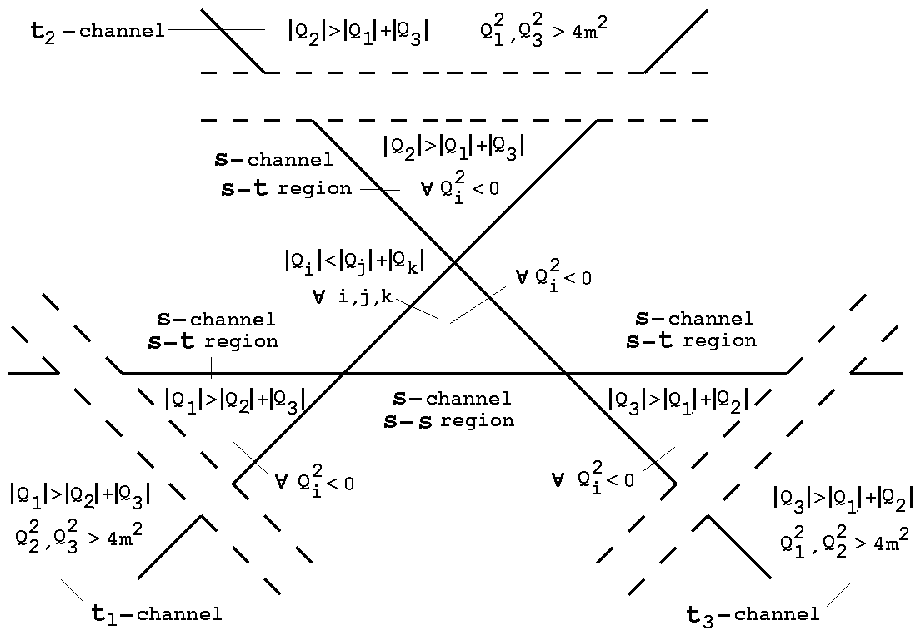}

\vspace{0.2in}

Fig.~D2 Physical Regions
\end{center}

We consider first the $t_3$-channel, illustrated in Fig.~D1, in which two
initial state particles, $3$ and $3'$, scatter into four final state
particles $2, 2', 3, 3'$. In this case $Q_1^2, Q_2^2 \geq 4m^2, ~\forall i$,
and $|Q_3| \geq |Q_1| + |Q_2|$. 
The frames ${\cal F}_i$, $i=1,2,3$, can be defined by requiring that 
$$
Q_i  = (Q_i,0,0, 0)  ~~~~~~~~~~~~~~~~
\eqalign{ P_i & = (mcosh \xi_i, 0,0, msinh \xi_i) \cr 
P'_i & = (mcosh \xi_i,0,0, - msinh \xi_i) 
}
\auto\label{f1}
$$
where $cosh\xi_i = Q_i /2m$. Clearly we could easily interchange the roles 
of $P_i$ and $P_i'$ by setting $\xi_i \to -\xi_i~\forall i$. As long as the theory is 
parity invariant, amplitudes can not depend on this choice.
For frame $\tilde{{\cal F}}_1$ we require that
$$
Q_1 = (Q_1,0,0,0) ~~~~~~~~~~~~~~~ 
\eqalign{ Q_2 & = (Q_2cosh\zeta_{21},0,0,Q_2sinh\zeta_{21}) 
\cr
Q_3  & = (Q_3cosh\zeta_{31},0,0,Q_3sinh\zeta_{31}) }
\auto\label{f2}
$$
where
$$
cosh\zeta_{21} = {Q_3^2 -Q^2_2 - Q_1^2 \over 2Q_1Q_2} ~, ~~~~
cosh\zeta_{31} = {Q_3^2 + Q^2_1 - Q_2^2 \over 2Q_1Q_3} 
\auto\label{f21}
$$
For the frames $\tilde{{\cal F}}_2$ and $\tilde{{\cal F}}_3$ we make the 
analagous requirements so that in $\tilde{{\cal F}}_3$, for example,
$$
Q_3 = (Q_3,0,0,0) ~~~~~~~~~~~~~~~ 
\eqalign{ Q_1 & = (Q_1cosh\zeta_{13},0,0,Q_1sinh\zeta_{13}) 
\cr
Q_2  & = (Q_2cosh\zeta_{23},0,0,Q_2sinh\zeta_{23}) }
\auto\label{f3}
$$
where 
$$
sinh \zeta_{13}~=~- sinh\zeta_{31}~=~ \biggl( { (Q_3^2 +Q^2_1 - Q_2^2)^2 - 
4Q_1^2 Q_3^2
 \over 4Q_1^2Q_2^2 } \biggr)^{{1 \over 2}} 
=~-{\lambda^{{1 \over 2}}(t_1,t_2,t_3) \over 2 ~\sqrt{t_1}~\sqrt{t_3} } 
\auto\label{f211}
$$
and 
$$
sinh \zeta_{23}~=~ {\lambda^{{1 \over 2}}(t_1,t_2,t_3) \over 
2 ~\sqrt{t_1}~\sqrt{t_3} }
\auto\label{f212}
$$
where $\lambda(t_1,t_2,t_3)$ is the familiar function
$$
\eqalign{&\lambda(t_1,t_2,t_3)~=~ 
t_1^2 + t_2^2 + t_3^2 - 2 t_1t_2 - 2 t_2t_3 - 2 t_3t_1 \cr
&=(\sqrt{t_1} + \sqrt{t_2} +\sqrt{t_3})(\sqrt{t_1} - \sqrt{t_2} -\sqrt{t_3})
(-\sqrt{t_1} + \sqrt{t_2} - \sqrt{t_3})(- \sqrt{t_1} - \sqrt{t_2} +\sqrt{t_3})
}
\auto\label{lam}
$$
Clearly we have to take opposite signs for $\lambda^{{1\over 2}}(t_1,t_2,t_3)$ 
in defining $sinh \zeta_{13}$ and $sinh \zeta_{23}$. Conversely we can 
reverse this sign by interchanging the form of $Q_1$ and $Q_2$ in the 
$t_3$-channel standard frames. In the next paragraph
we will discuss further the ambiguity in making this choice, together with the
remaining ambiguity in fixing the frames ${\cal F}_i$ and the frames
$\tilde{{\cal F}}_i$. It is linked, of course, to the ambiguity in the 
choice of the $sinh \xi_i$.

$\tilde{{\cal F}}_1$  and ${\cal F}_1$ are related by a Lorentz 
transformation $g_1$ that leaves $Q_1$ unchanged, i.e. $g_1$ belongs to
the little group of $Q_1$, which is $SO(3)$. We can parametrize $SO(3)$ in
the form 
$$ 
g_1=u_z(\mu_1)u_x(\theta_1)u_z(\nu_1)
~~~~~~~~~~ 0\leq \theta<\pi ~,~~ 0\leq \nu,\ \mu\leq 2\pi
\auto\label{sO3}
$$
where $u_z$ and $u_x$ are, respectively, rotations about the $z$ and $x$ 
axes. If we take $g_1$ to transform from ${\cal F}_1$ to $\tilde{{\cal 
F}}_1$, $g_2$ to transform from ${\cal F}_2$ to $\tilde{{\cal F}}_2$ 
and $g_3$ to transform from ${\cal F}_3$ to $\tilde{{\cal F}}_3$,
then we can absorb the $u_z(\mu_i)$ 
in our definition of the frames ${\cal F}_i$ so that, effectively, we set
$\mu_i = 0, ~i=1,2,3$. Apart from the choice of sign for $sinh\xi_i$, this 
removes the remaining 
ambiguity in the definition of the ${\cal F}_i$ frames after (\ref{f1})
is satisfied. Because 
the $u_z(\nu_i)$ commute with the boosts $a_z(\zeta_{ij})$ along the 
$z$-axis,  invariants can depend only on 
differences between the three $\nu_i$ - so that only two parameters are actually
involved. If we insist on both the parametrization (\ref{sO3}) and this last
commutativity property then the $\tilde{{\cal F}}_i$ frames are determined 
up to a reflection - an overall sign change for all the $sinh\zeta_{ij}$. 
Again, amplitudes can not depend on this choice of sign 
because of parity invariance. Nevertheless, the parity 
transformation that produces this overall sign change plays an important 
role in the discussion of Section 5.

In general, to calculate invariants we transform all the momenta involved from 
frames in which they take a simple form to a common frame where the invariant 
is most easily evaluated. For.example,
we transform $P_1$ from ${\cal F}_1$ to ${\cal F}_3$ via 
$\tilde{{\cal F}}_1$ and  $\tilde{{\cal F}}_3$ as follows. In $\tilde{{\cal
F}}_1$ 
$$
P_1 = (mcosh \xi_1,~ - msinh \xi_1 sin \theta_1 sin \nu_1, 
~-msinh \xi_1 sin \theta_1 cos\nu_1, 
~msinh \xi_1 cos \theta_1) 
\auto
$$
In $\tilde{{\cal F}}_3$
$$
\eqalign{P_1 = (&mcosh \xi_1cosh\zeta_{31} - 
msinh \xi_1 cos \theta_1 sinh\zeta_{31} , 
~-msinh \xi_1 sin \theta_1 sin \nu_1, \cr 
& - msinh \xi_1 sin \theta_1 cos\nu_1, 
~msinh \xi_1 cos \theta_1 cosh\zeta_{31} - 
mcosh \xi_1sinh\zeta_{31}) }
\auto
$$
In ${\cal F}_3$
$$
\eqalign{P_1 = (&mcosh \xi_1cosh\zeta_{31} - 
msinh \xi_1 cos \theta_1 sinh\zeta_{31}, 
 msinh \xi_1 sin \theta_1 sin (\nu_1 -\nu_3), \cr 
& ~msinh \xi_1 sin \theta_1 cos (\nu_1 - \nu_3)cos\theta_3 
- msinh \xi_1 cos \theta_1 cosh\zeta_{31}sin\theta_3 \cr
& ~- mcosh \xi_1sinh\zeta_{31}sin \theta_3, 
~- msinh \xi_1 sin \theta_1 cos (\nu_1 - \nu_3)sin \theta_3 \cr
& + msinh \xi_1 cos \theta_1 cosh\zeta_{31}cos \theta_3 -
mcosh \xi_1 sinh\zeta_{31}cos \theta_3)}
\auto
$$
Alternatively we can transform $P_1$ to $\tilde{{\cal F}}_2$ and to 
${\cal F}_2$ as follows. In $\tilde{{\cal F}}_2$
$$
\eqalign{P_1 = (&mcosh \xi_1cosh\zeta_{21} -
msinh \xi_1 cos \theta_1 sinh\zeta_{21} , 
~ - msinh \xi_1 sin \theta_1 sin \nu_1, \cr 
& - msinh \xi_1 sin \theta_1 cos\nu_1, 
~ msinh \xi_1 cos \theta_1 cosh\zeta_{21} -
mcosh \xi_1sinh\zeta_{21}) }
\auto
$$
In ${\cal F}_2$ 
$$
\eqalign{P_1 = (&mcosh \xi_1cosh\zeta_{21} + 
msinh \xi_1 cos \theta_1 sinh\zeta_{21}, 
~ - msinh \xi_1 sin \theta_1 sin (\nu_1 -\nu_2), \cr 
&msinh \xi_1 sin \theta_1 cos (\nu_1 - \nu_2)cos\theta_2 
+ msinh \xi_1 cos \theta_1 cosh\zeta_{21}sin\theta_2 \cr
& - mcosh \xi_1sinh\zeta_{21}sin \theta_2, 
~ - msinh \xi_1 sin \theta_1 cos (\nu_1 - \nu_2)sin \theta_2 \cr
& + msinh \xi_1 cos \theta_1 cosh\zeta_{21}cos \theta_2 -
mcosh \xi_1 sinh\zeta_{21}cos \theta_2)}
\auto
$$

From the above expressions for $P_1$ we can already 
calculate several invariants. In $\tilde{{\cal F}}_1$,
for example, $Q_3$ has the form (\ref{f2}) and so 
$$
P_1.Q_3 ~= ~mQ_3~\bigl[cosh \xi_1 cosh\zeta_{31} - sinh \xi_1 
sinh\zeta_{31}cos \theta_1\bigr] 
\auto\label{inv01}
$$
In ${\cal F}_3$, similarly, $P_3$ has the form (\ref{f1}) and so 
$$
\eqalign{ P_1.P_3  ~ = ~ &m^2~ \bigl[cosh \xi_1 cosh \xi_3cosh\zeta_{31} -  
sinh \xi_1 cosh \xi_3 sinh\zeta_{31}cos \theta_1  \cr
&-   sinh \xi_1 sinh \xi_3 sin \theta_1 sin \theta_3 cos (\nu_1 - \nu_3)
- sinh \xi_1 sinh \xi_3 cosh\zeta_{31} cos \theta_1 cos \theta_3 \cr 
& + cosh \xi_1 sinh \xi_3sinh\zeta_{31}cos \theta_3 \bigr]}
\auto\label{inv11}
$$
while, in ${\cal F}_2$, $P_2$ has the form (\ref{f1}) and so 
$$
\eqalign{ P_1.P_2 ~ = ~ &m^2~\bigl[ cosh \xi_1 cosh \xi_2 cosh\zeta_{21} - 
sinh \xi_1 cosh \xi_2  sinh\zeta_{21} cos \theta_1\cr
& - sinh \xi_1 sinh \xi_2 sin \theta_1 sin \theta_2 cos (\nu_1 - \nu_2) 
- sinh \xi_1 sinh \xi_2 cosh\zeta_{21} cos \theta_1 cos \theta_2 \cr 
& + cosh \xi_1 sinh \xi_2sinh\zeta_{21}cos \theta_2 \bigr]}
\auto\label{inv12}
$$
(\ref{inv11}) and (\ref{inv12}) differ only by the interchange of 
$1$ and $2$. It is straightforward to
calculate all other invariants in a similar manner. 
If we write $z_i = cos \theta_i$ and $u_{ij} = e^{i(\nu_i - \nu_j)}$ 
then we can take any two of the $u_{ij}$, together with the $z_i$ and the 
$t_i$, as eight independent variables. 

We see from the above formulae that a change of sign of $sinh\xi_1$ is
equivalent to a change of sign of both $cos\theta_1$ and $sin\theta_1$ 
($\theta \to \theta + \pi$). A change of sign of the $sinh\zeta_{ij}$ is
equivalent to a change of sign of all the $cos \theta_i$ which, in turn,
is equivalent to a change of sign of all the $sinh\xi_i$. 
It is also interesting to write (\ref{inv01}) and (\ref{inv11}) explicitly in 
terms of the $t_i$ and $z_1$, i.e.
$$
4~ P_1.Q_3~=~t_3 + t _1 -t_2 ~-~ 
\biggl({t_1 - 4m^2 \over t_1} \biggr)^{1 \over 2}
\lambda^{ 1 \over 2}(t_1,t_2,t_3) ~z_1 
\auto\label{inv33}
$$
and 
$$
\eqalign{ 8~P_1.P_3 ~&=~ t_3 + t _1 -t_2 ~-~\lambda^{ 1 \over 2}(t_1,t_2,t_3)
\biggl[{(t_1 - 4m^2)^{1 \over 2}  \over \sqrt{t_1} }z_1 
~- ~{(t_3 - 4m^2)^{1 \over 2}  \over \sqrt{t_3} } z_3 \biggr] \cr
&- ~(t_1 - 4m^2 )^{1 \over 2}(t_3 - 4m^2 )^{1 \over 2}
\biggl[(1-z_1^2)^{1 \over 2}(1-z_3^2)^{1 \over 2} \biggl(u_1 + {1 \over u_1}
\biggr) +  {t_3 + t_1 - t_2 \over \sqrt{t_1} \sqrt{ t_3} } z_1 
z_3 \biggl] \cr
&~ }
\auto\label{inv330}
$$
From these expressions we see that we will encounter analytic continuation 
problems at
the thresholds $t_i=4m^2$, at $t_i=0$, and at $\lambda(t_1,t_2,t_3)=0$. 
In particular, when $t_i < 0$ and also $\lambda(t_1,t_2,t_3) < 0$ the real 
relationship between the $z_i$ and the invariants is necessarily lost.
 
Consider now the $s$-channel in which $1, 2$ and $3$ are the three initial 
state particles and consider the $s-t$ region in which $Q_i^2 <0, \forall i$
and $|Q_3| \geq |Q_1| + |Q_2|$. 
The frames ${\cal F}_i$, $i=1,2,3$ are now defined by requiring that 
$$
Q_i  = (0,0, 0, q_i) ~~~~~~~~~~~~~~~~ 
\eqalign{ P_i & = (mcosh \xi_i, 0,0, msinh \xi_i) 
\cr 
P'_i & = (- mcosh \xi_i,0,0, msinh \xi_i) 
}
\auto
$$
where $sinh\xi_i = q_i /2m$ and 
$q_i = |Q_i| = [-t_i]^{{1\over 2}}$ (so that $sinh\xi_i
\equiv i ~cosh \xi_i$ if we consider the analytic continuation of
$ cosh\xi_i$ defined by (\ref{f1}) ). The obvious 
redefinition of the frame $\tilde{{\cal F}}_1$ is to require
$$
Q_1 = (0,0,0,q_1) ~~~~~~~~~~~~~~~ 
\eqalign{ Q_2 & = (q_2sinh\zeta_{21},0,0,q_2cosh\zeta_{21}) 
\cr
Q_3  & = (q_3sinh\zeta_{31},0,0,q_3cosh\zeta_{31}) }
\auto
$$
where
$$
cosh\zeta_{21} = {q_3^2 -q^2_2 - q_1^2 \over 2q_1q_2} ~, ~~~~
cosh\zeta_{31} = {q_3^2 + q^2_1 - q_2^2 \over 2q_1q_3} 
\auto
$$
These last expressions are simple analytic continuations of 
the expressions given in (\ref{f21}). The frames $\tilde{{\cal F}}_2$ and
$\tilde{{\cal F}}_3$ are redefined analagously. 
Note, however, that there is again an overall ambiguity in the 
choice of sign for the $sinh\zeta_{ij}$. Now the reflection involved is not a 
parity transformation since it applies to the time axis. If any of the  
$Q_i$ were timelike and associated with a particle state (as in a normal 
multi-regge production process) this sign would be determined. In the 
present case we will see that we must use both signs to fully cover the 
physical region. 

$\tilde{{\cal F}}_i$  and ${\cal F}_i$ are again related by a Lorentz 
transformation $g_i$ that leaves $Q_i$ unchanged, but now $g_i \in SO(2,1)$. 
Since the $Q_i$ triangle has a 
timelike component it is simplest to use the parametrization of  
$SO(2,1)$ that is closely related to that used above for $SO(3)$, i.e.
$$
g_1=u_z(\mu_1)a_x(\beta_1)u_z(\nu_1)
~~~~~~~~~~ - \infty \leq \beta< \infty ~~~ 0\leq \nu,\ \mu\leq 2\pi
\auto\label{s0211}
$$
where $a_x(\beta_1)$ is a boost along the $x$-axis. With this
parametrization, we can again choose the ${\cal F}_i$ such that $\mu_i =0, ~
i=1,2,3$ and the $u_z(\nu_i)$ commute with the boosts $a_z(\zeta_{ij})$. 

Repeating the transformation of  $P_1$ from ${\cal F}_1$ to ${\cal F}_3$ 
gives the following. 
In $\tilde{{\cal F}}_1$
$$
P_1 = (mcosh \xi_1cosh\beta_1, ~mcosh \xi_1 sinh \beta_1 cos\nu_1, 
~ - mcosh \xi_1 sinh \beta_1 sin \nu_1, 
~msinh \xi_1 ) 
\auto
$$
In $\tilde{{\cal F}}_3$
$$
\eqalign{P_1 = (&mcosh \xi_1cosh\beta_1 cosh \zeta_{31} - msinh \xi_1 sinh 
\zeta_{31}, ~ mcosh \xi_1 sinh \beta_1 cos\nu_1, \cr
& ~ - mcosh \xi_1 sinh \beta_1 sin \nu_1, 
~ msinh \xi_1 cosh \zeta_{31} - mcosh \xi_1cosh\beta_1 sinh \zeta_{31} ) }
\auto
$$
In ${\cal F}_3$
$$
\eqalign{P_1 = (& mcosh \xi_1cosh\beta_1 cosh \beta_3 cosh \zeta_{31} 
- msinh \xi_1 sinh \zeta_{31} cosh \beta_3 \cr
&- mcosh \xi_1 sinh \beta_1 sinh \beta_3 cos(\nu_1 - \nu_3), 
~ - mcosh \xi_1 cosh\beta_1 sinh \beta_3 cosh \zeta_{31} \cr 
& msinh \xi_1 sinh \zeta_{31} sinh \beta_3 
+ mcosh \xi_1 sinh \beta_1 cosh \beta_3 cos(\nu_1 - \nu_3), \cr
& - mcosh \xi_1 sinh \beta_1 sin (\nu_1 - \nu_3), 
~ msinh \xi_1 cosh \zeta_{31} - mcosh \xi_1 cosh\beta_1 sinh \zeta_{31} ) }
\auto
$$
Calculating in $\tilde{{\cal F}}_1$, we now obtain 
$$
P_1.Q_3~=~ mq_3~\bigl[ cosh \xi_1 sinh \zeta_{31}cosh\beta_1 - sinh \xi_1 cosh 
\zeta_{31}\bigr]
\auto\label{inv20}
$$
and in ${\cal F}_3$ (arranging terms to compare with (\ref{inv11}) )
$$
\eqalign{ P_1.P_3~=~& m^2~\bigl[ - sinh \xi_1sinh \xi_3 cosh \zeta_{31}
+ cosh \xi_1 
sinh \xi_3 sinh \zeta_{31} cosh\beta_1 \cr 
&- cosh \xi_1 cosh\xi_3 sinh \beta_1 sinh \beta_3 cos(\nu_1 - \nu_3) \cr
&+ cosh \xi_1cosh \xi_3 
cosh \zeta_{31} cosh \beta_1 cosh \beta_3  
 - sinh \xi_1 cosh\xi_3 sinh \zeta_{31} cosh \beta_3 \bigr] }
\auto\label{inv22}
$$
Comparing (\ref{inv20}) and (\ref{inv22}) with (\ref{inv01}) and
(\ref{inv11}) we see that, if we identify $cos\theta_i \leftrightarrow cosh
\beta_i = z_i$, the two sets of formulae are directly related by analytic 
continuation. All terms have changed sign as a result of 
$cosh\xi_i/sinh\xi_i ~\to ~i~sinh\xi_i/cosh\xi_i$ and $Q_i \to i~Q_i$, 
apart from that 
containing $sin\theta_1 sin\theta_2 $, which contains an extra minus sign via 
$sin \theta_i \to i~sinh \beta_i$.

In this last discussion we have effectively made the analytic continuation 
choice that the $sinh\zeta_{ij}$ do not change sign, yet we have emphasized 
that there is an overall sign ambiguity for these quantities. To see the 
significance of this ambiguity we note that (calculating in frame 
$\tilde{{\cal F}}_3$ for simplicity) 
$$
P_3.Q_1~=~ mq_1~\bigl[ cosh \xi_3 sinh \zeta_{13}cosh\beta_3 - sinh \xi_3 cosh 
\zeta_{13}\bigr]
\auto\label{inv201}
$$
and
$$
P_3.Q_2~=~ mq_2~\bigl[ cosh \xi_3 sinh \zeta_{23}cosh\beta_3 - sinh \xi_2 cosh 
\zeta_{23}\bigr]
\auto\label{inv202}
$$
where if we choose $sinh\zeta_{13}$ to be positive then we must 
choose $sinh\zeta_{23}$ to be negative. This in turn will imply that, for 
large $cosh\beta_3$, $P_3.Q_1$ is positive, while $P_3.Q_2$ is negative.
However, the part of the physical region we are discussing is completely 
symmetric with repect to $1$ and $2$. Therefore, to cover the full physical 
region, we must take both sign conventions for the $sinh\zeta_{ij}$. 
This would appear to prevent the full description of $s$-channel physical 
regions using angular variables defined by analytic continuation from the
$t_i$-channels since it implies, in particular, that we must choose both 
signs for $\lambda^{ 1 \over 2}(t_1,t_2,t_3) ~z_1$ in (\ref{inv33}).
Fortunately, as we remarked earlier, and can be seen directly from 
(\ref{inv20}),(\ref{inv22}),(\ref{inv201}), and (\ref{inv202}), 
changing the sign of the $sinh\zeta_{ij}$ is equivalent to changing the sign 
of the three $z_i= cosh \beta_i$. Therefore, to cover the  $s-t$ part of the
$s$-channel that we are discussing, using $z_i$ variables 
defined by anaytic continuation from a $t$-channel, we must use both 
$z_1,z_2,z_3, \geq 1$ and $z_1,z_2,z_3, \leq - 1$. This is a very important 
point for the discussion of dispersion theory and signature in the body of 
the paper.

Finally we consider the $s-s$ region of the same $s$-channel. In this case
the three $Q_i$ 
lie entirely in a spacelike plane so that $Q_i^2 <0 , \forall i$ and
 $|Q_i| \leq |Q_j| + |Q_k| ~~\forall
~i,j,k$. 
The ${\cal F}_i$ frames are again defined so that 
$$
Q_i  = (0,0, 0, q_i) ~~~~~~~~~~~~~~~~ 
\eqalign{ P_i & = (mcosh \xi_i, 0,0, msinh \xi_i)  \cr 
P'_i & = (- mcosh \xi_i,0,0, msinh \xi_i) }
\auto
$$
with $sinh\xi_i = q_i /2m$. However, the frame $\tilde{{\cal F}}_1$ 
is now defined so that
$$
Q_1 = (0,0,0,q_1) ~~~~~~~~~~~~~~~ 
\eqalign{ Q_2 & = (0,0,q_2sin\zeta_{21},q_2cos\zeta_{21})  \cr
Q_2 & = (0,0,q_3sin\zeta_{31},q_3cos\zeta_{31}) }
\auto
$$
where 
$$
cos\zeta_{21} = {q^2_2 + q_1^2 - q_3^2 \over 2q_1q_2} ~, ~~~~
cos\zeta_{31} = {q_3^2 + q^2_1 - q_2^2 \over 2q_1q_3} 
\auto
$$
Now there is a change of sign of $cos\zeta_{21}$ compared to the definition 
of $cosh \zeta_{21}$ in (\ref{f21}). Also the ambiguity in the choice of 
sign for the $\sin \zeta_{ij} = i \lambda^{1\over 2} (t_1,t_2,t_3)/2q_iq_j$
persists. 
The frames $\tilde{{\cal F}}_2$ and $\tilde{{\cal F}}_3$ are redefined 
analagously. 

$\tilde{{\cal F}}_i$  and ${\cal F}_i$ are again related by a Lorentz 
transformation $g_i \in SO(2,1) $. However, to proceed as in the previous 
cases, we have to use a different parametrization of $SO(2,1)$, i.e.
$$                              
g_i=u_z(\mu_i) a_y(\beta_i)a_x(\gamma_i)
~~~~~~~~~~~ -\infty<\beta_i,\,\gamma_i<\infty
~~~0\leq \mu_i\leq 2\pi
\auto
$$
where $a_x$ and $a_y$ are boosts in the $x-t$ and $y-t$ planes 
respectively. With this parametrization (provided we take $g_i$ to 
transform from ${\cal F}_i$ to $\tilde{{\cal F}}_i$) we can once again absorb 
the $u_z(\mu_i)$ 
in our definition of the frames ${\cal F}_i$ and also have the 
$a_x(\gamma_i)$ commute with the rotations $u_x(\zeta_{21})$ and 
$u_x(\zeta_{31})$.

Repeating, for a final time, the 
calculation of $P_1$ in the various frames. 
\newline In $\tilde{{\cal F}}_1$
$$
P_1 = (mcosh \xi_1cosh\beta_1 cosh \gamma_1,~ mcosh \xi_1 cosh \beta_1, 
~mcosh \xi_1 sinh \beta_1 sinh \gamma_1, 
~msinh \xi_1 ) 
\auto
$$
In $\tilde{{\cal F}}_3$
$$
\eqalign{P_1 = (&mcosh \xi_1cosh\beta_1 cosh \gamma_1, 
~mcosh \xi_1 cosh \beta_1 sinh \gamma_1, 
 ~m cos \zeta_{31} cosh \xi_1 sinh \beta_1 \cr
& - msin \zeta_{31}
sinh \xi_1, m sin \zeta_{31} cosh \xi_1 sinh \beta_1 sinh \gamma_1 - 
mcos \zeta_{31} sinh \xi_1 )} 
\auto
$$
In ${\cal F}_3$
$$
\eqalign{P_1 = (&mcosh \xi_1cosh\beta_1 cosh \beta_3 cosh (\gamma_1 - 
\gamma_3) - m sinh\beta_3 cos \zeta_{31} cosh \xi_1 sinh \beta_1 \cr
& + msinh \beta_3 sin \zeta_{31} sinh\xi_1, ~
 mcosh \xi_1 cosh \beta_1 sinh (\gamma_1 -\gamma_3),\cr
&- mcosh \xi_1cosh\beta_1 sinh \beta_3 cosh (\gamma_1 
 - \gamma_3) + m cosh\beta_3 cos \zeta_{31} cosh \xi_1 sinh \beta_1 \cr
& - mcosh \beta_3 sin \zeta_{31} sinh\xi_1,
~m sin \zeta_{31} cosh \xi_1 sinh \beta_1 - 
mcos \zeta_{31} sinh \xi_1 )} 
\auto
$$
The evaluation of invariants now gives, using $\tilde{{\cal F}}_1$, 
$$
P_1.Q_3 ~= ~mq_3~\bigl[- sin\zeta_{31} cosh\xi_1 sinh \beta_1 - cos 
\zeta_{31} sinh\xi_1 \bigr] 
\auto\label{inv31}
$$
and in ${\cal F}_3$, 
$$
\eqalign{ P_1.P_3  ~ = ~ &m^2~ \bigl[
cosh \xi_1 cosh\xi_3 cosh\beta_1 cosh \beta_3 cosh (\gamma_1 - 
\gamma_3) \cr
& - cosh \xi_1 cosh\xi_3 cos \zeta_{31}  sinh\beta_3 sinh \beta_1 
+ m  sinh\xi_1 cosh \xi_3 sin \zeta_{31} sinh \beta_3 \cr
& - cosh \xi_1sinh\xi_3 sin \zeta_{31}  sinh \beta_1 +
sinh \xi_1 sinh \xi_3 cos \zeta_{31} }
\auto\label{inv32}
$$

Now we see some more significant changes. Comparing (\ref{inv31}) with 
(\ref{inv01}) and (\ref{inv20}) we see that $cosh\beta_1$ has been replaced by 
$sinh\beta_1$ (in conjunction with $sinh\zeta_{31} \to sin\zeta_{31}$). 
We recognize that the change of sign of
$\lambda(t_1,t_2,t_3)$ produced by 
going from the $s-t$ to the $s-s$ region has, as anticipated, destroyed 
the real relationship between the $z_i$ defined in the $t_i$ channels 
and invariants of the form $P_i.Q_j$, so that now $z_i \leftrightarrow 
isinh \beta_i$. In Fig.~D3 we have shown the location
of the relevant physical regions in the $z_i$-planes, for the various values
of the $t_i$. 
\begin{center}
\leavevmode
\epsfxsize=4in
\epsffile{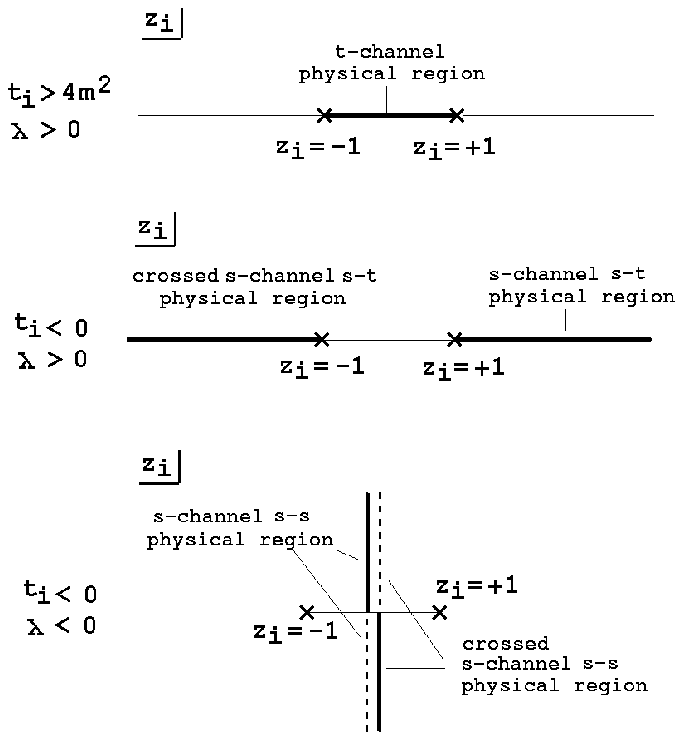}

Fig.~D3 Physical Regions in the $z_i$-planes

\end{center}
The $s-s$ part of one physical region fills the 
complete imaginary axis in each of the $z_i$-planes. However, the invariants 
also depend on $cosh\beta_i = \sqrt{z_i^2 -1}$, which should change 
sign as we go from one $s$-channel physical region to a crossed physical
region. This implies that, in the $s-s$ region, there are two physical
sheets for each $z_i$-plane separated by branch-cuts connecting the
branch-points at $z_i=\pm 1$. Crossing an incoming particle into an outgoing
particle takes us from one sheet to the other in the corresponding $z_i$
plane. Note that the same crossing can also be achieved by changing the sign
of $u_{ij}$ and $u_{ik}$, while leaving $\sqrt{z_i^2 - 1}$
unchanged. Therefore the second $z_i$-planes can alternatively be identified 
as the original $z_i$-plane but with a change of sign for $u_{ij}$ and 
$u_{ik}$. (Note that if we have chosen $u_1 = u_{31}= u^{-1}_{13}$ and 
$u_2=u_{23}$ as independent variables then changing the sign of $u_{13}$ and 
$u_{12} = u_2/ u_1$ corresponds to changing the sign of $u_1$ but not 
$u_2$.) This is important, of course, for the introduction of signature for
complex helicity continuations. Finally we note that changing the signs of
all the $sin \zeta_{ij}$ again corresponds to changing the signs of all the
$cosh\beta_i$. 

As we stated in Section 5, the asymptotic dispersion relation that we use
should be initially written in an $s-t$ region of the 
$s$-channels. It is straightforward to continue it directly to any of the 
$t_i$ channels. In the $s-s$ region it corresponds to  
using a combination of the upper and 
lower $z_i$ half-planes (from the two sheets).
Of course, that the $s-s$ physical region lies along the imaginary
$z_i$-axes is very important for discussing the phases obtained from the S-W
representation, particularly since it is only in this region that limits in
which the $u_{ij}$ are taken large (whether or not the $z_i$ are large)
are physical region limits.

\end{document}